\documentclass[a4paper,12pt]{article}
\usepackage[english]{babel}
\usepackage{jheppub2}
\usepackage{subfig}
\usepackage[T1]{fontenc} 
\usepackage{graphicx}
\usepackage{epsfig}
\usepackage{amsmath}
\usepackage{amssymb}
\usepackage{float}
\usepackage{placeins}
\usepackage{braket}
\usepackage{enumitem}
\allowdisplaybreaks[4]

\title{Low-energy theorem and OPE in the conformal window of massless QCD}

\author[a]{Marco Bochicchio}
\author[b,c]{Elisabetta Pallante}

\affiliation[a]{Physics Department, INFN Roma1, Piazzale Aldo Moro 2, 00185, Rome, Italy}
\affiliation[b]{Van Swinderen Institute for Particle Physics and Gravity, University of Groningen, 9747 AG, The Netherlands}
\affiliation[c]{Nikhef, Science Park, Amsterdam, The Netherlands}

\emailAdd{marco.bochicchio@roma1.infn.it}
\emailAdd{e.pallante@rug.nl}
\abstract{We develop a new technique, based on a low-energy theorem (LET) of NSVZ type derived in \cite{MBR}, for the nonperturbative investigation of SU($N$) QCD with $N_f$ massless quarks -- or, more generally, of massless QCD-like theories -- in phases where the beta function, $\beta(g)$, with $g=g(\mu)$ the renormalized gauge coupling, admits an isolated zero, $g_*$, in the infrared (IR) or ultraviolet (UV).
In the above phases the theory is either exactly conformal  invariant -- in the limit $g(\mu) \rightarrow g_*$ with $\mu \neq 0,+ \infty$ fixed -- or may be asymptotically conformal in the IR/UV -- as $g(\mu) \rightarrow g_*$ for $\mu \rightarrow 0^+$/$\mu \rightarrow +\infty $ with the RG-invariant scale $\Lrgi$ fixed. We point out that the LET sets constraints on 3-point correlators involving the insertion of $\Tr F^2$, its anomalous dimension $\gamma_{F^2}$, and the anomalous dimensions of multiplicatively renormalizable operators at $g_*$.
These constraints intertwine with either the exact conformal scaling or the aforementioned IR/UV asymptotics, which may or may not coincide with the IR/UV limit of the conformal scaling. Our new technical tool is the nonperturbative evaluation of the dimensionally regularized LET in the exactly conformal case. We also discuss how the LET for bare correlators is the rationale for the existence in massless QCD of the mysterious divergent contact term in the OPE of $\Tr F^2$ with itself discovered in perturbation theory in \cite{Z1,Z2} and computed to all orders in \cite{Z3}. Specifically, if $\gamma_{F^2} \neq  -3,-2,0,2,4,\cdots$, the divergent contact term in the rhs of the LET for the $2$-point correlator of $\Tr F^2$ has to match -- and we verify by direct computation that it actually does -- the divergence in the lhs due to the nontrivial anomalous dimension of $\Tr F^2$. Hence, remarkably, the additive renormalization due to the divergent contact term in the rhs is related by the LET to the multiplicative renormalization in the lhs, in such a way that a suitably renormalized version of the LET has no ambiguity for additive renormalization.}

\DeclareMathOperator{\Tr}{Tr}
\newcommand{\be}{\begin{equation}}
\newcommand{\ee}{\end{equation}}
\newcommand{\nn}{\nonumber}
\newcommand{\bea}{\begin{eqnarray}}
\newcommand{\eea}{\end{eqnarray}}
\newcommand{\bfig}{\begin{figure}}
\newcommand{\efig}{\end{figure}}
\newcommand{\bc}{\begin{center}}
\newcommand{\ec}{\end{center}}

\newcommand{\f}[2]{\frac{#1}{#2}}

\newcommand{\eps}{{\epsilon}}
\newcommand{\cO}{{\mathcal O}}
\newcommand{\cF}{{\mathcal F}}
\newcommand{\td}{{\tilde{d}}}
\newcommand{\tg}{{\tilde{g}}}
\newcommand{\tD}{{\tilde{\Delta}}}
\newcommand{\dr}{\text{dim.\,reg.}}
\newcommand{\Lrgi}{\Lambda_{\scriptscriptstyle{IR/UV}}}

\begin{document}
\maketitle
\flushbottom

\section{Introduction and physics motivations} \label{1}

In the present paper we develop a new technique for the nonperturbative investigation of SU($N$) QCD with $N_f$ massless quarks -- or, more generally, of massless QCD-like theories\footnote{By massless QCD-like theories we mean YM theories in $d=4$ dimensions that are massless to all perturbative orders.} -- in phases where the beta function, $\beta(g)$, with $g=g(\mu)$ the renormalized gauge coupling, admits an isolated\footnote{Of course, our methods also extend with minor modifications to the cases -- such as $\mathcal{N}=4$ SUSY YM theory -- where the beta function identically vanishes.} zero, $g_*$, possibly nontrivial, $g_*>0$, in the infrared (IR) or ultraviolet (UV).
In the above phases the theory is either exactly conformal -- in the limit $g(\mu) \rightarrow g_*$ with $\mu \neq 0,+ \infty$ fixed -- or may be asymptotically conformal in the IR/UV -- as $g(\mu) \rightarrow g_*$ for $\mu \rightarrow 0^+$/$\mu \rightarrow +\infty $ with the renormalization-group (RG) invariant scale $\Lrgi$ fixed.\par
Our approach is based on the interplay between conformal Yang-Mills (YM) theories in $d=4$ dimensions and a recently derived low-energy theorem (LET) \cite{MBR} of Novikov-Shifman-Vainshtein-Zakharov (NSVZ) type -- in not necessarily conformal YM theories --
that relates $n$-point correlators in the lhs to $n+1$ point correlators with the extra insertion of $\Tr F^2$ at zero momentum in the rhs.\par
Indeed, we point out that, while in abstract conformal field theories (CFTs) the data that determine $2$-point and $3$-point correlators are independent a priori, in $4$-dimensional conformal YM theories they are actually related by the LET above. \par
Therefore, our approach is threefold:\par
Firstly, we work out quantitatively the relation between $2$-point correlators and $3$-point correlators with the insertion of $\Tr F^2$ at zero momentum implied by the LET, both for the bare correlators and -- in massless QCD -- the renormalized ones. \par
Secondly, we study the physics implications of the LET in phases where the beta function admits a nontrivial zero in massless QCD or in massless QCD-like theories. They include both the conformal window and the asymptotically safe phase\footnote{The conformal window identifies an interval for $N_f$, where an IR zero of the beta function exists, with $N_f^{AF}$ its upper edge: For $N_f\geqslant N_f^{AF}$ the theory is no longer asymptotically free in the UV but it is so in the IR, and it may admit hypothetically a nontrivial UV zero of the beta function corresponding to a conjectural asymptotically safe phase.} -- if it exists -- of massless QCD or massless QCD-like theories. \par
In the conformal window, the beta function admits a nontrivial zero $g_*>0$ in the IR, in addition to the trivial zero at vanishing coupling in the UV. For a generic RG-flow trajectory with $g(\mu) \neq g_*$ labelled by $\Lambda_{\scriptscriptstyle{UV}}$, the theory is asymptotically free (AF) in the UV and, naively, asymptotically conformal in the IR, since $g(\mu) \rightarrow 0^+$ for $\mu \rightarrow +\infty$ and $g(\mu) \rightarrow g_*^{-}$ for $\mu \rightarrow 0^+$. Moreover, if instead the limit $g(\mu) \rightarrow g_*$ is taken for any $\mu \neq 0,+ \infty$ fixed, the theory is exactly conformal.\par
Similarly, in the asymptotically safe phase -- if it exists -- the theory is generically asymptotically free in the IR, naively asymptotically conformal in the UV, and exactly conformal if 
$g(\mu) \rightarrow g_*$ for any $\mu$ fixed.\par
By evaluating the LET for $2$-point correlators of scalar multiplicatively renormalizable operators in the lhs, we discover constraints -- on the corresponding $3$-point correlators with the insertion of $\Tr F^2$ in the rhs, its anomalous dimension $\gamma_{F^2}$, and the anomalous dimensions of the operators at $g_*$ --
that intertwine with the exact conformal symmetry. We will study elsewhere the LET in the asymptotically conformal case.\par
Relatedly, we work out the constraints from the LET in massless QCD on the existence of a nontrivial UV zero of the beta function in addition to the nontrivial IR zero in the conformal window and the possible merging of the two aforementioned zeroes.\par
Thirdly, in the above cases we also investigate the relation between the LET and the OPE, with special care about the occurrence in the rhs of the LET of contact terms, which have emerged in its previous applications that we recall in the following. \par
The LET applies to YM theories in $d=4$ dimensions -- not necessarily conformal -- and has two versions \cite{MBR}, one that involves the logarithmic derivative with respect to the gauge coupling in the lhs, and one that involves the logarithmic derivative with respect to the RG-invariant scale, $\Lambda_{\scriptscriptstyle{UV}}$, in YM theories AF in the UV.\par
The original motivation \cite{MBR} for working out the LET has been the study of the nonperturbative renormalization properties of large-$N$ confining massless QCD-like theories and, specifically, massless QCD \cite{H,V,Migdal,W}, as the second version of the LET -- combined with the nonperturbative renormalization properties \cite{MBR} of $\Lambda_{\scriptscriptstyle{QCD}}$ in the large-$N$ 't Hooft \cite{H} and Veneziano \cite{V} expansions -- controls \cite{MBR} the structure of the nonperturbative counterterms for the YM action in such expansions. \par
Moreover, the first version of the LET clarifies how the open/closed string duality may be implemented  in canonical  string models \cite{MBL} realizing perturbatively massless QCD-like theories.\par
In fact, in massless QCD-like theories an essential tool \cite{MBL,BB} to actually compute the rhs of the LET, corresponding to a $2$-point correlator of a multiplicatively renormalizable operator $O$ in the lhs, is the OPE of $O$ with $\Tr F^2$, 
both in perturbation theory \cite{MBL,BB} and in its RG-improved form \cite{MBL,BB}. \par
Finally, the second version of the LET furnishes an obstruction \cite{MBL} to the existence of canonical string models satisfying the open/closed string duality that would realize nonperturbatively 
the large-$N$ 't Hooft expansion of confining massless\footnote{The obstruction extends \cite{MBL} to the large-$N$ 't Hooft expansion of massive QCD and massive $\mathcal{N}=1$ SUSY QCD in the confining phase.} QCD-like theories. \par
Because of its importance, the LET has been verified in massless QCD for $O=\Tr F^2$ by comparing the scheme-independent divergences that occur in the lhs to order $g^2$ \cite{MBL} and $g^4$ \cite{BB} in perturbation theory \cite{Kataev,Z1,Z2,Z3} with the rhs evaluated \cite{MBL,BB} by means of the perturbative OPE \cite{Kataev,Z1,Z2,Z3} -- that includes the aforementioned contact terms \cite{Z1,Z2,Z3} -- of $\Tr F^2$ with itself. \par
Besides, it has been pointed out \cite{BB} that the logic above can be inverted and, by assuming the LET for the $2$-point correlator of $O$, the relevant contribution to order $g^2$ in the rhs of the LET for the OPE of $O$ with $\Tr F^2$ can be recovered \cite{BB} from the anomalous dimension of $O$ in the lhs. \par
In this respect, we should also mention the treatment of the OPE in \cite{F} and references therein that is closely related to the LET, but actually independent of the approach in \cite{MBR,BB} that is pursued in the present paper. \par
By summarizing, in the present paper we exploit the specific features of the LET in conformal YM theories in $d=4$ dimensions.\par
In the exactly conformal case $2$- and $3$-point scalar correlators are completely fixed -- up to contact terms -- by the conformal symmetry up to the overall normalization.
Therefore the rhs of the LET, which involves the $3$-point correlator at the zero of the beta function, may be computed nonperturbatively a priori. Yet, since the lhs of the LET involves the derivative of the $2$-point correlator at the zero, the ansatz for the general solution of the Callan-Symanzik (CS) equation in a neighborhood of the zero should be employed in the lhs. Then, the LET implies a number of physically relevant relations between the contributions to the lhs and rhs -- including the contact terms in the rhs evaluated by the OPE -- that we report in the following.

\section{Main results and conclusions} \label{2}

\subsection{Regularization, renormalization and LET} \label{2.1}

In the original formulation \cite{MBR} the first version (section \ref{1}) of the LET, with the canonical normalization (section \ref{4}) of the YM action \cite{BB}, involves the bare\footnote{Throughout the present paper we denote bare quantities by the subscript $_0$.} correlators and coupling constant, $g_0$, in $d=4$ Euclidean space-time:
\bea
\label{eq:LETCcoord}
&&\sum^{k=n}_{k=1} c_{O_k} \braket{O_1\cdots O_n}_0 +
\f{\partial}{\partial\log g_0} \braket{O_1\cdots O_n}_0 \nn\\
&&= \f{1}{2}\,\int\, \braket{O_1\cdots O_n F^2(x)}_0 
- \braket{O_1\cdots O_n}_0\braket{F^2(x)}_0\, d^4x
\eea
with $F^2 \equiv 2\Tr F^2$.
In fact, we only consider the applications of the LET for $n=2$ and $O_1=O_2=O$:
\bea \label{LET2}
&&2 c_{O} \braket{O(z)  O(0)}_0 +
\f{\partial}{\partial\log g_0} \braket{O(z) \cdots O(0)}_0 \nn\\
&&= \f{1}{2}\,\int\, \braket{O(z) O(0) F^2(x)}_0 
- \braket{O(z) O(0)}_0\braket{F^2(x)}_0\, d^4x
\eea
In general, the bare correlators are divergent and need regularization and, eventually, renormalization. Therefore, the first issue to apply nonperturbatively the LET to conformal YM theories is to find a regularization of the bare correlators and, correspondingly, of the LET. \par
Dimensional regularization in $\tilde d=4-2\epsilon$ dimensions is gauge invariant to all orders of perturbation theory. Besides, a version of the Callan-Symanzik (CS) equation (appendix \ref{CS1}) holds nonperturbatively in dimensional regularization allowing us to write down the ansatz for its solution (appendix \ref{CS1}):
\bea
\label{CSStilde00}
 \braket{O(z)O(0)}_{\tilde d=4-2\epsilon}=\f{\mathcal{ G}^{(O)}_2(\tilde g(z))}{|z|^{2 \tilde \Delta_{O_0}}}  Z^{(O)2}(\tilde g(z),g(\mu))
 \eea
in a neighborhood of an isolated zero of the beta function that is necessary to compute the logarithmic derivative in the lhs of the LET. 
Therefore, dimensional regularization is our natural choice.\par
The correlator in $\tilde d=4-2\epsilon$ dimensions differs from the correlator in $d=4$ dimensions (appendix \ref{CS0}):
\bea
\label{CSS}
 \braket{O(z)O(0)}_{d=4}=\f{\mathcal{G}^{(O)}_2(g(z))}{|z|^{2 \Delta_{O_0}}} Z^{(O)2}(g(z),g(\mu))
\eea
 by terms that vanish as $\epsilon \rightarrow 0$ (appendix \ref{CS1}).
 Hence, the LET for bare correlators in dimensional regularization reads (section \ref{5}):
\bea
\label{eq:LETCuv}
&&\sum^{k=n}_{k=1} c_{O_k} \braket{O_1\cdots O_n}_0 +
\f{\partial}{\partial\log {g}_0} \braket{O_1\cdots O_n}_0 \bigg|_{\dr}\nonumber \\
&&= \f{1}{2} \int\, \braket{O_1\cdots O_n F^2(x)}_0 - \braket{O_1\cdots O_n}_0\braket{F^2(x)}_0\, d^\td x \bigg|_{\dr}
\eea
where the integral over space-time in the rhs is defined by analytic continuation.\par
Given $\gamma_{F^2}=\gamma_{F^2}(g_*)$, with $g_*$ the nontrivial zero of the beta function in $d=4$ dimensions in the IR, we verify 
by explicit computation in the conformal window of massless QCD (section \ref{10}) the implications of the LET in dimensional regularization for $O=F^2$ in the exactly conformal case.\par
In dimensional regularization for the exceptional values\footnote{We exclude from the exceptional values the ones exceeding the unitarity bound (appendix \ref{B}).} $\gamma_{F^2}=-3,-2,0,2,4 \cdots$ the space-time integral in the rhs of the LET diverges as $\epsilon \rightarrow 0$ and its evaluation involves some subtle issues of computational nature (section \ref{2.1A}). \par
Hence, as an aside, for the exceptional values of $\gamma_{F^2}$ we adapt to the LET the conformal-invariant $(u,v_{F^2})$ scheme \cite{Skenderis} (appendix \ref{E}) that combines dimensional regularization in $\tilde d= 4+2u\epsilon$ dimensions with a deformation, $\tilde \Delta_{F^2_0} = 4+ (u+v_{F^2})\epsilon$, of the canonical dimension, $\Delta_{F^2_0}=4$, of the operator $F^2$, with $u,v_{F^2}$ real parameters.
For $u=v_{F^2}$ the $(u,v_{F^2})$ scheme reduces to dimensional regularization in $\tilde d= 4+2u\epsilon$ dimensions.
However, for $u \neq v_{F^2}$ gauge invariance is not obvious. In perturbation theory up to order $g^2$, where the theory is conformal for any $g$, we are able to reliably compute the lhs and rhs of the LET only in the $(0,v_{F^2})$ scheme that we actually find to be consistent with the LET (appendix \ref{E}).\par
Anyway, all of our physics results (section \ref{2.4}) and conclusions (section \ref{2.5}) are only based on the LET in the gauge-invariant dimensional regularization for nonexceptional $\gamma_{F^2}$ and, in the marginal case, by restricting the LET to $O=F^2$, where it turns out that the coefficient of the divergent space-time integral in the rhs of the LET actually vanishes, so that we do not encounter the subtle computational issues alluded to above.\par
In the lhs of the LET we may choose the operators at different points.
Hence, in the lhs we may ignore the possible occurrence of contact terms.
This is not the case for the rhs, where the insertion of $F^2$ is actually integrated over all space-time and, consequently, contact terms cannot be ignored.
Finally, in order to write down the renormalized version of the LET, it is necessary to express the bare correlators in terms of the renormalized ones. Specifically, by assuming that $F^2$ and $O$ are gauge-invariant multiplicatively renormalizable operators up to the mixing with operators that vanish by the equations of motion, we get from eq. \eqref{LET2} (section \ref{5}):
\bea
&&\braket{O(z)O(0)}\left( 2c_{O} +\f{\partial\log Z_O^{-2}}{\partial\log g_0}
+\f{\partial\log \braket{O(z)O(0)}}{\partial\log g_0}
\right)  \bigg|_{\dr}
\nonumber \\
&&\hspace{0.5truecm}=\f{1}{2}Z_{F^2}^{-1}\int \braket{O(z) O(0) F^2(x)}' - \braket{O(z)O(0)}\braket{F^2(x)} d^\td x  \bigg|_{\dr} \nonumber \\
&&\hspace{0.8truecm}+\f{1}{2} Z_O^{2} \int \braket{O(z)O(0)F^2(x)}_{0;\textrm{c.t.}} \, d^\td x  \bigg|_{\dr}
\eea
where $'$ denotes the correlator without the contact terms, while $0;\textrm{c.t.}$ denotes the contribution of the contact terms to the bare correlator, and (section \ref{5} and appendix \ref{C1}):
 \bea\label{eq:ZFZg}
 Z_{F^2}^{-1}(g,\eps) = 1-\f{\beta(g)}{\eps g}
 \eea
 
 \subsection{LET in conformal YM theories}  \label{2.1A}
 
To further proceed, we assume that $g=g(\mu) \rightarrow g_*$ with fixed $\mu$, so that the theory is exactly conformal in the above limit (appendix \ref{D}).
Correspondingly, the $2$-point correlator reduces to the conformal result (appendix \ref{D}):
\bea \label{co}
\f{\mathcal{G}^{(O)}_2(g(z))}{|z|^{2 \Delta_{O_0}}} Z^{(O)2}(g(z),g(\mu))\bigg|_{g \rightarrow g_*}
&=&\f{N_2(g_*)  \mu^{2\Delta_{O_0}-2\Delta_O}  }{|z|^{2\Delta_O}}
\eea
with $|z|=\sqrt {z^2}$ and $\Delta_{O_0}$, $\Delta_O=\Delta_{O_0}+\gamma_O$ the canonical and conformal dimensions of $O$ at $g_*$ respectively, thus implying (appendices \ref{CS0} and \ref{D}):
\bea
\mathcal{G}^{(O)}_2(g_*)=N_2(g_*)
\eea
The evaluation of the lhs of the LET for scalar operators $O$ by means of eq. \eqref{CSStilde00} yields (section \ref{7}):
\bea \label{lhsepsilon0}
\mbox{lhs} &=& \braket{O(z)O(0)}_{\tilde d=4-2\epsilon} \left( 2c_O -\f{2\gamma_O(g)}{\eps}+ \right. \nn\\
&& \left.   - \f{\beta(g,\epsilon)}{\eps} \left ( \f{\partial \log \mathcal{G}^{(O)}_2}{\partial \tilde g(z)}     \f{\beta(\tilde g(z),\epsilon)}{\beta(g,\epsilon)}   
  + 2    \f{\gamma_O(\tilde g(z)) - \gamma_O(g)}{\beta(g,\epsilon)} \right ) \right)\bigg|_{g \rightarrow g_*} \nn\\
  &=& \braket{O(z)O(0)}_{\tilde d=4-2\epsilon} \left( 2c_O -\f{2\gamma_O(g)}{\eps} \right. \nn\\
&& \left. - \left ( \f{\partial \log \mathcal{G}^{(O)}_2}{\partial \tilde g(z)}     \f{\beta(\tilde g(z),\epsilon)}{\epsilon}   
  + 2    \f{\gamma_O(\tilde g(z)) - \gamma_O(g)}{\epsilon} \right ) \right)\bigg|_{g \rightarrow  g_*}\nn\\
\eea
It turns out that the only term that may be divergent (section \ref{7}) as $\epsilon \rightarrow 0$ in the equation above is the second one in the rhs, in such a way that:
\bea
\label{CSStilde0}
-\f{ \gamma_O(g)}{\epsilon }  \braket{O(z)O(0)}_{\tilde d=4-2\epsilon}\bigg|_{g \rightarrow  g_*}=-\f{ \gamma_O(g)}{\epsilon}  \braket{O(z)O(0)}_{d=4}+ \mbox{finite terms} + \cdots\bigg|_{g \rightarrow  g_*} 
\eea
where the terms in the dots vanish as $\epsilon \rightarrow 0$, while the finite terms may only occur if $\gamma_O(g_*) \neq 0$ due to the mismatch of order $\epsilon$ between the correlators in $\tilde d=4-2\epsilon$ and $d=4$ dimensions.
Instead, in all the finite terms in eq. \eqref{lhsepsilon0} we may replace the correlator in $d=4$ dimensions with the one in $\tilde d=4-2\epsilon$ dimensions as $\epsilon \rightarrow 0$. \par
The evaluation of the rhs of the LET employs in the conformal limit:
 \bea
 \label{eq:CFT2300}
 \braket{O(z)O(0)F^2(x)}_{d=4}\bigg|_{g \rightarrow g_*}&=&\f{N_3(g_*) \mu^{2\Delta_{O_0}+\Delta_{F^2_0}-2\Delta_O - \Delta_{F^2}}   }{|z|^{2\Delta_O -\Delta_{F^2}} |x|^{\Delta_{F^2}}|x-z|^{\Delta_{F^2}}} 
\eea
up to contact terms, with $\Delta_{F^2_0}= 4$ the canonical and $\Delta_{F^2}=4+\gamma_{F^2}$ the conformal dimension of $F^2$ at $g_*$ and (appendix \ref{C}):
  \bea\label{eq:gamF}
 \gamma_{F^2}(g)=\beta'(g)-\f{\beta(g)}{g}
 \eea
where in the above equation $'$ is the partial derivative with respect to $g$. As a consequence the rhs of the LET reads:
\bea\label{eq:rhsRentilde0}
\mbox{rhs}=&&\f{1}{2}  Z_{F^2}^{-1} \int  
\f{N_3(g) \mu^{2\Delta_{O_0}+\Delta_{F^2_0}-2\Delta_O - \Delta_{F^2}}  }{|z|^{2\Delta_O -\Delta_{F^2}} } 
\f{1}
{|x|^{\Delta_{F^2}}|x-z|^{\Delta_{F^2}} } +\cdots d^\td x \bigg|_{g \rightarrow g_*}\nn\\
&&+\f{1}{2} Z_O^{2} \int \braket{O(z)O(0)F^2(x)}_{\tilde d=4-2\epsilon; 0;\textrm{c.t.}} \, d^\td x\bigg|_{g \rightarrow  g_*}
 \eea
where we skip in the dots terms that vanish as $\epsilon\rightarrow 0$.
The evaluation of the space-time integral in eq. \eqref{eq:rhsRentilde0}
yields:  
 \bea\label{eq:rhs230}
&&\f{1}{2}Z_{F^2}^{-1} \int  
\f{N_3(g) \mu^{2\Delta_{O_0}+\Delta_{F^2_0}-2\Delta_O - \Delta_{F^2}}  }{|z|^{2\Delta_O -\Delta_{F^2}} } 
\f{1}
{|x|^{\Delta_{F^2}}|x-z|^{\Delta_{F^2}} }  d^\td x \bigg|_{g \rightarrow g_*}\nn\\
&&=
\f{\mu^{2\Delta_{O_0}-2\Delta_O}  }{|z|^{2\Delta_O} } Z_{F^2}^{-1}\,
\f{1}{2} 
\f{N_3(g) \mu^{\Delta_{F^2_0}- \Delta_{F^2}}  }{|z|^{-\Delta_{F^2}} }(2\pi)^\td\, C_{\td,\f{\Delta_{F^2}}{2},\f{\Delta_{F^2}}{2}} |z|^{\td-2\Delta_{F^2}}\bigg|_{g \rightarrow g_*} \nn\\
&&=\braket{O(z)O(0)}_{d=4} Z_{F^2}^{-1}\,
\f{1}{2}\f{N_3(g)}{N_2(g)}
(2\pi)^\td\, C_{\td,\f{\Delta_{F^2}}{2},\f{\Delta_{F^2}}{2}} |z|^{-\gamma_{F^2}-2\epsilon} \mu^{-\gamma_{F^2}} \bigg|_{g \rightarrow  g_*}
\nn\\
\eea 
 where (appendix \ref{B}):
 \bea\label{eq:I}
I_{d,\Delta_1,\Delta_2}&\equiv& \int  \f{1}
{ |x|^{\Delta_1}|x-z|^{\Delta_2} }     d^dx \nn\\
&=&(2\pi)^d \,C_{d,\f{\Delta_1}{2},\f{\Delta_2}{2}}\, |z|^{d-\Delta_1-\Delta_2}
\eea
and:
\be
C_{d,\f{\Delta_1}{2},\f{\Delta_2}{2}}=\f{ 
\Gamma\big( \f{\Delta_1+\Delta_2-d}{2}\big)  
\Gamma\big( \f{d-\Delta_1}{2}\big)  \Gamma\big( \f{d-\Delta_2}{2}\big) 
}
{(4\pi)^{\f{d}{2}}\Gamma \big(\f{\Delta_1}{2}\big)\Gamma \big(\f{\Delta_2}{2}\big) \Gamma \big(d-\f{\Delta_1}{2}-\f{\Delta_2}{2}\big)}
\ee
The UV and IR properties of the integral in eq. \eqref{eq:rhs230}
only depend on the scaling dimension $\Delta_{F^2}=\Delta_{F^2_0}+\gamma_{F^2}$ and, specifically, on the anomalous dimension $\gamma_{F^2}$. In particular, the integral in eq. \eqref{eq:rhs230} is finite as $\epsilon \rightarrow 0$ for generic values of $\gamma_{F^2}$,
while it is divergent for the exceptional values $\gamma_{F^2}=-3,-2,0,2,4,\cdots$ (appendix \ref{B}).\par
If the integral is finite, its value as $\epsilon \rightarrow 0$ is independent of the correction of order $\epsilon$ to the scaling dimensions originating from the regularized integral defined in $\tilde d=4-2\epsilon$ dimensions (appendix \ref{B}). Therefore we expect that the terms in the dots in the integral in eq. \eqref{eq:rhsRentilde0}, which include nonconformal corrections of order $\epsilon$ as well, do not affect the limit of the rhs of the LET as $\epsilon \rightarrow 0$. \par
Yet, if the integral diverges as $\epsilon \rightarrow 0$, the divergence is sensitive to corrections of order $\epsilon$ to the scaling dimensions (appendix \ref{B}). Though these corrections to the scaling dimensions are known (appendix \ref{CS1}), by consistency also the nonconformal corrections of order $\epsilon$, which are not explicitly known a priori, should be included in the integrand. Therefore in the divergent case the exact evaluation of the integral involves a subtle computational issue (section \ref{7}).
However, the aforementioned nonconformal corrections of order $\epsilon$ are absent in the $(0,v_{F^2})$ scheme up to order $g^2$ in perturbation theory because in this scheme the theory is defined in $d=4$ dimensions and conformal for any $g$ (appendix \ref{E}).\par
The contribution of the contact terms (section \ref{2.3}) to the rhs of the LET is evaluated in dimensional regularization as $\epsilon \rightarrow 0$ by the coefficient:
\bea \label{OPE000}
{\mathcal{C}}_{O_0}(g,\eps)={\mathcal{C}}_{O_0,\textrm{div}}(g,\eps) +{\mathcal{C}}_{O_0,\textrm{finite}}(g,\epsilon)
\eea
of the delta function $\delta^{(4)}(x)$ in the short distance OPE (section \ref{6}):
\bea \label{OPE0}
&&\f{1}{2}  Z_O^{2}
\int \braket{O(z)O(0)F^2(x)}_{\tilde d=4-2\epsilon;0;\textrm{c.t.}} \, d^\td x  \nn\\
&&=  {\mathcal{C}}_{O_0}(g,\eps) \braket{O(z)O(0)}_{\tilde d=4-2\epsilon} + \cdots \nn\\
&&=  {\mathcal{C}}_{O_0}(g,\eps) \braket{O(z)O(0)}_{d=4} + \mbox{finite terms} +\cdots 
\eea
where the dots represent terms that vanish as $\epsilon \rightarrow 0$. The finite terms above only arise if ${\mathcal{C}}_{O_0}(g,\eps)$ contains the divergent contribution ${\mathcal{C}}_{O_0,\textrm{div}}(g,\eps)$ because of the mismatch of order $\epsilon$ between the correlators in $\tilde d=4-2\epsilon$ and $d=4$ dimensions. \par
Importantly, for $\gamma_{F^2}$ generic the space-time integral in the rhs of the LET is finite, so that the possible divergences in eqs. \eqref{CSStilde0} and \eqref{OPE0} must match. \par
Hence, the corresponding finite terms must match as well, so that they may be subtracted on both sides of the LET, which is then written only in terms of the correlators in $d=4$ dimensions as $\epsilon \rightarrow 0$, with no contribution of the aforementioned finite terms.
Thus, we analyze separately the following three cases:\par
 (I)  $\gamma_{F^2}=0$ in $d=4$ dimensions at $\beta(g_*)=0$.\par
 (II) $\gamma_{F^2}\neq 0$ nonexceptional in $d=4$ dimensions at $\beta(g_*)=0$. \par
 (III) $\gamma_{F^2}$ of order $\eps$ in $\td=4-2 \epsilon$ dimensions at $\beta(\tilde g_*,\eps)=0$.  \par

\subsection{Exact conformal scaling versus IR/UV asymptotics}  \label{2.2}

Depending on whether either  $g(\mu) \rightarrow g_*$ with $\mu \neq 0,+ \infty$ fixed or 
$g(\mu) \rightarrow g_*$ for $\mu \rightarrow 0^+$/$\mu \rightarrow +\infty $ with the renormalization-group (RG) invariant scale $\Lrgi$ fixed, the massless QCD-like theory either is exactly conformal or may be asymptotically conformal in the IR/UV (appendix \ref{D}).
Then, two questions naturally arise:\par
Firstly, whether the theory is actually asymptotically conformal (appendix \ref{D}) in the IR/UV for $\Lrgi$ fixed. \par
Secondly, should the answer to the above question be affirmative, whether (section \ref{2.4}) the IR/UV limit of the exactly conformal theory coincides with the aforementioned IR/UV conformal asymptotics, up to perhaps the overall normalization of the $2$- and $3$-point correlators. \par
In relation to the questions above, the main counterexample is a massless AF theory for $g_*=0$ in the UV.
While the theory is exactly conformal and free in the limit $g(\mu) \rightarrow 0$ for fixed $\mu$:
\bea \label{AA}
\langle O(z) O(0) \rangle &=& \f{\mathcal{G}^{(O)}_2(0)}{z^{2D}}
\eea
with $D$ the canonical dimension of the operator $O$,
the CS equation implies that for $ \Lambda_{\scriptscriptstyle{UV}}$ fixed the conformal scaling of the $2$-point correlators above
is multiplicatively corrected asymptotically in the UV by fractional powers of logarithms (appendix \ref{D}):
\bea \label{BB}
\langle O(z) O(0) \rangle &\sim& \f{\mathcal{G}^{(O)}_2(0)}{z^{2D}}
\left( \f{g(z)}{g(\mu)} \right)^{\f{2\gamma_0^{(O)}}{\beta_0}}
Z^{(O)\prime 2}(g(\mu)) \nn\\
&\sim& \f{\mathcal{G}^{(O)}_2(0)}{z^{2D}} 
\left( \f{1}{ g^2(\mu)2 \beta_0 \log( |z |  \Lambda_{\scriptscriptstyle{UV}})}\right)^{\f{\gamma_0^{(O)}}{\beta_0}}
Z^{(O)\prime 2}(g(\mu)) 
\eea
for operators with nonzero anomalous dimension,
so that the theory is not asymptotically conformal in the UV. Thus, in general, a theory needs not be asymptotically conformal for $\Lrgi$ fixed.\par
In fact, we answer the first question by demonstrating that for $\Lrgi$ fixed in the marginal case $\gamma_{F^2}=0$ (section \ref{I}) the theory is not asymptotically conformal in the IR/UV (appendix \ref{D}), as it is not in the AF case in the UV, while in the irrelevant case, $\gamma_{F^2}>0$ in the IR or $\gamma_{F^2}<0$ in the UV (section \ref{II}), the theory is asymptotically conformal in the IR/UV (appendix \ref{D}). \par
We also answer the second question by demonstrating that in the latter case the IR/UV asymptotics differs from the exactly conformal asymptotics
only by the normalization of the corresponding correlators (appendix \ref{D}).

\subsection{$3$-point correlators at zero momentum and contact terms} \label{2.3}

The conformal $3$-point correlators in the coordinate representation at different points in $d=4$ dimensions may be extended to the momentum representation by means of their Fourier transform or directly solving the conformal Ward identities in the momentum representation \cite{Skenderis}.\par
By limiting ourselves to the $3$-point correlator $\langle O O F^2 \rangle$ that occurs in the rhs of the LET, its Fourier transform exists for generic values of $\gamma_{F^2}$, but it is UV divergent\footnote{This divergence is classified in \cite{Skenderis} as semilocal.} for the exceptional values $\gamma_{F^2}=0,2,4 \cdots$ \cite{Skenderis} (appendix \ref{B}). \par 
Moreover, for the exceptional values $\gamma_{F^2}=-3,-2$, though the Fourier transform exists for generic values of the momentum, the zero momentum insertion of $F^2$ is actually IR divergent (appendix \ref{B}). In the theory regularized in $\td=4-2\epsilon$ dimensions the aforementioned UV and IR divergences manifest themselves as $\frac{1}{\epsilon}$ poles as $\epsilon \rightarrow 0$ (appendix \ref{B}). \par
Besides, in $d=4$ dimensions contact terms may arise as conformal anomalies in $2$- and $3$-point renormalized correlators \cite{Skenderis}.\par
In fact, in our concrete YM framework, the canonical dimension of the bare operator $F^2_0$ is always $4$, so that a $\delta^{(4)}$ may occur in the bare OPE with $F^2_0$ independently of the conformal dimension of the renormalized operator $F^2$. The contribution of such a contact term may be evaluated by the OPE in eq. \eqref{OPE0}, under the standard assumption that $O$ and $F^2$ belong to a complete orthonormal family of primary conformal  operators\footnote{This is a restriction on the mixing of $F^2$ with other dimension $4$ operators.}, since in this case only the term displayed in eq. \eqref{OPE0} survives. \par
Indeed, in massless QCD for $\gamma_{F^2}$ nonexceptional at the nontrivial zero, the LET requires that such a $\delta^{(4)}$ contact term should be present in general and, surprisingly, should be divergent for $\gamma_{O}\neq 0$ (section \ref{II}). \par
Remarkably, in massless QCD the occurrence of such a divergent contact term in the OPE of $F^2$ with itself\footnote{For $O=F^2$ in massless QCD, the only other operator that may contribute to the OPE in addition to $F^2$ itself is the quark Lagrangian density $\bar\psi \gamma_{\mu}D_{\mu} \psi$ that vanishes by the equations of motion, so that eq. \eqref{OPE0} applies without further assumptions, since the $2$-point correlator of $F^2$ with $\bar\psi \gamma_{\mu}D_{\mu} \psi$ is a contact term that vanishes for $z \neq 0$ in eq. \eqref{OPE0}.} has been discovered in \cite{Z1} and \cite{Z2} respectively to order $g^4$ and $g^6$ in perturbation theory and computed in a closed form (section \ref{II}) in terms
of the beta function and its first derivative in \cite{Z3} independently of the existence of a nontrivial zero of the beta function. \par
By explicit evaluation (section \ref{II}), the aforementioned divergent contact term survives in the limit $g \rightarrow g_*$ provided that $\gamma_{F^2} \neq 0$. \par Hence, the LET is the rationale for its existence in phases of massless QCD where the beta function has a nontrivial zero.
Yet, in \cite{Z3} it has been argued that it should be renormalized to zero. While in general this is incompatible with the LET, we determine the precise sense by which it is actually renormalized to zero -- somehow according to \cite{Z3} -- in a suitably renormalized version (section \ref{II}) of the LET in the conformal window of massless QCD.

\subsection{Physics implications of the LET} \label{2.4}

We evaluate the LET in the limit $g \rightarrow g_*$ in the cases (I), (II) and (III) (section \ref{2.1}), by assuming the exact conformal invariance, working out the corresponding constraints. 

\subsubsection{LET for $\gamma_{F^2}=0$ in $d=4$} \label{I}
 
By assumption $F^2$ is marginal at $g=g_*$. By eq. \eqref{eq:gamF}:
 \bea \label{gamma0}
 \gamma_{F^2}=\beta'(g_*)=0
\eea
In the marginal case, the theory is not asymptotically conformal for $\Lrgi$ fixed (section \ref{2.2}). Hence, we only consider the exactly conformal case and we choose $O=F^2$ in order to avoid the aforementioned computational issues in the rhs of the LET. Indeed, since $\gamma_{F^2}=0$ there is no divergence in eq. \eqref{CSStilde0}. Moreover, by direct computation and for the same reason (eq. \eqref{CC}), there is no divergent contact term in eq. \eqref{OPE0}. 
As a consequence, also the term involving the space-time integral in the rhs must be finite as $\epsilon \rightarrow 0$.
Yet, since the space-integral is actually divergent in the marginal case (section \ref{2.2}), the only possibility is that its coefficient $N_3(g_*)$ vanishes.
Therefore:
\bea \label{marginal}
\braket{F^2(z)F^2(0)F^2(x)}_{d=4} \bigg|_{g \rightarrow g_*}
=0
\eea
for $x\neq z \neq 0$ and $x \neq 0$.
Remarkably, eq. \eqref{marginal} coincides with the consistency condition \cite{Kadanoff, Kadanoff2} for the infinitesimal perturbation by a marginal operator at $g_*$.
Then, because of eq. \eqref{marginal} and the explicit evaluation of the lhs (section \ref{7.1}), the LET reads as $\epsilon \rightarrow 0$: 
 \bea
 \label{eq:LET_LR_0b}
&& \braket{F^2(z)F^2(0)}_{d=4} \left( 2c_{F^2}+  \left ( 1-\f{\beta(g)}{\eps g}\right )  \right. \nn\\
&&\left. \left(\f{\partial \log \mathcal{G}^{(F^2)}_2}{\partial \log g}   
  -2  g \frac{\partial\gamma_{F^2}}{\partial g} \log|z\mu|  \right)\right) \bigg|_{g \rightarrow g_*}+ \cdots
    \nn\\
&&={\mathcal{C}}_{O_0,\textrm{finite}}(g,\epsilon) \braket{F^2(z)F^2(0)}_{d=4}+ \cdots
\eea 
Since no logarithmic term may occur in the rhs, it follows that:
\bea \label{gamma0'}
g\f{\partial\gamma_{F^2}}{\partial g} \bigg|_{g=g_*}=g_*\beta^{\prime\prime}(g_*)
=0
\eea
Hence, at the nontrivial zero $g_*> 0$:
\bea \label{ggamma10}
\gamma'_{F^2}=\beta^{\prime\prime}(g_*)=0
\eea
under the assumption:
\bea \label{ggamma0}
\gamma_{F^2}=\beta'(g_*)=0
\eea
that suffice to rule out (section \ref{10.4}) a model of merging \cite{Kaplan} in the conformal window of massless QCD.\par

\subsubsection{LET for $\gamma_{F^2}\neq 0$ nonexceptional in $d=4$} \label{II}

By eq. \eqref{eq:gamF}:
 \bea
 \gamma_{F^2}=\beta'(g_*)\neq 0
 \eea
so that the operator $F^2$ is either irrelevant/relevant for $\gamma_{F^2} >0$ or relevant/irrelevant for $\gamma_{F^2} <0$ in the IR/UV. 
In the exactly conformal case the LET reads as $\epsilon \rightarrow 0$:
\bea\label{eq:LETgamnotzero00}
&&\braket{O(z)O(0)}_{\td=4-2\epsilon} \left( 2c_O -\f{2\gamma_O(g)}{\eps}+  \left ( 1-\f{\beta(g)}{\eps g}\right )  \right. \nn\\
&&\left. \left(\left(\f{\partial \log \mathcal{G}^{(O)}_2}{\partial \log g}+\f{2}{\beta_0^*} g  \frac{\partial\gamma_O}{\partial g}\right)  |z\mu|^{-\beta_0^*}
      -\f{2}{\beta_0^*} g \frac{\partial\gamma_O}{\partial g} \right)\right) \bigg|_{g \rightarrow g_*} +\cdots
    \nn\\
&&=\braket{O(z)O(0)}_{d=4} Z_{F^2}^{-1}
\f{1}{2}\f{N_3(g)}{N_2(g)}
\pi^{2}
 \f{ 
\Gamma\big( {2+\gamma_{F^2}}\big)  
\Gamma\big( \f{ -\gamma_{F^2}}{2}\big)^2 
}
{\Gamma \big(2+\f{\gamma_{F^2}}{2}\big)^2
 \Gamma \big( -\gamma_{F^2}\big)}|z\mu|^{-\gamma_{F^2}} \bigg|_{g \rightarrow g_*}+\cdots
\nn\\
&&~~+\f{1}{2} Z_O^{2} \int \braket{O(z)O(0)F^2(x)}_{\tilde d=4-2\epsilon;0;\textrm{c.t.}} \, d^\td x \bigg|_{g \rightarrow g_*}
\eea
with $Z^{-1}_{F^2}(g_*)=\left(1-\f{\beta(g_*)}{\eps g_*}\right)=1$.
Hence, for $\gamma_{F^2}$ nonexceptional, the renormalized version of the LET reads as $\epsilon \rightarrow 0$:
\bea\label{eq:LETgamnotzero0}
&&\braket{O(z)O(0)}_{d=4} \Bigg( 2c_O -\f{2\gamma_O(g)}{\eps}- {\mathcal{C}}_{O_0,\textrm{div}}(g,\eps)  -\f{2}{\beta_0^*} g \frac{\partial\gamma_O}{\partial g} 
+ \bigg(\f{\partial \log \mathcal{G}^{(O)}_2}{\partial \log g} +\f{2}{\beta_0^*} g  \frac{\partial\gamma_O}{\partial g}\bigg)  \nn\\
&&|z\mu|^{-\beta_0^*}
 \Bigg) \bigg|_{g \rightarrow g_*}+\cdots
=\braket{O(z)O(0)}_{d=4} \Bigg({\mathcal{C}}_{O_0,\textrm{finite}}(g,0)
+\f{1}{2}\f{N_3(g)}{N_2(g)}
\pi^{2}
\nn\\&&
 \f{ 
\Gamma\big( {2+\gamma_{F^2}}\big)  
\Gamma\big( \f{ -\gamma_{F^2}}{2}\big)^2 
}
{\Gamma \big(2+\f{\gamma_{F^2}}{2}\big)^2
 \Gamma \big( -\gamma_{F^2}\big)}
 |z\mu|^{-\gamma_{F^2}}\Bigg) \bigg|_{g \rightarrow g_*}+\cdots
 \eea
 with:
\bea
-\f{2\gamma_O(g)}{\eps}
 - {\mathcal{C}}_{O_0,\textrm{div}}(g,\eps) \bigg|_{g \rightarrow g_*}=0
 \eea
thus demonstrating that the additive renormalization due to the contact term in the rhs should be related to the anomalous dimension in the lhs that originates from the multiplicative renormalization, so that the above renormalized version of the LET has no ambiguity for additive renormalization. 
Remarkably, in the conformal window of massless QCD we verify (section \ref{10}) the equation above for the special case $O=F^2$,
thanks to the all-order computation of $ {\mathcal{C}}_{F^2_0,\textrm{div}}(g,\eps)$ in \cite{Z3} in terms of the beta function (appendix \ref{A}):
\bea \label{CC}
{\mathcal{C}}_{F^2_0,\textrm{div}}(g_*,\eps)
&=& -\f{4}{\eps}\left(\f{1}{2}g\f{\partial}{\partial g} \left(\f{\beta(g)}{g} \right)  - \f{\beta(g)}{g}\right)\bigg|_{g \rightarrow g_*}\nn\\
&=& -\f{{2}\beta'(g_*)}{\eps}\nn\\
&=&-\f{2\gamma_{F^2}}{\eps}
\eea
Thus, we come to the conclusion that, in massless QCD for $\gamma_{F^2}\neq 0$ nonexceptional, a divergent contact term must occur in the rhs of the LET to compensate for the
divergence in the lhs due to the anomalous dimension of $F^2$. \par

\subsubsection{LET for $\gamma_{F^2}$ of order $\epsilon$ in $\tilde d=4- 2 \epsilon$} \label{III}

Massless QCD in $\tilde d=4-2\epsilon$ dimensions at the Wilson-Fisher zero $\tilde g_*$ \cite{WF}:
\bea \label{WF0}
\beta(\tilde g_*,\epsilon)=- \epsilon \tilde g_*  + \beta(\tilde g_*)=0
\eea
has been introduced \cite{Braun1,Braun2} in order to efficiently compute anomalous dimensions to higher orders in perturbation theory.
The positivity of $\tilde g_*$ in eq. \eqref{WF0} requires $\beta(\tilde g_*)>0$. Hence, for small $\tilde g_*$, $\beta_0 <0$ in eq. \eqref{beta00} that may only occur above the conformal window. 
Indeed, in \cite{Braun1,Braun2} eq. \eqref{WF0} is solved perturbatively in an infinitesimal right neighborhood of the upper edge of the conformal window, so that 
$\tilde g_*^2$ is of order $\epsilon$ (section \ref{7.3}). \par
Therefore, despite $\beta_0 <0$, $\beta(g,\epsilon)$ is negative for small $g$ in $\tilde d=4-2\epsilon$ dimensions so that the theory flows from the UV to the Wilson-Fisher IR zero at $\tilde g_*>0$, where the LET reduces exactly to (section \ref{7.3}):
\be \label{III0}
2c_O-\f{2\gamma_O(g)}{\eps}\bigg|_{g \rightarrow \tilde g_*}={\mathcal{C}}_{O_0}(g,\eps)\bigg|_{g \rightarrow \tilde g_*}
\ee
with ${\mathcal{C}}_{O_0}(g,\eps)$ in eq. \eqref{OPE0}. \par
Surprisingly, in this case no constraint on the space-time structure of the renormalized correlators arises from the LET because $Z_{F^2}^{-1}(\tilde g_*,\epsilon)=0$ by eq. \eqref{WF0}.\par
In massless QCD at the Wilson-Fisher zero
only the divergent part of ${\mathcal{C}}_{F^2_0}(\tilde g_*,\eps)$ survives (appendix \ref{A}), which reads by eq. \eqref{CC}:
\bea  \label{IIIF0}
{\mathcal{C}}_{F^2_0,\textrm{div}}(\tilde g_*,\eps)
&=& -\f{4}{\epsilon}\left(\f{1}{2}g\f{\partial}{\partial g} \left(\f{\beta(g)}{g} \right)  - \f{\beta(g)}{g}\right)\bigg|_{g \rightarrow \tilde g_*}\nn\\
&=& -\f{4}{\epsilon}\left(\f{1}{2} \gamma_{F^2} - \eps \right)\nn\\
&=&-\f{2}{\epsilon}\gamma_{F^2}+4
\eea
By means of eq. \eqref{IIIF0} we verify, as in the case (II), 
eq. \eqref{III0} for $O=F^2$, since $c_{F^2}=2$ (section \ref{4.2}).\par

\subsection{Conclusions} \label{2.5}

We summarize our main conclusions under the standard assumption that $O$ and $F^2$ are multiplicatively renormalizable scalar primary operators belonging to a complete orthonormal basis at the isolated nontrivial zero of the beta function $g_*>0$.\par
(1) If $F^2$ is marginal at $g_*>0$, the theory is exactly conformal in the limit $g(\mu) \rightarrow g_*$ with fixed $\mu\neq 0,+\infty$ (appendix \ref{D}), but it is not asymptotically conformal for $g(\mu) \rightarrow g_*$ with $\Lrgi$ fixed that necessarily implies $\mu \rightarrow 0^+/ \mu \rightarrow +\infty$ (appendix \ref{D}), in analogy (section \ref{2.2}) with the exact conformal symmetry for $g(\mu)=g_*=0$ with fixed $\mu$ and the asymptotic freedom in the UV for $g(\mu)\neq g_*=0$ with fixed $\Lambda_{\scriptscriptstyle{UV}}$. The case (1) may occur in the conformal window or in the asymptotically safe phase of massless QCD, if it exists.\par
(2) If $\gamma_{F^2}$ is nonexceptional, we verify (section \ref{II}) nonperturbatively by explicit computation by means of the OPE the matching of the divergences as $\epsilon \rightarrow 0$ in the lhs and rhs of the LET for $O=F^2$ in the conformal window of massless QCD.\par
(3) Moreover, in the above case we work out explicitly the matching between the lhs and rhs of the LET for an operator $O$.\par
(4) If $F^2$ is irrelevant in the IR/UV and $\gamma_{F^2}$ nonexceptional at $g_*>0$, the theory is asymptotically conformal for $g(\mu) \rightarrow g_*$ with $\Lrgi$ fixed (appendix \ref{D}) that may occur in the IR in the conformal window of massless QCD or in the UV in the asymptotically safe phase of massless QCD, if it exists. \par
(5) If the theory is conformal and $F^2$ is marginal at $g_*>0$, the LET implies $\gamma'_{F^2}(g_*)=0$ (section \ref{I}). \par
(6) The constraint above suffices to rule out a model of merging (section \ref{10.4}) in the conformal window of massless QCD that involves the merging of two simple zeroes into a double zero of the beta function. \par

\section{Plan of the paper} \label{3}

In section \ref{4} we recall the LET for bare correlators both with the Wilsonian and canonical normalization of the action.\par
In section \ref{5} we work out the canonical version of the LET in dimensional regularization.\par
In section \ref{6} we evaluate the contribution of the contact terms to the LET.\par
In section \ref{7} we specialize the LET to conformal YM theories and evaluate the contact term that contributes to the rhs of the LET by the OPE.\par
In section \ref{8} we work out the physics implications of the LET by matching the lhs with the rhs
and we write down a suitably renormalized version of the LET. \par
In section \ref{9A} we outline the LET in asymptotically conformal YM theories. \par
In section \ref{10} we verify the LET nonperturbatively for $O=F^2$ in massless QCD and discuss its further physics implications. \par
In the appendices \ref{C}, \ref{A}, \ref{B} we provide ancillary computations, respectively the anomalous dimension of $F^2$ in terms of the beta function, the contact term in the OPE of $F^2$ with itself and the space-time integral of conformal $3$-point correlators in the rhs of the LET in the $(u,v_{F^2})$ regularization. \par
In appendices \ref{CS0} and \ref{CS1} we study the structure of the solution of the CS equation in $d=4$ and $\tilde d=4-2\epsilon$ dimensions respectively. \par
In appendix \ref{D} we compute the IR/UV asymptotics of $2$-point correlators of a multiplicatively renormalizable operator $O$ for  $\Lrgi$ fixed, including $O=F^2$, and study under which conditions the IR/UV limit of the conformal theory for $g(\mu) \rightarrow g_*$ with fixed $\mu$
coincides with the aforementioned asymptotics -- up to perhaps the normalization of the correlators. We also compute the renormalized multiplicative factor, $Z^{(F^2)}(g(z),g(\mu))$, of $F^2$ in the coordinate representation in a closed form in terms of the beta function.\par
In appendix \ref{E} we study the LET in the $(u,v_{F^2})$ regularization and demonstrate that it is satisfied to order $g^2$ in perturbation theory in the $(0, v_{F^2})$ scheme.

\section{LET for bare correlators} \label{4}

According to \cite{MBR} the LET for bare correlators applies to YM theories in $d=4$ dimensions. 

\subsection{LET with the Wilsonian normalization of the action}
\label{4.1}

The LET for bare correlators with the Wilsonian normalization of the action has been derived in \cite{MBR} starting from the Euclidean path-integral formulation of SU($N$) YM theories in $d=4$ dimensions:
\be
\label{eq:PI}
\braket{ \cO_1\cdots\cO_n}_0=\f{\int\,\cO_1\cdots\cO_n \,e^{-\f{N}{2g_0^2}\int \,\Tr\cF^2d^4x+\cdots} }{\int\,e^{-\f{N}{2g_0^2}\int \,\Tr\cF^2d^4x+\cdots}}
\ee 
where $g_0$ is the bare 't Hooft coupling, $g_0^2=g^2_{0 YM} N$, the trace, $\Tr$, is in the fundamental representation,
$\cO_i$, $i=1,\cdots ,n$, are a set of local bare operators that are independent of $g_0$, $\Tr \mathcal{F}^2 \equiv \Tr (\mathcal{F}_{\mu\nu}\mathcal{F}_{\mu\nu})$ with the sum over repeated indices understood, and $\mathcal{F}_{\mu\nu} = \partial_{\mu}A_{\nu} - \partial_{\nu} A_{\mu} + i [A_{\mu},A_{\nu}]$. \par
In eq. \eqref{eq:PI} we write explicitly the only term in the action that depends on $g_0$ and thus enters the derivation of the LET. Indeed, by deriving eq. \eqref{eq:PI} with respect to $-1/g_0^2$, we immediately arrive at the LET in the coordinate representation, which reads \cite{MBR}:
\be
\label{eq:LETWcoord}
\f{\partial}{\partial\log g_0} \braket{\cO_1\cdots\cO_n}_0 = \f{N}{2g_0^2}\,\int\, \braket{\cO_1\cdots\cO_n \cF^2(x)}_0 - \braket{\cO_1\cdots\cO_n}_0\braket{\cF^2(x)}_0\, d^4x
\ee
where we define $\cF^2\equiv 2\Tr\cF^2$.

\subsection{LET with the canonical normalization of the action} \label{4.2}

It is useful for our purposes to rewrite \cite{BB} the LET for the canonically normalized YM action. 
Thus, we rescale the gauge fields in the functional integral in eq. \eqref{eq:PI}  
by the factor $\frac{g_0}{\sqrt N}$. 
Formally, the rescaling does not affect the v.e.v. of the operators, as it is just a change of variables. \par
Therefore, defining after the rescaling $\frac{g_0^2}{N} \Tr F^2= \Tr \mathcal{F}^2$ and $(\frac{g_0}{\sqrt N})^{c_{O_k}} O_k= \mathcal{O}_k$ for some $c_{O_k}$\footnote{For example, $c_{F^2}=2$.}, with $F_{\mu\nu} = \partial_{\mu}A_{\nu} - \partial_{\nu} A_{\mu} + i \f{g_0}{\sqrt N}[A_{\mu},A_{\nu}]$ and $O_k$ now dependent on $g_0$ but canonically normalized, we obtain the identity \cite{BB}:
\bea 
\label{res}
\braket{ \mathcal{O}_1 \cdots \mathcal{O}_n}_0 = \prod^{k=n}_{k=1}(\frac{g_0}{\sqrt N})^{c_{O_k}}   \braket{ O_1\cdots O_n }_{0}
\eea
where the expectation value in the lhs is defined with the Wilsonian normalization in eq. \eqref{eq:PI} and in the rhs with the canonical normalization, i.e., after the aforementioned rescaling of the gauge fields in the functional integral in eq. \eqref{eq:PI}.\par
By employing eq. \eqref{res}, the LET in eq. \eqref{eq:LETWcoord}
 is rewritten in terms of canonically normalized bare local operators \cite{BB}:
\bea
\label{eq:LETCcoord_sec4}
&&\sum^{k=n}_{k=1} c_{O_k} \braket{O_1\cdots O_n}_0 +
\f{\partial}{\partial\log g_0} \braket{O_1\cdots O_n}_0 \nn\\
&&\hspace{0.5truecm}= \f{1}{2}\,\int\, \braket{O_1\cdots O_n F^2(x)}_0 
- \braket{O_1\cdots O_n}_0\braket{F^2(x)}_0\, d^4x
\eea
where we define $F^2\equiv 2\Tr F^2$.

\section{LET in dimensional regularization} \label{5}

In \cite{MBL,BB} the LET has been applied to massless QCD in perturbation theory by employing a hard-cutoff regularization of the space-time integral in the rhs.
However, this regularization of the LET is not compatible in general with gauge invariance, but for the leading -- scheme-independent -- logarithms to order $g^2$ \cite{MBL,BB} and the square of the logarithms to order $g^4$ \cite{BB}. \par
Instead, in the present paper we aim to derive exact implications of the LET for conformal YM theories.
Hence, we need a manifestly gauge- and conformal-invariant regularization of the LET.
Dimensional regularization is the obvious choice. \par

\subsection{Dimensional regularization and renormalization of YM theories}
\label{5.1}
In dimensional regularization the space-time dimension 
is shifted as:
\be
d\to \td=d-2\eps
\ee
with $\eps$ infinitesimal and positive, and $d=4$ in the present paper. Then, the canonical dimension of the bare operators entering the action is implied by dimensional analysis, by requiring the action in $\td$ dimensions to be dimensionless. \par
Specifically, the YM action with the canonical normalization in $\td$ dimensions formally reads:
\be\label{eq:caction}
S= \f{1}{2} \int \Tr F^2_0\,d^\td x
\ee
with the bare operator $\Tr F^2_0=\Tr F_{\mu\nu}F_{\mu\nu}$. Given that $[d^\td x] = -\td$, where $[.]$ stands for the mass dimension of its argument, a dimensionless action\footnote{We employ the natural units $\hbar=c=1$.}, $[S]=0$, implies:
\be
[\Tr F^2_0]\equiv\tD_{F^2_0}=\td
\ee
Since $F_{\mu\nu}=\partial_\mu A_{\nu} -\partial_\nu A_{\mu} +i\f{g_0}{\sqrt{N}}[A_{\mu}, A_{\nu}]$, the canonical dimension of the bare gauge field, $A_{\mu}$, is:
\bea \label{F}
[A_{\mu}]&=&\f{\tD_{F^2_0}}{2}-1\nn\\\
&=&\f{\td}{2}-1\nn\\
&\overset{d=4}{=}&1-\eps
\eea
Moreover, given $[\partial_\mu]=1$, eq. \eqref{F} in turn implies:
\bea\label{eq:g0DR}
[g_0]&=&-2[A_\mu ]+\f{\td}{2} \nn\\
 &=&2-\f{\td}{2}\nn\\
 &\overset{d=4}{=}&\eps
\eea
for the bare gauge coupling.
We now recall a few relations, in dimensionally regularized YM theories, needed in order to rewrite the LET in terms of the renormalized coupling and correlators. \par
According to eq. \eqref{eq:g0DR}, for $\td=4-2\eps$ the bare gauge coupling, $g_0$, has canonical dimension $[g_0]=\eps$, so that the dimensionless renormalized coupling $g$ is related to $g_0$ as follows:
\bea\label{eq:gg0}
g_0=Z_g(g,\eps) \mu^\eps g
\eea
where the power of the renormalization scale, $\mu$, carries the dimension of $g_0$, $Z_g$ is the (dimensionless) renormalization factor for the coupling, and we employ an $\overline{MS}$-like renormalization scheme, where the renormalization factor $Z_{O}$ for a multiplicatively renormalizable operator $O$ is a series of pure poles in $\eps$. \par
Besides, in $\td=4-2\eps$ dimensions the beta function for the gauge coupling reads: 
\be\label{eq:betaeps}
\beta(g,\epsilon )=-\epsilon g +\beta(g)
\ee
where:
\bea
\beta (g,\epsilon )=\f{dg}{d\log\mu} 
\eea
and:
\bea
\beta(g) &=& -g\,\f{d\log Z_g}{d\log\mu}\nn\\
&=&\f{dg}{d\log\mu}\bigg|_{\eps=0}
\eea
 is the beta function in $d=4$ dimensions. The renormalized coupling is related to the bare one by:
 \bea\label{eq:gg0rel0}
  \f{\partial\log g}{\partial\log g_0} &=& \left ( 1+ \f{\partial\log Z_g}{\partial\log g}\right )^{-1}\nn\\
  &=&\left ( 1- \f{\beta(g)}{\eps g}\right )\nn\\
  &=&Z_{F^2}^{-1}(g,\eps)
  \eea
  according to eq. \eqref{ZF2}. Eq. \eqref{eq:gg0rel0}
  is obtained by taking the logarithmic derivative of eq. \eqref{eq:gg0} and equating eq. \eqref{eq:betaeps} to:
  \bea
  \beta(g,\eps)&=& -\eps g \left ( 1+ \f{\partial\log Z_g}{\partial\log g}\right )^{-1}
  \eea
  Moreover, given the anomalous dimension, $\gamma_O$, of a multiplicatively renormalizable operator, $O=Z_O O_0$:
  \be\label{gamma}
  \gamma_O(g) = - \f{d\log Z_O}{d\log\mu}
  \ee
we get: 
  \bea\label{eq:Zrel0}
  \f{\partial\log Z_O^{-2}}{\partial\log g_0}&=&-2\f{\partial\log Z_O}{\partial\log g_0}\nn\\
  &=&-2\f{d\log Z_O}{d\log\mu}\f{d\log\mu}{d\log g}\f{\partial\log g}{\partial\log g_0}\nn\\
  &=&2\gamma_O(g)\left(\f{1}{g}\f{d g}{d\log\mu} \right)^{-1}\left ( 1+ \f{\partial\log Z_g}{\partial\log g}\right )^{-1}\nn\\
  &=&2\gamma_O(g)\left(\f{\beta(g,\eps)}{g} \right)^{-1}\left ( 1- \f{\beta(g)}{\eps g}\right )\nn\\
  &=&-2\f{\gamma_O(g)}{\eps}\left ( 1- \f{\beta(g)}{\eps g}\right )^{-1}\left ( 1- \f{\beta(g)}{\eps g}\right )\nn\\
   &=&-2\f{\gamma_O(g)}{\eps}
  \eea
by employing eqs. \eqref{gamma} and \eqref{eq:gg0rel0}.

\subsection{LET for bare correlators in dimensional regularization} \label{5.2}

The generalization to $\td$ space-time dimensions of the LET with the Wilsonian normalization in eq. \eqref{eq:LETWcoord} is conveniently written in terms of the dimensionless bare coupling $\bar{g}_0=g_0\mu^{\td/2-2}$, according to eq. \eqref{eq:g0DR}:
\be
\label{eq:LETWcoordd}
\f{\partial}{\partial\log \bar{g}_0} \braket{\cO_1\cdots\cO_n}_0 = \f{N}{2\bar{g}_0^2}\mu^{2-\f{\td}{2}}\,\int\, \braket{\cO_1\cdots\cO_n \cF^2(x)}_0 - \braket{\cO_1\cdots\cO_n}_0\braket{\cF^2(x)}_0\, d^\td x
\ee
The residual $\mu$ dependence in the rhs is due to the fact that $[\cF^2_0]=4$. \par
Repeating the steps from the Wilsonian formulation in eq. \eqref{eq:LETWcoord} to the canonical one in eq. \eqref{eq:LETCcoord_sec4}, we obtain in $\td$ dimensions:
\bea
\label{eq:LETCcoordd}
&&\sum^{k=n}_{k=1} c_{O_k} \braket{O_1\cdots O_n}_0 +
\f{\partial}{\partial\log {g}_0} \braket{O_1\cdots O_n}_0 \nn\\
&&\hspace{0.5truecm}= \f{1}{2}\,\int\, \braket{O_1\cdots O_n F^2(x)}_0 
- \braket{O_1\cdots O_n}_0\braket{F^2(x)}_0\, d^\td x
\eea
where no residual dependence on the scale $\mu$ appears in the rhs, since $[F^2_0]=\td$.

\subsection{LET for renormalized correlators in dimensional regularization} \label{5.3}

To derive the implications of the LET for conformal YM theories, it is convenient to rewrite eq. \eqref{eq:LETCcoordd} in terms of renormalized correlators. \par
In the present paper, we consider the simplest case where no operator mixing occurs under renormalization up to operators that vanish by the equations of motion, and we focus on eq. \eqref{eq:LETCcoordd} with $n=2$ and 
$O_1=O_2= O$. Eq. \eqref{eq:LETCcoordd} then reads:
\bea
\label{eq:LETn2bare0}
&&2c_O \braket{O(z)O(0)}_0 +
\f{\partial}{\partial\log {g}_0} \braket{O(z)O(0)}_0 \bigg|_{\dr} \nn\\
&&\hspace{0.5truecm}= \f{1}{2}\, \int \braket{O(z) O(0) F^2(x)}_0 
- \braket{O(z)O(0)}_0\braket{F^2(x)}_0
 d^\td x \bigg|_{\dr}
\eea
We assume as well that $F^2$ is multiplicatively renormalizable up to operators that vanish by the equations of motion. 
This assumption is satisfied in massless QCD by limiting ourselves to gauge-invariant correlators, where $F^2$ only mixes with the quark Lagrangian density that vanishes by the equations of motion.\par
The bare 1-, 2- and 3-point correlators are thus related to the renormalized ones by a multiplicative renormalization up to contact terms. Hence, up to contact terms, we obtain:
\bea
\label{eq:RvsB0}
\braket{F^2(x)}_0&=&Z_{F^2}^{-1}(g,\eps)\braket{F^2(x)}\nn\\
\braket{O(z)O(0)}_0&=&Z_O^{-2}(g,\eps)\braket{O(z)O(0)}\nn\\
\braket{O(z)O(0)F^2(x)}_0&=&Z_O^{-2}(g,\eps)Z_{F^2}^{-1}(g,\eps)\braket{O(z)O(0)F^2(x)}
\eea
where the renormalization factors $Z_O$ and $Z_{F^2}$ are functions of the renormalized coupling $g$ and the parameter of the dimensional regularization $\epsilon$. \par
The lhs and rhs of the regularized LET in eq. \eqref{eq:LETn2bare0} can now be rewritten in terms of the renormalized 1-, 2- and 3-point correlators  by means of eq. \eqref{eq:RvsB0}.\par
We rewrite the lhs of eq. \eqref{eq:LETn2bare0} without contact terms, since once for all we assume $z\neq 0$:
 \bea
 \label{eq:lhs_gen0}
\mbox{lhs}&=& 2c_O\, Z_O^{-2}\braket{O(z)O(0)} +
g_0\f{\partial}{\partial {g}_0} Z_O^{-2}\braket{O(z)O(0)}
+Z_O^{-2}g_0\f{\partial}{\partial {g}_0} \braket{O(z)O(0)}\bigg|_{\dr}\nn\\
&=& Z_O^{-2}\braket{O(z)O(0)}\left( 2c_O +\f{\partial\log Z_O^{-2}}{\partial\log g_0}
+\f{\partial\log \braket{O(z)O(0)}}{\partial\log g_0}
\right) \bigg|_{\dr}\nn\\
&=& Z_O^{-2}\braket{O(z)O(0)}\left( 2c_O - \f{2\gamma_O(g)}{\eps}+\left ( 1-\f{\beta(g)}{\eps g}\right ) \f{\partial\log \braket{O(z)O(0)}}{\partial\log g}\right)\bigg|_{\dr}
\eea
where in the last line we have employed:
\bea
\frac{\partial}{\partial\log g_0} =\frac{\partial \log g}{\partial\log g_0} \frac{\partial}{\partial\log g}
\eea
and eqs. \eqref{eq:gg0rel0} and \eqref{eq:Zrel0}.\par
We rewrite the rhs of eq. \eqref{eq:LETn2bare0} by means of eq. \eqref{eq:RvsB0}:
\bea\label{eq:rhs_gen0}
\mbox{rhs}&=&\f{1}{2}\int \braket{O(z) O(0) F^2(x)}_0 - \braket{O(z)O(0)}_0\braket{F^2(x)}_0 d^\td x \bigg|_{\dr}\nn\\
&=&\f{1}{2}Z_O^{-2}Z_{F^2}^{-1}\int \braket{O(z) O(0) F^2(x)}' - \braket{O(z)O(0)}\braket{F^2(x)} d^\td x \bigg|_{\dr}\nn\\
&&+\f{1}{2} \int \braket{O(z)O(0)F^2(x)}_{0;\textrm{c.t.}} \, d^\td x \bigg|_{\dr}
\eea
where the ${}^\prime$ in the second line denotes the correlator at different points. The LET for renormalized correlators follows by equating the lhs and rhs after multiplying both by the factor $Z_O^{2}$:
\bea \label{LEET}
\mbox{lhs}=\mbox{rhs}
\eea

\section{LET and contact terms in the OPE} \label{6}

Both in the exactly and asymptotically conformal cases, under the standard assumptions (section \ref{2.3}) the contact terms that contribute to the rhs of the LET are computed by the OPE of the product of $O$ with $F^2$. \par
In both cases, the only term of the OPE that may contain the contact terms contributing to the rhs of the LET is:
\be\label{eq:OPEb}
O_0(x)F_0^2(0) =   \cdots + C_{1}^{(O_0,F^2_0)}(x)O_0(0) + \cdots
\ee
for the bare operators, and:
\be\label{eq:OPEr}
O(x)F^2(0) = \cdots + C_1^{(O,F^2)}(x)O(0) + \cdots
\ee
for the renormalized ones,
where the bare operators $O_0, F^2_0$ are related to the multiplicatively renormalized operators $O, F^2$ by:
\bea\label{eq:BRZ}
O_0(x)&=&Z_O^{-1} O(x)\nonumber\\
F^2_0(x)&=& Z_{F^2}^{-1}F^2(x)
\eea
Correspondingly, the coefficients of the renormalized and bare OPE, in eqs. \eqref{eq:OPEr} and \eqref{eq:OPEb} respectively, are related by means of eq. \eqref{eq:BRZ}: 
\be
C_{1}^{(O_0,F^2_0)}(x)=Z_{F^2}^{-1}C_{1}^{(O,F^2)}(x)
\ee
We conveniently separate the contact terms from the part of the renormalized OPE coefficient at nonzero separation $C_{1}^{(O,F^2)'}(x)$:
\be\label{eq:C10OF}
C_{1}^{(O,F^2)}(x)={\mathcal{C}}_{O}(g,\eps)\delta^{(4)}(x)+C_{1}^{(O,F^2)'}(x)
\ee
As a consequence, the contact terms that contribute to the bare 3-point correlator in the rhs of the LET read as $\epsilon \rightarrow 0$:
\be\label{eq:ct00}
\braket{O(z)O(0)F^2(x)}_{0;\textrm{c.t.}} ={\mathcal{C}}_{O_0}(g,\eps)\braket{O(z)O(0)}_0(\delta^{(4)}(x)+\delta^{(4)}(x-z))+\cdots
\ee
according to eq. \eqref{eq:OPEb}, where the dots represent terms that vanish as $\epsilon \rightarrow 0$ and
${\mathcal{C}}_{O_0}(g,\eps)$ admits the expansion:
\bea\label{eq:bareC}
{\mathcal{C}}_{O_0}(g,\eps)&=&{\mathcal{C}}_{O_0,\textrm{div}}(g,\eps)+{\mathcal{C}}_{O_0,\textrm{finite}}(g,\eps)\nn\\
&=&Z_{F^2}^{-1}{\mathcal{C}}_{O,\textrm{div}}(g,\eps)+Z_{F^2}^{-1}{\mathcal{C}}_{O,\textrm{finite}}(g,\eps)
\eea
into a pole and a finite part. Hence:
\bea
&&\f{1}{2} \int \braket{O(z)O(0)F^2(x)}_{0;\textrm{c.t.}} \, d^\td x \bigg|_{\dr}
= {\mathcal{C}}_{O_0}(g,\eps) \braket{O(z)O(0)}_0 + \cdots
\eea

\section{LET in conformal YM theories}
\label{7}
We evaluate nonperturbatively eqs. \eqref{eq:lhs_gen0} and \eqref{eq:rhs_gen0} in conformal YM theories. The conformal theory is obtained taking the limit $g=g(\mu) \rightarrow g_*$ keeping $\mu$ fixed (appendix \ref{D}) in $d=4$ dimensions. We also consider the conformal theory  in $\tilde d=4-2\epsilon$ dimensions in the limit $g \rightarrow \tilde g_*$ with $\beta(\tilde g_*,\epsilon)=0$. \par
The following simplifications occur: 
$\braket{F^2(x)}_0$ and its renormalized version $\braket{F^2(x)}$ vanish in eq. \eqref{eq:rhs_gen0}. 
Besides, the renormalized 2- and 3-point correlators in $d=4$ dimensions are known in terms of the conformal dimensions of the operators, up to the overall normalization and contact terms. Thus, we get: 
 \bea
 \label{eq:CFT23}
 \braket{O(z)O(0)}_{d=4}\bigg|_{g \rightarrow g_*}&=&\f{N_2(g_*)  \mu^{2\Delta_{O_0}-2\Delta_O}  }{|z|^{2\Delta_O}}
 \eea
 and:
 \bea
 \label{eq:CFT23001}
 \braket{O(z)O(0)F^2(x)}_{d=4}\bigg|_{g \rightarrow g_*}&=&\f{N_3(g_*)\mu^{2\Delta_{O_0}+\Delta_{F^2_0}-2\Delta_O - \Delta_{F^2}}}{|z|^{2\Delta_O -\Delta_{F^2}} |x|^{\Delta_{F^2}}|x-z|^{\Delta_{F^2}}}
 \eea
where $N_2(g_*),N_3(g_*)$ are constant dimensionless normalization factors, $\Delta_O$ is the conformal dimension of the renormalized operator $O$ in $d=4$ dimensions, $\Delta_O = \Delta_{O_0}+\gamma_O$, with canonical dimension $\Delta_{O_0}$ and anomalous dimension $\gamma_O=\gamma_O(g_*)$ and, with analogous notation, $\Delta_{F^2}$ is the conformal dimension of $F^2$, with $\Delta_{F^2}=\Delta_{F^2_0}+\gamma_{F^2}$ and $\gamma_{F^2}=\gamma_{F^2}(g_*)$. 
The theory in $\tilde d=4-2 \epsilon$ dimensions differs from the theory in $d=4$ dimensions by terms $\mathcal{R}_2(\epsilon),\mathcal{R}_3(\epsilon)$ that vanish as $\epsilon \rightarrow 0$ :
 \bea
 \label{eq:CFT23t}
 \braket{O(z)O(0)}_{\td=4-2\epsilon}&=&\braket{O(z)O(0)}_{d=4}+\mathcal{R}_2(\epsilon)
  \eea
  and:
 \bea
 \label{eq:CFT2300t}
 \braket{O(z)O(0)F^2(x)}_{\td=4-2\epsilon}&=&\braket{O(z)O(0)F^2(x)}_{d=4}+\mathcal{R}_3(\epsilon)
 \eea
Firstly, we evaluate the rhs of the LET in eq. \eqref{eq:rhs_gen0} in dimensional regularization as $\epsilon \rightarrow 0$ by employing the $3$-point correlator in eq. \eqref{eq:CFT23001} in $d=4$ dimensions and skipping in the dots the contribution of $\mathcal{R}_3(\epsilon)$ (section \ref{2.1A}):
\bea\label{eq:rhsRentilde}
\mbox{rhs}=&&\f{1}{2} Z_{F^2}^{-1}  \int
\f{N_3(g) \mu^{2\Delta_{O_0}+\Delta_{F^2_0}-2\Delta_O - \Delta_{F^2}}  }{|z|^{2\Delta_O -\Delta_{F^2}} } 
 \f{1}
{|x|^{\Delta_{F^2}}|x-z|^{\Delta_{F^2}} } +\cdots d^\td x \bigg|_{g \rightarrow g_*}\nn\\
&&+\f{1}{2} Z_O^{2}\int \braket{O(z)O(0)F^2(x)}_{d=4;0;\textrm{c.t.}} \, d^\td x\bigg|_{g \rightarrow  g_*}+\mbox{finite terms}+\cdots
 \eea
where $\Delta_{F^2_0}=4$ in $d=4$ space-time dimensions.
We compute the space-time integral in the first line of eq. \eqref{eq:rhsRentilde} by means of (appendix \ref{B}):
\bea\label{eq:I1}
I_{d,\Delta_1,\Delta_2}&\equiv& \int  \f{1}
{ |x|^{\Delta_1}|x-z|^{\Delta_2} }     d^dx \nn\\
&=&(2\pi)^d \,C_{d,\f{\Delta_1}{2},\f{\Delta_2}{2}}\, |z|^{d-\Delta_1-\Delta_2}
\eea
where:
\be
C_{d,\f{\Delta_1}{2},\f{\Delta_2}{2}}=\f{ 
\Gamma\big( \f{\Delta_1+\Delta_2-d}{2}\big)  
\Gamma\big( \f{d-\Delta_1}{2}\big)  \Gamma\big( \f{d-\Delta_2}{2}\big) 
}
{(4\pi)^{\f{d}{2}}\Gamma \big(\f{\Delta_1}{2}\big)\Gamma \big(\f{\Delta_2}{2}\big) \Gamma \big(d-\f{\Delta_1}{2}-\f{\Delta_2}{2}\big)}
\ee
Hence, by the substitutions $d\to\td$ and $\Delta_1=\Delta_2\to \Delta_{F^2}$, the integral in the first line of eq. \eqref{eq:rhsRentilde} reads:
\bea\label{eq:rhsRen2}
&&\f{1}{2}Z_{F^2}^{-1}  
\f{N_3(g) \mu^{2\Delta_{O_0}+\Delta_{F^2_0}-2\Delta_O - \Delta_{F^2}}  }{|z|^{2\Delta_O -\Delta_{F^2}} } I_{\td,\Delta_{F^2},\Delta_{F^2}}\bigg|_{g  \rightarrow g_*}
\nn\\
&&=\f{1}{2} Z_{F^2}^{-1}  
\f{N_3(g) \mu^{2\Delta_{O_0}+\Delta_{F^2_0}-2\Delta_O - \Delta_{F^2}}  }{|z|^{2\Delta_O -\Delta_{F^2}} } 
(2\pi)^\td\, C_{\td,\f{\Delta_{F^2}}{2},\f{\Delta_{F^2}}{2}} |z|^{\td-2\Delta_{F^2}}\bigg|_{g  \rightarrow  g_*}
\eea
with:
 \be\label{eq:C}
 C_{\td,\f{\Delta_{F^2}}{2},\f{\Delta_{F^2}}{2}} = \f{ 
\Gamma\big( \f{2\Delta_{F^2}-\td}{2}\big)  
\Gamma\big( \f{\td-\Delta_{F^2}}{2}\big)^2 
}
{(4\pi)^{\f{\td}{2}}\Gamma \big(\f{\Delta_{F^2}}{2}\big)^2
 \Gamma \big(\td-\Delta_{F^2}\big)}
 \ee
 The last line in eq. \eqref{eq:rhsRen2} may be rewritten in terms of the $2$-point correlator in eq. \eqref{eq:CFT23}:
 \bea\label{eq:rhs23}
 &&
\f{\mu^{2\Delta_{O_0}-2\Delta_O}  }{|z|^{2\Delta_O} } Z_{F^2}^{-1}\,
\f{1}{2} 
\f{N_3(g) \mu^{\Delta_{F^2_0}- \Delta_{F^2}}  }{|z|^{-\Delta_{F^2}} }(2\pi)^\td\, C_{\td,\f{\Delta_{F^2}}{2},\f{\Delta_{F^2}}{2}} |z|^{\td-2\Delta_{F^2}}\bigg|_{g \rightarrow g_*}
\nn\\
&&=\braket{O(z)O(0)}_{d=4} Z_{F^2}^{-1}\,
\f{1}{2}\f{N_3(g)}{N_2(g)}
(2\pi)^\td\, C_{\td,\f{\Delta_{F^2}}{2},\f{\Delta_{F^2}}{2}} |z |^{-\gamma_{F^2}-2\epsilon}|\mu |^{-\gamma_{F^2}}  \bigg|_{g \rightarrow  g_*}
\eea 
The UV and IR properties of eq. \eqref{eq:rhs23}
only depend as $\epsilon \rightarrow 0$ on the conformal dimension $\Delta_{F^2}=\Delta_{F^2_0}+\gamma_{F^2}$ and, specifically, on the anomalous dimension $\gamma_{F^2}$. 
Therefore, we analyze separately the following three cases:\par
 (I)  $\gamma_{F^2}=0$ in $d=4$ dimensions at $\beta(g_*)=0$.\par
 (II) $\gamma_{F^2}\neq 0$ nonexceptional in $d=4$ dimensions at $\beta(g_*)=0$. \par
 (III) $\gamma_{F^2}$ of order $\eps$ in $\td=4-2 \epsilon$ dimensions at $\beta(\tilde g_*,\eps)=0$.  \par
Secondly, we evaluate the lhs of the LET in eq. \eqref{eq:lhs_gen0} as $\epsilon \rightarrow 0$.
While the $2$-point conformal correlator at $g_*$ may be evaluated by eq. \eqref{eq:CFT23}, its logarithmic derivative involves the $2$-point correlator in a neighborhood of $g_*$, where the theory is not conformal for an isolated zero of the beta function. \par
Therefore, in order to compute the logarithmic derivative in the lhs of the LET, we employ the ansatz for the exact solution of the CS equation for the $2$-point renormalized correlator in $\td=4-2\epsilon$ dimensions (appendix \ref{CS1}) up to contact terms. We eventually relate (section \ref{2.1A}) the computation above to the $2$-point correlator in $d=4$ dimensions as $\epsilon \rightarrow 0$.
In $d=4$ dimensions (appendix \ref{CS0}):
\bea
\label{CSS1}
 \braket{O(z)O(0)}_{d=4}=\f{\mathcal{G}^{(O)}_2(g(z))}{|z|^{2 \Delta_{O_0}}} Z^{(O)2}(g(z),g(\mu))
\eea
that reduces to eq. \eqref{eq:CFT23} for $g(\mu) \rightarrow g_*$ with fixed $\mu$ (appendix \ref{D}):
\bea \label{co1}
\f{\mathcal{G}^{(O)}_2(g(z))}{|z|^{2 \Delta_{O_0}}} Z^{(O)2}(g(z),g(\mu))\bigg|_{g \rightarrow g_*}
&=&\f{N_2(g_*)  \mu^{2\Delta_{O_0}-2\Delta_O}  }{|z|^{2\Delta_O}}
\eea
thus implying (appendices \ref{CS0} and \ref{D}):
\bea
\mathcal{G}^{(O)}_2(g_*)=N_2(g_*)
\eea
In $\td=4-2\epsilon$ dimensions (appendix \ref{CS1}):
\bea
\label{CSStilde}
 \braket{O(z)O(0)}_{\tilde d=4-2\epsilon}=\f{\mathcal{ G}^{(O)}_2(\tilde g(z))}{|z|^{2 \tilde \Delta_{O_0}}}  Z^{(O)2}(\tilde g(z),g(\mu))
\eea
where now the running coupling $\tilde g(z)=g(z,\epsilon)$ depends explicitly on $\epsilon$, since the RG flow is determined by the beta function $\beta(g,\epsilon)=-\epsilon g+ \beta(g)$ in $\tilde d=4-2 \epsilon$ dimensions. \par 
We compute the logarithmic derivative in the lhs of the LET in the following two alternative ways (section \ref{2.1A}).\par
Employing eq. \eqref{eq:CFT23t}, by skipping in the dots the contribution of $\mathcal{R}_2(\epsilon)$, the logarithmic derivative of the renormalized correlator in $\td=4-2\epsilon$ dimensions reads as $\epsilon \rightarrow 0$:
\bea
&&g\f{\partial}{\partial {g}} \braket{O(z)O(0)}_{\td=4-2\epsilon}=g\f{\partial}{\partial {g}}\left (\f{\mathcal{G}^{(O)}_2(g(z))}{|z|^{2 \Delta_{O_0}}} Z^{(O)2}(g(z),g(\mu))
  \right )+\cdots\nn\\
  &&=\f{\mathcal{G}^{(O)}_2(g(z))}{|z|^{2 \Delta_{O_0}}} Z^{(O)2}(g(z),g(\mu))  
  \left ( \f{\partial \log \mathcal{G}^{(O)}_2}{\partial g(z)}  \f{\partial g(z)}{\partial \log g}    
  + 2    \f{\partial g(z)}{\partial \log g} \f{\gamma_O(g(z))}{\beta(g(z))}\right.
  \nn\\
  &&~~~~\left. -2    \f{\partial g(\mu)}{\partial \log g} \f{\gamma_O(g(\mu))}{\beta(g(\mu))} \right ) +\cdots\nn\\
  &&=\braket{O(z)O(0)}_{d=4}  
    g(\mu) \left ( \f{\partial \log \mathcal{G}^{(O)}_2}{\partial g(z)}     \f{\beta(g(z))}{\beta(g(\mu))}   
  + 2    \f{\beta(g(z))}{\beta(g(\mu))} \f{\gamma_O(g(z))}{\beta(g(z))}\right.\nn\\
  &&~~~~\left. -2  \f{\gamma_O(g(\mu))}{\beta(g(\mu))} \right ) +\cdots
\eea
where we have employed (appendix \ref{CS0}):
\bea
\f{\partial g(z)}{\partial g(\mu)} =\f{\beta(g(z))}{\beta(g(\mu))}
\eea
and (appendix \ref{CS0}):
\bea
 \f{\partial Z^{(O)2}(g(z),g(\mu)) }{\partial \log g(\mu)}&=& 2\left(\f{\partial g(z)}{\partial \log g(\mu)} \f{\gamma_O(g(z))}{\beta(g(z))} - \f{\partial g(\mu)}{\partial \log g(\mu)} \f{\gamma_O(g(\mu))}{\beta(g(\mu))}\right)\nn\\
 &&Z^{(O)2}(g(z),g(\mu)) 
\eea 
Hence, eq. \eqref{eq:lhs_gen0} for $g \rightarrow g_*$ and $\epsilon \rightarrow 0$ reads (section \ref{2.1A}):
 \bea
 \label{eq:lhsRen}
\mbox{lhs}
&=& \braket{O(z)O(0)}_{d=4}\left( 2c_O -\f{2\gamma_O(g)}{\eps}+\left ( 1-\f{\beta(g)}{\eps g}\right ) \f{\partial\log \braket{O(z)O(0)}}{\partial\log g}\right)
\bigg|_{g \rightarrow g_*}\nn\\
&+&\mbox{finite terms}+\cdots
\nn\\
&=&\braket{O(z)O(0)}_{d=4} \left( 2c_O -\f{2\gamma_O(g)}{\eps}+ g \left ( 1-\f{\beta(g)}{\eps g}\right )  \right. \nn\\
&&\left. \left ( \f{\partial \log \mathcal{G}^{(O)}_2}{\partial g(z)}     \f{\beta(g(z))}{\beta(g)}   
  + 2  \f{\gamma_O(g(z))}{\beta(g)} -2  \f{\gamma_O(g)}{\beta(g)} \right ) \right)\bigg|_{g \rightarrow g_*} \nn\\
&+&\mbox{finite terms}+\cdots
  \eea
Alternatively, we evaluate eq. \eqref{eq:lhs_gen0} in $\tilde d=4-2\epsilon$ dimensions directly by means of eq. \eqref{CSStilde}. The analog of:
\bea
 \f{\partial \log \mathcal{G}^{(O)}_2}{\partial g(z)}     \f{\beta(g(z))}{\beta(g)}   
  + 2    \f{\beta(g(z))}{\beta(g)} \f{\gamma_O(g(z))}{\beta(g(z))} -2  \f{\gamma_O(g)}{\beta(g)}  \bigg|_{g \rightarrow g_*}
\eea
in $\tilde d=4-2\epsilon$ dimensions is:
\bea
 \f{\partial \log \mathcal{G}^{(O)}_2}{\partial \tilde g(z)}     \f{\beta(\tilde g(z), \epsilon)}{\beta(g,\epsilon)}   
  + 2    \f{\beta(\tilde g(z),\epsilon)}{\beta(g,\epsilon)} \f{\gamma_O(\tilde g(z))}{\beta(\tilde g(z),\epsilon)} -2  \f{\gamma_O(g)}{\beta(g,\epsilon)}  \bigg|_{g \rightarrow  g_*}
\eea
where we have employed (appendix \ref{CS1}):
\bea
\f{\partial \tilde g(z)}{\partial g(\mu)} =\f{\beta(\tilde g(z),\epsilon)}{\beta(g(\mu),\epsilon)}
\eea
The lhs of the LET yields:
\bea \label{lhsepsilon}
\mbox{lhs} &=& \braket{O(z)O(0)}_{\tilde d=4-2\epsilon} \left( 2c_O -\f{2\gamma_O(g)}{\eps}+ \right. \nn\\
&& \left.   - \f{\beta(g,\epsilon)}{\eps} \left ( \f{\partial \log \mathcal{G}^{(O)}_2}{\partial \tilde g(z)}     \f{\beta(\tilde g(z),\epsilon)}{\beta(g,\epsilon)}   
  + 2    \f{\gamma_O(\tilde g(z)) - \gamma_O(g)}{\beta(g,\epsilon)} \right ) \right)\bigg|_{g \rightarrow g_*} \nn\\
  &=&  \braket{O(z)O(0)}_{\tilde d=4-2\epsilon} \left( 2c_O -\f{2\gamma_O(g)}{\eps} \right. \nn\\
&& \left. - \left ( \f{\partial \log \mathcal{G}^{(O)}_2}{\partial \tilde g(z)}     \f{\beta(\tilde g(z),\epsilon)}{\epsilon}   
  + 2    \f{\gamma_O(\tilde g(z)) - \gamma_O(g)}{\epsilon} \right ) \right)\bigg|_{g \rightarrow  g_*}\nn\\
\eea
We compare the two computations in eqs. \eqref{eq:lhsRen} and \eqref{lhsepsilon}. It turns out that the only divergent term as $\epsilon \rightarrow 0$ in the lhs of the LET is the one where the factor $-\f{2\gamma_O(g)}{\eps}$ occurs. Since this term multiplies the correlator in $d=4$ dimensions in eq. \eqref{eq:lhsRen} and the correlator in $\tilde d=4-2\epsilon$ dimensions in eq. \eqref{lhsepsilon}, the two computations may differ by finite terms (section \ref{2.1A}) as $\epsilon \rightarrow 0$ because the correlators in $d=4$ and $\tilde d=4-2\epsilon$ dimensions may differ by contributions of order $\epsilon$. \par 
In the case (II), since the space-time integral of the $3$-point correlator in the rhs of the LET is finite, a divergent term as $\epsilon \rightarrow 0$ equal to the one in the lhs must arise from the contribution of the contact term in the rhs that multiplies the corresponding correlator either in $d=4$ or $\tilde d=4-2\epsilon$ dimensions. \par
Therefore, in the case (II) also the finite terms in the lhs and rhs of the LET are actually equal, so that the final form of the LET may be written in terms of the correlators in $d=4$ dimensions as $\epsilon \rightarrow 0$ (section \ref{2.1A}). \par
Instead, in the case (I) the space-time integral of the $3$-point correlator is actually divergent as $\epsilon \rightarrow 0$, so that this divergence mixes with the possible divergence arising from the contact terms in the rhs, and it matters whether the LET is written in terms of the correlators in $d=4$ or $\tilde d=4-2\epsilon $ dimensions. As a consequence, in the case (I) the LET theorem should only be written in terms of the correlators in $\tilde d=4-2\epsilon$ dimensions, including the $3$-point correlator in the rhs (section \ref{2.1A}).

\subsection{LET for $\gamma_{F^2}=0$ in $d=4$} \label{7.1} 

For $\beta(g_*)=0$, the requirement that $\gamma_{F^2}=\gamma_{F^2}(g_*)$ vanishes implies by eq. \eqref{eq:gamF}:
\be
\gamma_{F^2}=\beta'(g_*)=\beta_0^*=0
\ee
so that $g_*$ is a stationary point of the beta function and the conformal dimension of $F^2$ is $\Delta_{F^2}=\Delta_{F^2_0}=d=4$. \par
We get in eq. \eqref{lhsepsilon} as $\epsilon \rightarrow 0$:
\bea
&&- \f{\partial \log \mathcal{G}^{(O)}_2}{\partial \tilde g(z)}     \f{\beta(\tilde g(z),\epsilon)}{\epsilon}   
  - 2    \f{\gamma_O(\tilde g(z)) - \gamma_O(g)}{\epsilon} \bigg|_{g \rightarrow  g_*} \nn\\
&&= - \f{\partial \log \mathcal{G}^{(O)}_2}{\partial \tilde g(z)}     \f{-\epsilon \tilde g(z)}{\epsilon}   
  - 2    \f{\frac{\partial\gamma_O}{\partial g}(\tilde g(z) - g)}{\epsilon} + \cdots\bigg|_{g \rightarrow  g_*} \nn\\
&&=  \f{\partial \log \mathcal{G}^{(O)}_2}{\partial \tilde g(z)} \tilde g(z)
  - 2    \f{\frac{\partial\gamma_O}{\partial g}\eps g_*\log|z\mu|}{\epsilon} + \cdots\bigg|_{g \rightarrow  g_*} \nn\\
&&=  \f{\partial \log \mathcal{G}^{(O)}_2}{\partial \tilde g(z)} g_*
  - 2    \frac{\partial\gamma_O}{\partial g} g_*\log|z\mu|+ \cdots\bigg|_{g \rightarrow  g_*} 
\eea
where we have employed eq. \eqref{m_eps}.
We obtain the very same result for $\beta_0^*=0$, $\beta_1^*=\frac{1}{2}\beta''(g_*) \neq 0$ from the derivative of the correlator in $d=4$ dimensions as $\epsilon \rightarrow 0$:
\bea
&& \f{\partial \log \mathcal{G}^{(O)}_2}{\partial g(z)}     \f{\beta(g(z))}{\beta(g)}   
  + 2   \f{\gamma_O(g(z))  -\gamma_O(g)}{\beta(g)}   \bigg|_{g \rightarrow g_*} \nn\\
&& =\f{\partial \log \mathcal{G}^{(O)}_2}{\partial g(z)}     \f{\beta_1^*(g(z)-g_*)^2}{\beta_1^*(g-g_*)^2}   
  + 2   \f{\gamma_1^{(O)*}((g(z)-g_*)-(g-g_*))}{\beta_1^*(g-g_*)^2}   \bigg|_{g \rightarrow g_*} \nn\\
  &&=\f{\partial \log \mathcal{G}^{(O)}_2}{\partial g}   
  -2   \f{\gamma_1^{(O)*}\beta_1^*(g-g_*)^2 \log|z\mu|}{\beta_1^*(g-g_*)^2}   \bigg|_{g \rightarrow g_*} \nn\\
    &&=\f{\partial \log \mathcal{G}^{(O)}_2}{\partial g}   
  -2   \frac{\partial\gamma_O}{\partial g} \log|z\mu|   \bigg|_{g \rightarrow g_*}
\eea
where we have employed eq. \eqref{D13_rgi_small}.
Either way, we get for the LET as $\epsilon \rightarrow 0$:
 \bea
 \label{eq:LET_LR_0b1}
&& \braket{O(z)O(0)}_{d=4} \left( 2c_O -\f{2\gamma_O(g)}{\eps}+  \left ( 1-\f{\beta(g)}{\eps g}\right )  \right. \nn\\
&&\left. \left(\f{\partial \log \mathcal{G}^{(O)}_2}{\partial \log g}   
  -2  g \frac{\partial\gamma_O}{\partial g} \log|z\mu|  \right)\right) \bigg|_{g \rightarrow g_*} 
+ \mbox{finite terms}+\cdots
    \nn\\
&&= \braket{O(z)O(0)}_{d=4} Z_{F^2}^{-1}\,
\f{1}{2}  \f{N_3(g)}{N_2(g)}
\f{1}{|z|^{-4}}
\int\,\f{1}{|x|^4 |x-z|^4}\, + \cdots \, d^{4-2\epsilon} x
\bigg|_{g \rightarrow g_*}\nn\\
&&\,\,\,\,\,\,+ \mbox{finite terms}+\cdots
\nn\\
&&\,\,\,\,\,\,+\f{1}{2}   Z_O^{2} \int \braket{O(z)O(0)F^2(x)}_{d=4;0;\textrm{c.t.}} \, d^{4} x \bigg|_{g \rightarrow g_*}+  \mbox{finite terms}+\cdots
\eea 
with $Z^{-1}_{F^2}(g_*)=\left(1-\f{\beta(g_*)}{\eps g_*}\right)=1$. Unless $N_3(g_*)=0$, the rhs of the LET is UV divergent\footnote{Its divergence is actually semilocal according to the classification in \cite{Skenderis}.} as $\epsilon \rightarrow 0$ due to the short-distance singularities at $x\sim 0$ and $x\sim z$ (appendix \ref{B}) of the integrand in the first line of the rhs. \par
It turns out (appendix \ref{E}) that also the divergent part of the space-time integral in the rhs depends on corrections of order $\epsilon$ to the anomalous dimension of $F^2$ in $\tilde d= 4-2\epsilon$ dimensions. As a consequence, in the applications in the present paper (section \ref{10}) we only consider the case $O=F^2$, where $\gamma_O =\gamma_{F^2}=0$, $N_3(g_*)=0$ and no ambiguity about divergences and finite terms may arise (section \ref{2.1A}).

\subsection{LET for $\gamma_{F^2}\neq 0$ nonexceptional in $d=4$} \label{7.2}

For a nonvanishing anomalous dimension of $F^2$:
\bea
\gamma_{F^2}&=&\beta'(g_*)=\beta_0^* \neq 0
 \eea
the conformal dimension of $F^2$ is $\Delta_{F^2}=4+\gamma_{F^2}$.
For $\beta_0^* \neq 0$ and $\gamma_1^{(O)*}=\frac{\partial\gamma_O}{\partial g}(g_*) \neq 0$  (appendix \ref{D}), we get in eq. \eqref{eq:lhsRen}:
\bea
&& \f{\partial \log \mathcal{G}^{(O)}_2}{\partial g(z)}     \f{\beta(g(z))}{\beta(g)}   
  + 2   \f{\gamma_O(g(z))  -\gamma_O(g)}{\beta(g)}   \bigg|_{g \rightarrow g_*} \nn\\
&& =\f{\partial \log \mathcal{G}^{(O)}_2}{\partial g(z)}     \f{\beta_0^*(g(z)-g_*)}{\beta_0^*(g-g_*)}   
  + 2   \f{\gamma_1^{(O)*}((g(z)-g_*)-(g-g_*))}{\beta_0^*(g-g_*)}   \bigg|_{g \rightarrow g_*} \nn\\
  &&=\f{\partial \log \mathcal{G}^{(O)}_2}{\partial g}   |z\mu|^{-\beta_0^*}
  +2   \f{\gamma_1^{(O)*}(g-g_*)   (|z\mu|^{-\beta_0^*}-1)}{\beta_0^*(g-g_*)}   \bigg|_{g \rightarrow g_*} \nn\\
    &&=\left(\f{\partial \log \mathcal{G}^{(O)}_2}{\partial g} +\f{2}{\beta_0^*}   \frac{\partial\gamma_O}{\partial g}\right)  |z\mu|^{-\beta_0^*}
      - \f{2}{\beta_0^*}   \frac{\partial\gamma_O}{\partial g} \bigg|_{g \rightarrow g_*}
\eea
where we have employed eq. \eqref{LL}.
Equivalently, we get in eq. \eqref{lhsepsilon}:
\bea
&&- \f{\partial \log \mathcal{G}^{(O)}_2}{\partial \tilde g(z)}     \f{\beta(\tilde g(z),\epsilon)}{\epsilon}   
  - 2    \f{\gamma_O(\tilde g(z)) - \gamma_O(g)}{\epsilon} \bigg|_{g \rightarrow  g_*} \nn\\
&&= - \f{\partial \log \mathcal{G}^{(O)}_2}{\partial \tilde g(z)}     \f{-\epsilon \tilde g(z)+\beta_0^*(\tilde g(z)-g_*)}{\epsilon}   
  - 2    \f{\frac{\partial\gamma_O}{\partial g}(\tilde g(z) - g)}{\epsilon} + \cdots\bigg|_{g \rightarrow  g_*} \nn\\
&&=  -\f{\partial \log \mathcal{G}^{(O)}_2}{\partial \tilde g(z)}  \f{-\epsilon g_*+\beta_0^*(\f{\eps g_*}{\beta_0^*}
 (1-|z\mu|^{-\beta_0^*}+\cdots))}{\epsilon}  \nn\\
 &&~~~   - 2    \f{\frac{\partial\gamma_O}{\partial g} \f{\eps g_*}{\beta_0^*}
 (1-|z\mu|^{-\beta_0^*}+\cdots)
}{\epsilon} + \cdots\bigg|_{g \rightarrow  g_*} \nn\\
    &&=\left(\f{\partial \log \mathcal{G}^{(O)}_2}{\partial g} +\f{2}{\beta_0^*}   \frac{\partial\gamma_O}{\partial g}\right)  |z\mu|^{-\beta_0^*}
      - \f{2}{\beta_0^*}   \frac{\partial\gamma_O}{\partial g} + \cdots\bigg|_{g \rightarrow g_*}
\eea
where we have employed eq. \eqref{notm_eps}.\par
Eq. \eqref{eq:rhs23} yields for the rhs of the LET without contact terms:
 \bea\label{eq:rhs_gamma}
\mbox{rhs}|_{\textrm{no c.t.}}
&=&\braket{O(z)O(0)}_{d=4}Z_{F^2}^{-1}
\f{1}{2}\f{N_3(g)}{N_2(g)}
(2\pi)^{\td}\, C_{\td,\f{\Delta_{F^2}}{2},\f{\Delta_{F^2}}{2}}
 |z |^{-\gamma_{F^2}-2\epsilon}|\mu |^{-\gamma_{F^2}} \bigg|_{g\to g_*} +\cdots \nn\\
&=&\braket{O(z)O(0)}_{d=4}Z_{F^2}^{-1}
\f{1}{2}\f{N_3(g)}{N_2(g)}
\pi^{2-\eps}
\f{ 
\Gamma\big( {2+\eps+\gamma_{F^2}}\big)  
\Gamma\big( \f{-2\epsilon -\gamma_{F^2}}{2}\big)^2 
}
{  \Gamma \big(2 +\f{\gamma_{F^2}}{2}\big)^2
 \Gamma \big( -2\epsilon-\gamma_{F^2}\big)}
 \nn\\
&& |z |^{-\gamma_{F^2}-2\epsilon}|\mu |^{-\gamma_{F^2}} \bigg|_{g\to g_*}+\cdots 
 \eea
 with $Z^{-1}_{F^2}(g_*)=\left(1-\f{\beta(g_*)}{\eps g_*}\right)=1$. 
 Hence, the LET for $\gamma_{F^2}\neq 0$ nonexceptional reads as $\epsilon \rightarrow 0$:
\bea\label{eq:LETgamnotzero0}
&&\braket{O(z)O(0)}_{d=4} \Bigg( 2c_O -\f{2\gamma_O(g)}{\eps}-\f{2}{\beta_0^*} g \frac{\partial\gamma_O}{\partial g} 
+ \bigg(\f{\partial \log \mathcal{G}^{(O)}_2}{\partial \log g} +\f{2}{\beta_0^*} g  \frac{\partial\gamma_O}{\partial g}\bigg)  \nn\\
&&|z\mu|^{-\beta_0^*}
 \Bigg) \bigg|_{g \rightarrow g_*}+\cdots
=\braket{O(z)O(0)}_{d=4} \Bigg({\mathcal{C}}_{O_0,\textrm{div}}(g,\eps)+{\mathcal{C}}_{O_0,\textrm{finite}}(g,0)
+\f{1}{2}\f{N_3(g)}{N_2(g)}
\pi^{2}
\nn\\&&
 \f{ 
\Gamma\big( {2+\gamma_{F^2}}\big)  
\Gamma\big( \f{ -\gamma_{F^2}}{2}\big)^2 
}
{\Gamma \big(2+\f{\gamma_{F^2}}{2}\big)^2
 \Gamma \big( -\gamma_{F^2}\big)}
 |z\mu|^{-\gamma_{F^2}}\Bigg) \bigg|_{g \rightarrow g_*}+\cdots
 \eea

\subsection{LET for $\gamma_{F^2}$ of order $\eps$ in $\tilde d=4-2\epsilon$} \label{7.3}
\label{russians}

According to \cite{Braun1,Braun2}, it is convenient to perform perturbative computations of the anomalous dimensions in massless QCD to higher loops
in dimensional regularization at the Wilson-Fisher zero $\tilde g_*>0$ of the beta function below: 
\be \label{6.200}
0=\beta(\tilde g_*,\eps)=-\eps \tilde g_* +\beta(\tilde g_*)
\ee
where massless QCD in $\td=4-2\epsilon$ dimensions is exactly conformal (appendix \ref{D}).
Given $\beta(g)=-\beta_0 g^3+O(g^5)$, eq. \eqref{6.200} admits a positive solution for arbitrarily small $g$ only for $\beta_0<0$. Indeed, eq. \eqref{6.200} yields:
\be\label{eq:g2eps}
\tilde g_*^2=-\f{\eps}{\beta_0}+O(\eps^2)
\ee
and the positivity of $\tilde g_*^2$ implies $\beta_0<0$. 
For massless QCD in $d=4$ dimensions the latter condition occurs above the conformal window, i.e., for $N_f>N_f^{AF}$ (section \ref{1}). \par
In $\td=4-2\eps$ dimensions, the term $-\eps g$ in $\beta(g,\eps)$ renders $\beta(g,\eps)$ negative for small $g$, and the zero in eq. \eqref{6.200} can be thought of as an analog of the Wilson-Fisher IR zero for QED in $4-2\epsilon$ dimensions\footnote{Indeed, $\beta_0$ in $4$-dimensional QED is negative as well.}. Hence, the theory has a trivial zero in the UV and a nontrivial one in the IR. \par
 The anomalous dimension of $F^2$ reads according to eq. \eqref{gF2}:
 \bea
 \gamma_{F^2}(\tilde g_*)&=&\beta'(\tilde g_*)-\f{\beta(\tilde g_*)}{\tilde g_*}\nn\\
 &=&\beta'(\tilde g_*)-\eps
 \eea
that yields perturbatively: 
\bea
\gamma_{F^2}(\tilde g_*)&=& -2\beta_0g^2-4\beta_1g^4+\cdots \big|_{g=\tilde g_*}\nn\\
&=&-2\beta_0 \left( -\f{\eps}{\beta_0}\right )+O(\tilde g_*^4)\nn\\
&=&2\eps+O(\eps^2)
\eea
We are now interested in the LET at the zero in eq. \eqref{6.200}.
Correspondingly, we employ the ansatz for the solution of the CS equation in $\tilde d=4-2 \epsilon$ dimensions in eq. \eqref{CSStilde}.
In the limit $g(\mu) \rightarrow \tilde g_*$ with fixed $\mu$, the theory in $\tilde d=4-2 \epsilon$ dimensions is conformal:
\bea \label{cotil}
\f{\mathcal{ G}^{(O)}_2(\tilde g(z))}{|z|^{2 \tilde \Delta_{O_0}}}  Z^{(O)2}(\tilde g(z),g(\mu))   \bigg|_{g \rightarrow \tilde g_*}
&=&\f{N_2(\tilde g_*)  \mu^{2\tilde \Delta_{O_0}-2 \tilde\Delta_O}  }{|z|^{2\tilde \Delta_O}}
\eea 
thus implying:
\bea \label{last0}
\mathcal{ G}^{(O)}_2(\tilde g_*)= N_2(\tilde g_*)
\eea
where we have consistently defined the dimensional regularization of eq. \eqref{eq:CFT23} up to contact terms:
\bea
\braket{O(z)O(0)}_{\tilde d=4-2\epsilon}\big|_{g \rightarrow \tg*}= 
\f{N_2(\tg_*)\mu^{2\tD_{O_0}-2\tD_O}  }{|z|^{2\tD_O} } 
\eea
The lhs of the LET yields:
\bea
\mbox{lhs} &=& \braket{O(z)O(0)}_{\tilde d=4-2\epsilon} \left( 2c_O -\f{2\gamma_O(g)}{\eps}+ \right. \nn\\
&& \left.   - \f{\beta(g,\epsilon)}{\eps} \left ( \f{\partial \log \mathcal{G}^{(O)}_2}{\partial \tilde g(z)}     \f{\beta(\tilde g(z),\epsilon)}{\beta(g,\epsilon)}   
  + 2    \f{\gamma_O(\tilde g(z)) - \gamma_O(g)}{\beta(g,\epsilon)} \right ) \right)\bigg|_{g \rightarrow \tilde g_*} \nn\\
  &=&  \braket{O(z)O(0)}_{\tilde d=4-2\epsilon} \left( 2c_O -\f{2\gamma_O(g)}{\eps} \right. \nn\\
&& \left. - \left ( \f{\partial \log \mathcal{G}^{(O)}_2}{\partial \tilde g(z)}     \f{\beta( \tilde g(z),\epsilon)}{\epsilon}   
  + 2    \f{\gamma_O(\tilde g(z)) - \gamma_O(g)}{\epsilon} \right ) \right)\bigg|_{g \rightarrow \tilde g_*}\nn\\
 &=& \braket{O(z)O(0)}_{\tilde d=4-2\epsilon} \left( 2c_O -\f{2\gamma_O(g)}{\eps} \right)\bigg|_{g \rightarrow  \tilde g_*}
\eea
since:
\bea
 \f{\beta(\tilde g(z),\epsilon)}{\epsilon} \bigg|_{g \rightarrow \tilde g_*}=0
 \eea
 and:
 \bea
\f{\gamma_O(\tilde g(z)) - \gamma_O(g)}{\epsilon} \bigg|_{g \rightarrow \tilde g_*}=0
\eea
 for fixed $\epsilon$. By employing eq. \eqref{eq:Cuv} for $u=v=-1$, we get:
\bea
\mbox{rhs}|_{\textrm{no c.t.}}&=&
\braket{O(z)O(0)}_{\tilde d=4-2\epsilon} Z_{F^2}^{-1}
\f{1}{2}\f{N_3(g)}{N_2(g)}
\pi^{2-\eps}
\f{ 
\Gamma\big( {2-\eps+\gamma_{F^2}}\big)  
\Gamma\big( \f{ -\gamma_{F^2}}{2}\big)^2 
}
{\Gamma \big(2-\eps+\f{\gamma_{F^2}}{2}\big)^2
 \Gamma \big( -\gamma_{F^2}\big)} 
 |z\mu|^{-\gamma_{F^2}} \bigg|_{g \rightarrow  \tg_*}\nn\\
 &=&0
  \eea
that vanishes because: 
\be\label{ZFwf}
 Z_{F^2}^{-1}(\tilde g_*,\eps) = \left (1-\f{\beta(\tilde g_*)}{\eps \tilde g_*}\right )=0
 \ee
Therefore: 
\bea\label{eq:LETgamma_eps}
&&\braket{O(z)O(0)}_{\tilde d=4-2\epsilon}\left( 2c_O 
-\f{2\gamma_O(g)}{\eps}
\right)\bigg|_{g \rightarrow \tilde g_*}\nn\\
&&~~~~~=\f{1}{2} Z_O^{2} \int \braket{O(z)O(0)F^2(x)}_{\tilde d=4-2\epsilon;0;\textrm{c.t.}} \, d^\td x \bigg|_{g \rightarrow  \tilde g_*} 
\eea
where now the above equality holds for any small $\epsilon$.

\section{Physics implications of the LET by matching the lhs with the rhs} \label{8}

The LET implies the matching of the lhs with the corresponding terms in the rhs. 
The matching and its physics implications depend on the value of $\gamma_{F^2}$. 
We perform the matching term by term for the cases (II) and (III) in this section.
The case (I) is considered in section \ref{10} for $O=F^2$ and in appendix \ref{E} 
to order $g^2$ in perturbation theory.

\subsection{$\gamma_{F^2}\neq 0$ nonexceptional in $d=4$} 
\label{8.2}

The LET is given by eq. \eqref{eq:LETgamnotzero0} as $\epsilon \rightarrow 0$.

\subsubsection{Matching the power-like terms}

The matching of the power-like terms in the lhs and rhs of eq. \eqref{eq:LETgamnotzero0} leads to the constraint:
\bea
&&\f{\partial \log \mathcal{G}^{(O)}_2}{\partial \log g} +\f{2}{\beta_0^*} g  \frac{\partial\gamma_O}{\partial g}
    \bigg|_{g \rightarrow g_*}
    =\f{1}{2}\f{N_3(g)}{N_2(g)}
\pi^{2}
 \f{ 
\Gamma\big( {2+\gamma_{F^2}}\big)  
\Gamma\big( \f{ -\gamma_{F^2}}{2}\big)^2 
}
{\Gamma \big(2+\f{\gamma_{F^2}}{2}\big)^2
 \Gamma \big( -\gamma_{F^2}\big)} \bigg|_{g \rightarrow g_*}
\nn\\
\eea

\subsubsection{Matching the $1/\eps$ terms}
\label{sec:pole2} 
The $1/\eps$ contribution in the lhs of eq. \eqref{eq:LETgamnotzero0} may only be matched by the UV divergent contact term in the rhs. 
By means of eqs. \eqref{eq:ct00} and \eqref{eq:bareC}, the matching implies:
\be\label{eq:C1div}
-\f{2\gamma_O(g)}{\eps}\bigg|_{g \rightarrow g_*}={\mathcal{C}}_{O_0,\textrm{div}}(g,\eps)\bigg|_{g \rightarrow g_*}
\ee

\subsubsection{Matching the finite terms}

Similarly, the finite contribution proportional to the $2$-point correlator in the lhs may only be matched by a finite contact term in the rhs. By means of eqs. \eqref{eq:ct00} and \eqref{eq:bareC}, the matching implies:
\be\label{eq:C1finite}
2c_O -\f{2}{\beta_0^*} g{\f{\partial \gamma_O}{\partial g}}\bigg|_{g  \rightarrow g_*}= {\mathcal{C}}_{O_0,\textrm{finite}}(g,0)\bigg|_{g  \rightarrow g_*}
\ee
by employing eq. \eqref{eq:gg0rel0} and $Z_{F^2}^{-1}  \rightarrow 1$ for $g  \rightarrow g_*$.\par

\subsubsection{Renormalized LET with $\gamma_{F^2}\neq 0$ nonexceptional}

By summarizing, the multiplicatively renormalized correlator $\braket{OOF^2}$ at a nontrivial zero of the beta function in massless QCD with $\gamma_{F^2}$ nonexceptional  in $d=4$ dimensions reads:
 \bea
 \braket{O(z)O(0)F^2(x)}_{d=4}\bigg|_{g \rightarrow g_*}&=& \f{N_3(g)\mu^{2\Delta_{O_0}+\Delta_{F^2_0}-2\Delta_O - \Delta_{F^2}}}{|z|^{2\Delta_O -\Delta_{F^2}} |x|^{\Delta_{F^2}}|x-z|^{\Delta_{F^2}}}
 \nn\\
&&\hspace{-4.0cm} +\left(-\f{2\gamma_O(g)}{\eps}+ 2c_O  -\f{2}{\beta_0^*}g{\f{\partial \gamma_O}{\partial g}}  \right) \braket{O(z)O(0)}(\delta^{(4)}(x)+\delta^{(4)}(x-z))\bigg|_{g \rightarrow g_*}
 \eea
with $z\neq 0$. Remarkably, the correlator needs an extra infinite additive renormalization, i.e., it is not made finite by just the multiplicative renormalization. 
However, the additive renormalization may be fixed unambiguously so that both sides of the LET are finite.
Hence, for $\gamma_{F^2}$ nonexceptional, the renormalized version of the LET reads as $\epsilon \rightarrow 0$:
\bea\label{eq:LETgamnotzero00}
&&\braket{O(z)O(0)}_{d=4} \Bigg( 2c_O -\f{2\gamma_O(g)}{\eps}- {\mathcal{C}}_{O_0,\textrm{div}}(g,\eps)  -\f{2}{\beta_0^*} g \frac{\partial\gamma_O}{\partial g}
+ \bigg(\f{\partial \log \mathcal{G}^{(O)}_2}{\partial \log g}
\nn\\
&&~~ +\f{2}{\beta_0^*} g  \frac{\partial\gamma_O}{\partial g}\bigg)  |z\mu|^{-\beta_0^*}
 \Bigg) \bigg|_{g \rightarrow g_*}
   =\braket{O(z)O(0)}_{d=4} \Bigg({\mathcal{C}}_{O_0,\textrm{finite}}(g,0)
   \nn\\&&~~~+
\f{1}{2}\f{N_3(g)}{N_2(g)}
\pi^{2}
 \f{ 
\Gamma\big( {2+\gamma_{F^2}}\big)  
\Gamma\big( \f{ -\gamma_{F^2}}{2}\big)^2 
}
{\Gamma \big(2+\f{\gamma_{F^2}}{2}\big)^2
 \Gamma \big( -\gamma_{F^2}\big)}|z\mu|^{-\gamma_{F^2}}\Bigg) \bigg|_{g \rightarrow g_*}
 \eea
 where $\beta_0^*=\gamma_{F^2}$ and:
\bea
-\f{2\gamma_O(g)}{\eps}
 - {\mathcal{C}}_{O_0,\textrm{div}}(g,\eps) \bigg|_{g \rightarrow g_*}=0
 \eea

\subsection{ $\gamma_{F^2}$ of order $\eps$ in $\tilde d= 4-2\epsilon$}
\label{sec:Matchingeps}
 
The matching of the lhs and rhs in eq. \eqref{eq:LETgamma_eps}, which corresponds to massless QCD at the Wilson-Fisher zero of $\beta(g, \eps)$ in $\tilde d=4-2\eps$ dimensions, implies:
\be\label{eq:ceps}
2c_O-\f{2\gamma_O(g)}{\eps}\bigg|_{g  \rightarrow \tg_*}= {\mathcal{C}}_{O_0}(g,\eps)\bigg|_{g  \rightarrow \tg_*}
\ee
with ${\mathcal{C}}_{O_0}(g,\eps)$ the coefficient in eq. \eqref{eq:bareC} in dimensional regularization. Surprisingly, only the constraint on the contribution of the contact terms survives, while no constraint on the space-time structure of the correlators occurs.

\section{LET in asymptotically conformal YM theories} \label{9A}

The above versions of the LET have been obtained taking the limit $g(\mu) \rightarrow g_*$ keeping $\mu$ fixed that results in a conformal YM theory (appendix \ref{D}).\par
Yet, we may also take the limit $g(\mu) \rightarrow g_*$ keeping $\Lrgi$ fixed. In this case necessarily $\mu \rightarrow 0^+/+\infty$ and the resulting YM theory may be asymptotically conformal in the IR/UV. 
It turns out that for $F^2$ marginal the theory is not asymptotically conformal in the IR/UV (appendix \ref{D}), but it is asymptotically 
conformal for $F^2$ irrelevant (appendix \ref{D}).\par
Therefore, in the latter case we may try to evaluate the LET asymptotically in the IR/UV. Indeed, the LET holds independently of the conformal limit in the following form:
 \bea
&&\braket{O(z)O(0)}\left( 2c_O - \f{2\gamma_O(g)}{\eps}+\left ( 1-\f{\beta(g)}{\eps g}\right ) \f{\partial\log \braket{O(z)O(0)}}{\partial\log g}\right)\bigg|_{\dr} \nn\\
&&=\f{1}{2}Z_{F^2}^{-1}\int \braket{O(z) O(0) F^2(x)}' - \braket{O(z)O(0)}\braket{F^2(x)} d^\td x \bigg|_{\dr}\nn\\
&&\,\,\,\,\,\,+\f{1}{2} Z_O^{2} \int \braket{O(z)O(0)F^2(x)}_{0;\textrm{c.t.}} \, d^\td x \bigg|_{\dr}
\eea
where now by eq. \eqref{CSS0_IR}:
\be
 \langle O(z) O(0) \rangle \sim\f{\mathcal{G}^{(O)}_2(g_*)}{z^{2D}} 
 (|z|\Lrgi)^{-2\gamma_0^{(O)*}}\left( \f{1}{g_*-g(\mu)}\right )^{\f{2\gamma_0^{(O)*}}{\beta_0^*}} Z^{(O)'2}(g_*-g(\mu))
\ee
Interestingly, up to the overall normalization, the IR/UV asymptotics coincides with the exactly conformal case. In fact, in the asymptotically conformal case it is more natural to write the lhs of the LET in terms of the derivative with respect to the RGI scale $\Lrgi$ \cite{MBR} (section \ref{1}), a task that we will consider elsewhere.

\section{LET for $O=F^2$ in massless QCD}
\label{10}

The choice $O=F^2$ in the LET is especially interesting, since the OPE of $F^2$ with itself has been extensively investigated in massless QCD \cite{Z1,Z2,Z3} -- not necessarily inside the conformal window -- and an UV divergent contact term has been discovered \cite{Z1} to two loops, i.e., to order $g^4$ in perturbation theory, and to higher loops \cite{Z2}.\par
Remarkably, a closed form for this contact term has been obtained in \cite{Z3} to all orders of perturbation theory.\par
The rationale for the existence of the divergent contact term discovered in \cite{Z1} is somehow mysterious.
We demonstrate that, in the conformal window of massless QCD, the LET is actually the aforementioned rationale.\par
Besides, the exact computation in \cite{Z3} allows us to perform the following powerful test of both the LET and \cite{Z3} for $O=F^2$, with $Z^{-1}_{F^2}=1$ at $g_*$, $2c_{F^2}=4$, $\Delta_{F^2}= 4+\gamma_{F^2}$ in eq. \eqref{eq:LETgamnotzero0}. \par
The contribution of the contact term in the rhs of eq. \eqref{eq:LETgamnotzero0} is computed as follows. By eqs. \eqref{eq:OPEr} and \eqref{eq:OPEb}, 
the relevant terms in the OPE of $F^2$ with itself read:
\bea\label{eq:OPEus}
F^2(x)F^2(0)&=& \cdots + C_{1}^{(F^2,F^2)}(x) F^2(0)+ \cdots
\eea
and correspondingly for the bare operator $F_0^2$:
\bea\label{eq:OPEusb}
F_0^2(x)F_0^2(0)&=& \cdots + C_{1}^{(F^2_0,F^2_0)}(x) F_0^2(0)+ \dots
\eea
with: 
\be
C_{1}^{(F^2_0,F^2_0)}(x)= Z_{F^2}^{-1}C_{1}^{(F^2,F^2)}(x) 
\ee
By eqs. \eqref{eq:C10OF}, \eqref{eq:ct00} and \eqref{eq:bareC},
${\mathcal{C}}_{F^2,\textrm{div}}$ reads: 
\be \label{Coeff}
C_{1}^{(F^2,F^2)}(x)= \left( {\mathcal{C}}_{F^2,\textrm{div}}(g,\eps)+{\mathcal{C}}_{F^2,\textrm{finite}}(g,\eps)  \right)\delta^{(4)}(x)+ C_{1}^{(F^2,F^2)\prime}(x)
\ee
The computation in \cite{Z3} implies in dimensional regularization (appendix \ref{A}):
\bea\label{eq:ZLtous}
 {\mathcal{C}}_{F^2,\textrm{div}}(g,\eps)&=& -\f{4}{\eps}\left(1-\f{\beta(g)}{g\eps} \right)^{-1}\left(\f{1}{2}g\f{\partial}{\partial g} \left(\f{\beta(g)}{g} \right)  - \f{\beta(g)}{g}\right)
\eea
Therefore:
\bea\label{eq:C0pQCD}
{\mathcal{C}}_{F^2_0,\textrm{div}}(g,\eps)
&=& -\f{4}{\eps}\left(\f{1}{2}g\f{\partial}{\partial g} \left(\f{\beta(g)}{g} \right)  - \f{\beta(g)}{g}\right)
\eea
Eq. \eqref{eq:C0pQCD} yields:
\bea\label{eq:c1DR}
{\mathcal{C}}_{F^2_0,\textrm{div}}(g_*,\eps)
&=& -\f{4}{\eps}\left(\f{1}{2}g\f{\partial}{\partial g} \left(\f{\beta(g)}{g} \right)  - \f{\beta(g)}{g}\right)\bigg|_{g  \rightarrow g_*}\nn\\
&=&-\f{2}{\eps}\left(g\f{\partial}{\partial g} \left(\f{\beta(g)}{g} \right) \right)\bigg|_{g  \rightarrow g_*} \nn\\
&=&-\f{2\gamma_{F^2}}{\eps}
\eea
according to eq. \eqref{gF2}. We distinguish the following three cases.

\subsection{$\gamma_{F^2}=0$ in $d=4$}
\label{10.1}

Since $\gamma_{F^2}=0$ the divergent part of the lhs of the LET vanishes. Moreover, by the computation above, also the divergent part of the contact term in the rhs vanishes. Consequently, also $N_3(g_*)=0$, since it multiplies a divergent integral. Hence, the LET reads:
 \bea
 \label{eq:LET_LR_0b2}
&& \braket{F^2(z)F^2(0)}_{d=4} \left( 2c_{F^2}+  \left ( 1-\f{\beta(g)}{\eps g}\right )  \right. 
\nn\\&&
\left. \left(\f{\partial \log \mathcal{G}^{(F^2)}_2}{\partial \log g}   
  -2  g \frac{\partial\gamma_{F^2}}{\partial g} \log|z\mu|  \right)\right) \bigg|_{g  \rightarrow g_*} + \cdots 
    \nn\\
&&={\mathcal{C}}_{O_0,\textrm{finite}}(g,0) \braket{F^2(z)F^2(0)}_{d=4}\bigg|_{g  \rightarrow g_*}+ \cdots 
\eea 
that implies:
\be\label{eq:statio}
g\f{\partial\gamma_{F^2}}{\partial g}\bigg|_{g=g_*} =0
\ee 
because the is no logarithmic term in the rhs matching the one in the lhs of eq. \eqref{eq:LET_LR_0b2}.
Interestingly, the above equation translates into a constraint on the first and second derivatives of the beta function at the zero:
\bea\label{eq:statioc}
0&=&g\f{\partial}{\partial g}\left(g\f{\partial}{\partial g}\left(\f{\beta(g)}{g} \right)  \right)\bigg|_{g=g_*}\nn\\
&=&g\f{\partial}{\partial g}\left(\beta'(g)-\f{\beta(g)}{g}\right)\bigg|_{g=g_*}\nn\\
&=&g\beta^{\prime\prime}(g)-  g\f{\partial}{\partial g}\left(\f{\beta(g)}{g} \right) \bigg|_{g=g_*}\nn\\
&=&g_*\beta^{\prime\prime}(g_*)-\beta'(g_*)
\eea
Hence, for $g_*>0$:
\be\label{eq:fs0}
\gamma_{F^2}=\beta'(g_*)=\beta^{\prime\prime}(g_*)=0
\ee 
Consequently, the Taylor expansion of $\beta(g)$ reads:
\bea\label{eq:Taylor}
\beta(g)&=&\beta(g_*)+\beta'(g_*)(g-g_*)+\f{1}{2} \beta^{\prime\prime}(g_*)(g-g_*)^2 +\f{1}{6}\beta^{\prime\prime\prime}(g_*)(g-g_*)^3\nn\\
&&+O((g-g_*)^4)\nn\\
&=&\f{1}{6}\beta^{\prime\prime\prime}(g_*)(g-g_*)^3+O((g-g_*)^4)
\eea

\subsection{$\gamma_{F^2}\neq 0$ nonexceptional in $d=4$}
\label{10.2}
Remarkably, for $\gamma_{F^2}\neq 0$ the value implied by the LET for ${\mathcal{C}}_{F^2_0,\textrm{div}}$ in eq. \eqref{eq:C1div} for $O=F^2$  actually coincides with eq. \eqref{eq:c1DR}. \par

\subsection{$\gamma_{F^2}$ of order $\eps$ in $\td =4-2\eps$}
\label{10.3}
Since:
\bea \label{WF}
\beta(\tg_*,\epsilon)=- \epsilon  \tg_* + \beta(\tg_*)=0
\eea
by the above equation:
\bea \label{s0}
Z^{-1}_{F^2}(g,\epsilon) \bigg|_{g \rightarrow \tg_*}=1-\frac{\beta(g)}{\epsilon g} \bigg|_{g \rightarrow \tg_*}=0
\eea
and the relation between $F^2_0$ and $F^2$ is singular at $g=\tg_*$ despite the theory is defined in $\tilde d=4-2\epsilon$ dimensions.
By eq. \eqref{s0} the LET reduces exactly to (section \ref{7.3}):
\be \label{eq:III}
4-\f{2\gamma_{F^2}(g)}{\eps}\bigg|_{g \rightarrow \tg_*}={\mathcal{C}}_{F^2_0}(g,\eps)\bigg|_{ g \rightarrow \tg_*}
\ee
since $c_{F^2}=2$ (section \ref{4.2}).
Interestingly, eq. \eqref{s0} also implies that
only the divergent part of ${\mathcal{C}}_{F^2_0}(\tg_*,\eps)$ survives. 
Indeed:
\bea
{\mathcal{C}}_{F^2_0}(g,\eps)= Z^{-1}_{F^2}(g,\epsilon) {\mathcal{C}}_{F^2,\textrm{div}}(g,\eps)+Z^{-1}_{F^2}(g,\epsilon){\mathcal{C}}_{F^2,\textrm{finite}}(g,\eps)
\eea
and ${\mathcal{C}}_{F^2,\textrm{div}}(g,\eps)$ has a pole at $g=\tg_*$ (appendix \ref{A}) that is cancelled by the zero of $Z^{-1}_{F^2}(g,\epsilon)$ in eq. \eqref{s0}, while
${\mathcal{C}}_{F^2,\textrm{finite}}(g,\eps)$ may not have such a pole because, being finite, it does not contain the powers of $1/\epsilon$ that would arise in the expansion of the pole $\left(1-\frac{\beta(g)}{\epsilon g}\right)^{-1}$. Therefore:
\bea
{\mathcal{C}}_{F_0^2,\textrm{finite}}(\tg_*,\eps)
= Z^{-1}_{F^2}(\tg_*,\epsilon){\mathcal{C}}_{F^2,\textrm{finite}}(\tg_*,\eps)=0
\eea
and (appendix \ref{A}):
\bea  \label{IIIF}
{\mathcal{C}}_{F_0^2,\textrm{div}}(\tg_*,\eps)
&=& -\f{4}{\epsilon}\left(\f{1}{2}g\f{\partial}{\partial g} \left(\f{\beta(g)}{g} \right)  - \f{\beta(g)}{g}\right)\bigg|_{g=\tg_*}\nn\\
&=& -\f{4}{\epsilon}\left(\f{1}{2} \gamma_{F^2} - \eps \right)\nn\\
&=&-\f{2}{\epsilon}\gamma_{F^2}+4
\eea
according to the LET in eq. \eqref{eq:III}.

\subsection{On the merging of two nontrivial zeroes} \label{10.4}

\subsubsection{A short summary of merging}

It has been conjectured \cite{Kaplan} that in massless QCD there exists a nontrivial UV zero of the beta function in addition to the Banks-Zaks (BZ)
IR zero in the conformal window \cite{BZ}, as a way to explain the disappearance -- as $x=\f{N_f}{N}$ varies -- of the conformal window at its lower edge for $x=x_{c}$. 
In fact, the proposed merging \cite{Kaplan} of an UV and IR zero of the QCD beta function appears to allow for
two different but related scenarios -- (1) and (2) -- summarized as follows (section VI of \cite{Kaplan}):\par
(1)"If the picture emerging from the previous examples also holds for QCD, then conformality is lost when the 
BZ fixed point annihilates with another UV fixed point. Therefore, we predict that when $x$ is slightly larger 
than $x_{c}$, QCD has an UV fixed point, in addition to the IR fixed point (and the free UV fixed point). The UV fixed point, called QCD$^*$, is a different CFT compared to the usual IR fixed point; for example, the dimension of the operator $\bar \psi \psi$ should be different between the two fixed points.
What is the nature of QCD$^*$? It could be that the $\beta$-function for the QCD gauge coupling simply has an unstable fixed point at strong coupling." \par
While for $\mathcal{N}=1$ SUSY QCD the exclusion of the scenario (1) -- and of asymptotic safety -- has been established by means of the a-theorem and the constraints on the mass anomalous dimension $\gamma_m$ \cite{scalar},
the same conclusion could not be reached for massless QCD by the same methods, though some supporting arguments \cite{scalar} have been given in the Veneziano limit of QCD.
In fact, in SU($N$) $\mathcal{N}=1$ SUSY QCD the disappearance of the BZ zero at the lower edge of the conformal window may be explained by assuming Seiberg duality \cite{Seiberg} to the same theory but with a different number of colors $N'=N_f-N$.\par
The scenario (1) has also been investigated for QCD in perturbation theory \cite{DiPietro}. We investigate it nonperturbatively in the present paper in the light of the LET (section \ref{MLET}).\par
(2) "However, this picture implicitly assumes that the set of relevant operators in QCD$^*$ consist of just kinetic terms for the gauge fields and fermions, as is the case at weak coupling. At strong coupling other operators could be relevant as well, and guided by our defect QFT example of Sec. V, it is natural to consider the possibility that a chirally symmetric four-fermion operator is relevant in QCD$^*$
\bea \label{QCD00}
\mathcal{L}_{QCD^*} = \mathcal{L}_{QCD} -c(\bar \psi \gamma_{\mu} t^a \psi)^2 
\eea
and that the unstable fixed point exists at some value $(g_*, c_*)$ in the two-dimensional space
of couplings. ... This is essentially the picture advocated by Gies and Jaeckel \cite{Gies}".\par
In the scenario (2) QCD$^*$ is an extension of QCD that contains in its defining Lagrangian at the cutoff scale an extra coupling $c$ in addition to $g$\footnote{If the Lagrangian of QCD$^*$ in eq. \eqref{QCD00} is interpreted as the local part of the effective Lagrangian of QCD obtained integrating out its UV degrees of freedom, then $c$ is necessarily a function of $g$ and we are back to the scenario (1).} and, therefore, the QCD renormalized LET of the present paper does not apply to it. Indeed, in the scenario (2) the link of QCD with QCD$^*$ is considerably weaker than in the scenario (1), since QCD$^*$ would be a CFT different from QCD that is only conjectured to flow to the BZ zero of QCD in the IR.
Moreover, in the scenario (2) the merging of the two zeroes is conjectured to actually occur in the beta function for the coupling $c$ \cite{Kaplan} and not for a flow only involving the gauge coupling $g$.\par
An example of the above merging, but in a framework different from QCD and QCD$^*$, where the role of $(\bar \psi \gamma_{\mu} t^a \psi)^2 $ in the Lagrangian of QCD$^*$ is played by a double-trace operator involving scalar fields, has been realized in a concrete weakly coupled gauge theory in $d=4$ dimensions \cite{Benini}.

\subsubsection{Constraints on merging from the LET} \label{MLET}

In the scenario (1) of \cite{Kaplan} above, it is conjectured that there exist two nontrivial zeroes of $\beta(g)$, $g_-$ and $g_+$,
with $g_-$ the BZ IR zero and $g_+$ a new zero in the UV that merges with $g_-$ for a certain critical value $x_{c}$. \par
We demonstrate momentarily that the model of merging for $\beta(g)$ in the scenario (1) of \cite{Kaplan} contradicts the implications of the LET for $F^2$ marginal in eq. \eqref{eq:fs0}  and, therefore, it is excluded by the LET in the conformal window of massless QCD. \par
The beta function \cite{Kaplan}:
\be\label{eq:merging}
\beta(a,x)=\f{da}{d\log\mu}=(x-x_c) -(a -a_c)^2
\ee
has two real simple zeroes $a_\pm=a_c\pm\sqrt{x-x_c}$ for $x>x_c$. They approach each other as $x\to x_c$ and coalesce at $x=x_c$, where $a_+=a_-=a_c$. \par
By setting $x=\f{N_f}{N}$
and $a=g^2$, eq. \eqref{eq:merging} has been conjectured \cite{Kaplan} to hold for QCD in the scenario (1) in the conformal window $x\gtrsim x_c$ close to its lower edge\footnote{The analysis and conclusions do not change by taking $a=g$.} $x_c=(\f{N_f}{N})_c$. 
For $x=x_c$, eq. \eqref{eq:merging} satisfies:
\bea
&&\beta(a_c,x_c)=\beta^\prime(a_c,x_c)=0\nn\\
&&\beta^{\prime\prime}(a_c,x_c)=-2<0
\eea
where $'$ is the partial derivative with respect to $a$, which in turn, by means of the relation $\beta(a,x)=2g\beta(g,x)$ and eq. \eqref{gamma0} for $\gamma_{F^2}$ implies:
\bea\label{eq:mema}
&&\gamma_{F^2}(g_c)=\beta^\prime(g_c,x_c)=0\nn\\
&&\beta^{\prime\prime}(g_c,x_c)=2g_c \beta^{\prime\prime}(a_c,x_c)=-4g_c    <0
\eea
where $'$ is the partial derivative with respect to the argument, $g$ or $a$, of the corresponding functions.
Hence, in the merging model \cite{Kaplan} $F^2$ is inevitably marginal at the zero with $\beta^{\prime\prime}(g_c,x_c)<0$, i.e., the zero is a maximum of the beta function.
On the contrary, the LET for $F^2$ marginal, i.e., $\gamma_{F^2}=0$ at $g=g_c$, implies that also the second derivative of the beta function vanishes at the zero, according to eq. \eqref{eq:fs0} that contradicts eq. \eqref{eq:mema}. Hence, the LET excludes the realization in massless QCD of the model in eq. \eqref{eq:merging}. \par
To evade the constraint in eq. \eqref{eq:fs0} due to the LET,
we consider a generalization of eq. \eqref{eq:merging} of the type:
\be\label{eq:merging_g}
\beta(a,x)=(x-x_c) -(a -a_c)^{2n}
\ee
for integer $n>1$, in a neighborhood of $x_c$, with $x\gtrsim x_c$. Indeed, the latter model realizes the merging of the two real zeroes:
\bea
 a_\pm=a_c\pm (x-x_c)^{\f{1}{2n}}
 \eea
 at $x=x_c$, and it also fulfils eq. \eqref{eq:fs0} at the merging point:
\be
\beta(a_c,x_c)=\beta^\prime(a_c,x_c)=\beta^{\prime\prime}(a_c,x_c)=0
\ee
so that for $n >1$ the two real zeroes occur together with multiple complex zeroes that merge at the merging point. In this case, the LET does not seem to rule out the generalized merging models above.

\section{Acknowledgments}

We would like to thank Marco Serone for a clarifying discussion about the nature of QCD$^*$.

\appendix

\section{Anomalous dimension of $F^2$} 
\label{C}

We recall two methods to compute the anomalous dimension of the canonically normalized operator $\Tr F^2=\frac{1}{2}F^2$ in massless QCD and $\mathcal{N}=1$ SUSY YM theory.\par
The first one follows from the multiplicative renormalization of $\Tr F^2$ in $\overline{MS}$-like schemes. 
The second one follows from the RG invariance of the trace anomaly, which in $d=4$ dimensions \cite{Spiridonov} reads:
\be\label{trace-anom-pg}
T_{\alpha \alpha} = \f{\beta(g)}{g}\Tr F^2
\ee
The two methods are not really independent. Indeed, in $\tilde d=4-2\epsilon$ dimensions:
\be
T_{\alpha \alpha} = - \epsilon \Tr F^2_0
\ee
with $\Tr F^2_0$ the bare operator. Expressing the rhs of the equation above in terms of the renormalized operator $\Tr F^2$ in $\overline{MS}$-like schemes, we obtain:
\bea
T_{\alpha \alpha} &=& - \epsilon Z^{-1}_{F^2}(g,\epsilon) Z_{F^2}(g,\epsilon) \Tr F^2_0 \nonumber \\
 &=& - \epsilon Z^{-1}_{F^2}(g,\epsilon) \Tr F^2 \nonumber \\
 &=& - \epsilon \left ( 1-\f{\beta(g)}{g\epsilon}\right ) \Tr F^2 \nonumber \\
 &=& \f{\beta(g,\epsilon)}{g} \Tr F^2 \nonumber \\
 \eea
with:
\bea
\f{dg}{d\log\mu}=\beta(g,\epsilon)= -\epsilon g+\beta(g)
\eea
the beta function in $\tilde d$ dimensions.
Hence, the trace anomaly is a consequence of the multiplicative renormalization of $\Tr F^2$. 
In turn, the anomalous dimension of $\Tr F^2$ follows from the multiplicative renormalization of $\Tr F^2$ (appendix \ref{C1}).
Finally, the anomalous dimension of $\Tr F^2$ also follows from the trace anomaly (appendix \ref{C2}).

\subsection{Multiplicative renormalization of $F^2$ in $\overline{MS}$-schemes}\label{C1}

In massless QCD and $\mathcal{N}=1$ SUSY YM theory, by limiting ourselves to gauge-invariant correlators, $\Tr F^2$ is multiplicatively renormalizable in $\overline{MS}$-like schemes up to gauge-invariant operators proportional to the equations of motion:
\be
\Tr F^2=Z_{F^2}(g,\epsilon) \Tr F^2_0
\ee
with $Z_{F^2}(g,\epsilon)$ \cite{Spiridonov,Z1}:
\be\label{ZF2}
Z_{F^2}(g,\epsilon)=1+g\f{\partial}{\partial g}\log Z_g(g,\epsilon) = \left ( 1-\f{\beta(g)}{g\epsilon}\right )^{-1}
\ee
where $Z_g(g,\epsilon)$ is the renormalization factor for the gauge coupling in dimensional regularization:
\bea
g_0=Z_g \mu^\epsilon g(\mu)
\eea
and $\beta(g) = -g\,\f{d\log Z_g}{d\log\mu}$.
Hence, by employing $\f{d}{d\log\mu} = \beta(g,\epsilon )\f{\partial}{\partial g}+\f{\partial}{\partial\log\mu}$ and eq. \eqref{ZF2}, the anomalous dimension of $\Tr F^2$ reads:  
\be\label{gF2}
\gamma_{F^2}(g)=-\f{d\log Z_{F^2}}{d\log\mu}=-\beta(g,\epsilon)\f{\partial\log Z_{F^2}}{\partial g}=g\f{\partial}{\partial g}\left (\f{\beta(g,\epsilon)}{g}\right )
= g\f{\partial}{\partial g}\left (\f{\beta(g)}{g}\right )
\ee
From eq. (\ref{ZF2}) we also obtain $Z_{F^2}(g,\epsilon)$ and $Z_g(g,\epsilon)$ to two-loops in dimensional regularization:
\bea\label{Zp}
Z_{F^2}(g,\epsilon)&=& 1-\f{\beta_0g^2}{\epsilon}-
\f{\beta_1g^4}{\epsilon} + \f{\beta_0^2g^4}{\epsilon^2} +O(g^6)
\eea
and:
\bea 
Z_g(g,\epsilon) &=& 1-\f{\beta_0g^2}{2\epsilon}-\f{\beta_1 g^4}{4\epsilon} +\f{3\beta_0^2g^4}{8\epsilon^2} +O(g^6)
\eea
It follows:
\bea \label{Zp1}
Z_g^2(g,\epsilon) &=& 1-2\f{\beta_0g^2}{2\epsilon}-2\f{\beta_1 g^4}{4\epsilon}+2\f{3\beta_0^2g^4}{8\epsilon^2}+\left(\f{\beta_0g^2}{2\epsilon} \right)^2+O(g^6)\nonumber\\
&=&
 1-\f{\beta_0g^2}{\epsilon}-\f{\beta_1 g^4}{2\epsilon} +\f{\beta_0^2g^4}{\epsilon^2} +O(g^6)
\eea
Thus, as it has been observed in \cite{Z1}, $Z_{F^2}$ coincides with $Z_g^2$ only to one loop. \par

\subsection{Anomalous dimension of $F^2$ from the RG-invariance of the trace anomaly} \label{C2}

Eq. (\ref{gF2}) also follows \cite{MBM,scalar} by comparing the CS equation for the trace anomaly with the one for 
 $\Tr F^2$.
We report essentially the derivation in \cite{MBM}. The CS equation for the trace anomaly reads:
\bea
\label{CS-traceanom}
&&\left(x\cdot\f{\partial}{\partial x} + \beta(g)\f{\partial}{\partial g} +4 \right)  \left(\f{\beta(g)}{g}\Tr F^2 (x)\right)  = 0 
\eea
because of its RG invariance.
It follows:
\bea
 \left(x\cdot\f{\partial}{\partial x} +
 \beta(g)\f{\partial}{\partial g} +4 + \beta(g)\f{\partial}{\partial g}\log \left( \f{\beta(g)}{g}\right)
\right) \Tr F^2 (x)=0
\eea
Hence:
\bea\label{A10}
 \left(x\cdot\f{\partial}{\partial x} 
+ \beta(g)\f{\partial}{\partial g} +4+
g\f{\partial}{\partial g}\left( \f{\beta(g)}{g}\right) \right) \Tr F^2 (x)  = 0
\eea
Comparing eq. (\ref{A10}) with the CS equation for $\Tr F^2$:
  \be
\label{CS-F2}
\left(x\cdot\f{\partial}{\partial x} + \beta(g)\f{\partial}{\partial g} + 4 + \gamma_{F^2}(g)\right) \Tr F^2 (x) = 0
\ee
implies:
\be\label{gF22}
\gamma_{F^2}(g)=g\f{\partial}{\partial g}\left (\f{\beta(g)}{g}\right )
\ee
which coincides with eq. (\ref{gF2}).
Then, from the two-loop beta function: 
\bea
\beta(g) = -\beta_0g^3-\beta_1g^5+\cdots
\eea
it follows the anomalous dimension to two loops in $\overline{MS}$-like schemes:
\be\label{gF2p}
\gamma_{F^2}(g)=-2\beta_0g^2-4\beta_1g^4 + \cdots
\ee
The one-loop coefficient, $\gamma_0=-2\beta_0$, is universal, i.e., renormalization-scheme independent, as it generally holds for the one-loop coefficient of the anomalous dimension of a canonically normalized operator. \par
The two-loop coefficient, $\gamma_1=-4\beta_1$, is renormalization-scheme dependent, and it may be modified \cite{MB00} by a finite multiplicative renormalization of $\Tr F^2$ reducing to the identity for $g=0$, as opposed to a reparametrization of $g$ that would not affect the second coefficient of the beta function $\beta_1$ (appendix  \ref{DRG10}).

\section{Contact terms from the OPE for $F^2$ in \cite{Z1,Z3}}
\label{A}

Firstly, we convert the notation in \cite{Z1,Z3} to the one in the present paper.\par
The Euclidean version of the renormalized operator $O_1$ in \cite{Z1,Z3} is:  
\bea\label{eq:O1F2}
O_1&=&-\f{1}{2g_{YM}^2}\Tr{{\cal F}}_{\mu\nu}{{\cal F}}_{\mu\nu}\nn\\
&=&-\f{N}{2g^2}\Tr{{\cal F}}_{\mu\nu}{{\cal F}}_{\mu\nu}\nn\\
&=&-\f{N}{2g^2}\f{g^2}{N}\Tr{F}_{\mu\nu}{F}_{\mu\nu}\nn\\
&=&-\f{1}{2}\Tr{F}_{\mu\nu}{F}_{\mu\nu}\nn\\
&=&-\f{1}{4}F^2
\eea
where we have employed $g^2=Ng_{YM}^2$ and the conversion from ${{\cal F}}_{\mu\nu}$ in the Wilsonian normalization to ${F}_{\mu\nu}$ in the canonical normalization (section \ref{4}). \par
The Euclidean version of the OPE in \cite{Z1,Z3} for the renormalized $O_1$  reads (appendix B of \cite{BB}):
\be\label{b2}
O_1(x)O_1(0)=C_{0CZ}^{(S)}(x) +\left(C_{1CZ}^{(S)\prime}(x) +\delta^{(4)}(x) \left(\f{Z_{11}^L}{Z_{11}}+ \cdots \right)\right) O_1(0)+ \cdots
\ee
where $'$ indicates the contribution for $x \neq 0$ and only the divergent part of the contact term is displayed in the above equation, with:
 \be
 Z_{11}=\left(1-\f{\beta(\alpha_s)}{\eps} \right)^{-1}
 \ee
 and:
\bea\label{eq:Z11}
\f{Z_{11}^L}{Z_{11}}&=& \f{1}{\eps}\left(1-\f{\beta(\alpha_s)}{\eps} \right)^{-1}\alpha_s^2\f{\partial}{\partial\alpha_s}\left(  \f{\beta(\alpha_s)}{\alpha_s}    \right)\nn\\
&=&\f{1}{\eps}\left(1-\f{\beta(\alpha_s)}{\eps} \right)^{-1}\left(\alpha_s\beta'(\alpha_s)-\beta(\alpha_s)    \right)
\eea
computed in \cite{Z3} in a closed form in dimensional regularization.

The beta function $\beta(\alpha_s)$ in \cite{Z1,Z3} is related to $\beta(g)$ by: 
\bea
\beta(\alpha_s)&=&\f{d\log\alpha_s}{d\log\mu^2}
=\f{1}{2}\f{1}{\alpha_s}\f{\partial\alpha_s}{\partial g}\f{dg}{d\log\mu}\nn\\
&=&\f{1}{2}\f{4\pi N}{g^2}\f{2g}{4\pi N}\f{dg}{d\log\mu}
=\f{1}{g}\f{dg}{d\log\mu}\nn\\
&=&\f{\beta(g)}{g}
\eea
where $\alpha_s=g_{YM}^2/4\pi = g^2/4\pi N$ and $g^2=Ng_{YM}^2$.
Hence, by employing:
\bea
\alpha_s^2\f{\partial}{\partial\alpha_s}\left(  \f{\beta(\alpha_s)}{\alpha_s}    \right)&=&
\f{1}{2}\f{1}{4\pi N}g^3\f{\partial}{\partial g}\left( 4\pi N\f{\beta(g)}{g^3}    \right)
=\f{1}{2}g^3\f{\partial}{\partial g}\left( \f{\beta(g)}{g^3}    \right)\nn\\
&=&\f{1}{2}g^3\f{\partial}{\partial g}\left( \f{1}{g^2}\f{\beta(g)}{g}    \right)
=\f{1}{2}g\f{\partial}{\partial g}\left( \f{\beta(g)}{g}    \right) -\f{\beta(g)}{g}  \nn\\
&=& \f{1}{2} \gamma_{F^2}(g)-\f{\beta(g)}{g}  
\eea
eq. \eqref{eq:Z11} reads: 
\bea\label{eq:Z11usFP}
\f{Z_{11}^L}{Z_{11}}&=& \f{1}{\eps}\left(1-\f{\beta(g)}{g\eps} \right)^{-1}\left(\f{1}{2}g\f{\partial}{\partial g} \left(\f{\beta(g)}{g} \right)  - \f{\beta(g)}{g}\right)
\eea
By employing $F^2=-4O_1$ in eq. \eqref{eq:O1F2} and comparing the OPE in eq. \eqref{eq:OPEus} with eq. \eqref{b2}, we obtain the relation between the coefficients in both the notations:
 \bea
 C_1^{(F^2,F^2)}(x)&=&-4\left(C_{1CZ}^{(S)\prime}(x) +\delta^{(4)}(x) \left(\f{Z_{11}^L}{Z_{11}}+\cdots \right) \right)
 \eea
As a consequence, the divergent part of the coefficient of the contact term in eq. \eqref{Coeff} is given by:
\bea\label{eq:35us}
 {\mathcal{C}}_{F^2,\textrm{div}}(g,\eps)&=&
-4\f{Z_{11}^L}{Z_{11}}
\eea
that coincides with eq. \eqref{eq:ZLtous}.

\section{$I_{\td,{\Delta_{F^2}},{\Delta_{F^2}}}$ in the $(u,v_{F^2})$ scheme}
\label{B}

\subsection{Computation of  $I_{\td,{\Delta_{F^2}},{\Delta_{F^2}}}$}
\label{A1}
The space-time integral that appears in the rhs of the LET and in the first line of eq. \eqref{eq:rhsRentilde} is a special case of the general integral \cite{Skenderis}:
\bea\label{eq:Igen}
I_{d,\Delta_1,\Delta_2}&=& \int  \f{1}
{ |x|^{\Delta_1}|x-z|^{\Delta_2} }     d^dx \nn\\
&=&(2\pi)^d \,C_{d,\f{\Delta_1}{2},\f{\Delta_2}{2}}\, |z|^{d-\Delta_1-\Delta_2}
\eea
with:
\be\label{eq:Cgen}
C_{d,\f{\Delta_1}{2},\f{\Delta_2}{2}}=\f{ 
\Gamma\big( \f{\Delta_1+\Delta_2-d}{2}\big)  
\Gamma\big( \f{d-\Delta_1}{2}\big)  \Gamma\big( \f{d-\Delta_2}{2}\big) 
}
{(4\pi)^{\f{d}{2}}\Gamma \big(\f{\Delta_1}{2}\big)\Gamma \big(\f{\Delta_2}{2}\big) \Gamma \big(d-\f{\Delta_1}{2}-\f{\Delta_2}{2}\big)}
\ee
Specifically, we are interested in $I_{\td,\Delta_{F^2},\Delta_{F^2}}$ in eq. \eqref{eq:rhsRentilde}, obtained for $d\to \td$ and $\Delta_1=\Delta_2\to \Delta_{F^2}$ in eqs. \eqref{eq:Igen} and \eqref{eq:Cgen}:
\bea\label{eq:Iapp}
I_{\td,\Delta_{F^2},\Delta_{F^2}}
&=& \int  \f{1}
{ |x|^{\Delta_{F^2} }|x-z|^{\Delta_{F^2}} }     d^\td x \nn\\
&=&(2\pi)^\td \,C_{\td,\f{\Delta_{F^2}}{2},\f{\Delta_{F^2}}{2}} 
\, |z|^{\td-2\Delta_{F^2}}
\eea
with:
\be\label{eq:Capp}
 C_{\td,\f{\Delta_{F^2}}{2},\f{\Delta_{F^2}}{2}} = \f{ 
\Gamma\big( \f{2\Delta_{F^2}-\td}{2}\big)  
\Gamma\big( \f{\td-\Delta_{F^2}}{2}\big)^2 
}
{(4\pi)^{\f{\td}{2}}\Gamma \big(\f{\Delta_{F^2}}{2}\big)^2
 \Gamma \big(\td-\Delta_{F^2}\big)}
 \ee
where $\td = d-2\eps= 4-2\eps$ and $\Delta_{F^2}=d+\gamma_{F^2}=4+\gamma_{F^2}$, so that for the integral in eq. \eqref{eq:Iapp} we obtain:
 \bea\label{eq:I_DR}
I_{\td,d+\gamma_{F^2},d+\gamma_{F^2}}
&=& \int  \f{1}
{ |x|^{d+\gamma_{F^2} }|x-z|^{d+\gamma_{F^2}} }     d^\td x \nn\\
&=&(2\pi)^\td \,C_{\td,\f{d+\gamma_{F^2}}{2},\f{d+\gamma_{F^2}}{2}} 
\, |z|^{\td-2d-2\gamma_{F^2}}\nn\\
&=&(2\pi)^{4-2\eps} \,C_{4-2\eps, 2+\f{\gamma_{F^2}}{2},2+\f{\gamma_{F^2}}{2}} 
\, |z|^{-4-2\eps-2\gamma_{F^2}}
\eea
with:
\bea\label{eq:C_DR1}
C_{\td,\f{d+\gamma_{F^2}}{2},\f{d+\gamma_{F^2}}{2}}  &=& \f{ 
\Gamma\big(d+\gamma_{F^2} -\f{\td}{2}\big)  
\Gamma\big( \f{\td-d-\gamma_{F^2}}{2}\big)^2 
}
{(4\pi)^{\f{\td}{2}}\Gamma \big(\f{d+\gamma_{F^2}}{2}\big)^2
 \Gamma \big(\td-d-\gamma_{F^2}\big)}
 \eea
 which reads:
 \bea\label{eq:C_DR2}
 C_{4-2\eps, 2+\f{\gamma_{F^2}}{2},2+\f{\gamma_{F^2}}{2}} &=&
 \f{ 
\Gamma\big( {2+\eps+\gamma_{F^2}}\big)  
\Gamma\big( \f{-2\epsilon -\gamma_{F^2}}{2}\big)^2 
}
{(4\pi)^{2-\eps}  \Gamma \big(2 +\f{\gamma_{F^2}}{2}\big)^2
 \Gamma \big( -2\epsilon-\gamma_{F^2}\big)}
 \eea
In order to treat all the cases at once, it is convenient to regularize the integral in eq. \eqref{eq:Iapp} in the conformal $(u,v_{F^2})$ scheme (appendix \ref{E}), where $\td=4+2u\eps$ and $\tD_{F^2}=4+(u+v_{F^2})\eps+\gamma_{F^2}$. Setting for brevity $v_{F^2}=v$, it reads in the $(u,v)$ scheme:
\bea\label{eq:Iuv}
I_{\td,\tD_{F^2},\tD_{F^2}}
&=& \int  \f{1}
{ |x|^{\tD_{F^2} }|x-z|^{\tD_{F^2}} }     d^\td x \nn\\
&=&(2\pi)^\td \,C_{\td,\f{\tD_{F^2}}{2},\f{\tD_{F^2}}{2}} 
\, |z|^{\td-2\tD_{F^2}}\nn\\
&=&(2\pi)^{4+2u\eps} \,C_{4+2u\eps, \f{4+(u+v)\eps +\gamma_{F^2}}{2},\f{4+(u+v)\eps +\gamma_{F^2}}{2}} 
\, |z|^{-4-2v\eps-2\gamma_{F^2} }
\eea
with:
\bea\label{eq:Cuv}
&&\hspace{-1.2truecm}C_{4+2u\eps, \f{4+(u+v)\eps +\gamma_{F^2}}{2},\f{4+(u+v)\eps +\gamma_{F^2}}{2}} =
 \f{ 
\Gamma\big( {2+v\eps+\gamma_{F^2}}\big)  
\Gamma\big( \f{(u-v)\eps -\gamma_{F^2}}{2}\big)^2 
}
{(4\pi)^{2+u\eps}\Gamma \big(2+\f{(u+v)\eps +\gamma_{F^2}}{2}\big)^2
 \Gamma \big((u-v)\eps -\gamma_{F^2}\big)}
 \eea
The integral in eq. \eqref{eq:Iapp} corresponds to $u=-1$ and $v=1$.
We anticipate the results of the analysis below.
For $\gamma_{F^2}$ nonexceptional $C$ in eq. \eqref{eq:Cuv}
is finite as $\epsilon \rightarrow 0$.
For the values $\gamma_{F^2}=-3,-2,0,2,4,\cdots$ --  which we refer to as exceptional, where we exclude from the exceptional values the ones exceeding the unitarity bound -- the ratio of gamma functions in eq. \eqref{eq:Cuv} has a simple pole as $\eps \rightarrow 0$, with the negative values of $\gamma_{F^2}$ corresponding to an IR divergence and the nonnegative ones to an UV divergence. \par
The divergence or finiteness of $I_{\td,\tD_{F^2},\tD_{F^2}}$ as $\epsilon \rightarrow 0$ in eq. \eqref{eq:Iuv} is determined by the way the rhs of eq. \eqref{eq:Cuv} depends on $\eps$ and $\gamma_{F^2}$.
Taking into account the unitarity bound on the scalar operator $F^2$, i.e., $\Delta_{F^2}\geqslant 1$ in $d=4$ dimensions\footnote{In a CFT the unitarity bound for a scalar operator of scaling dimension $\Delta$ is $\Delta(\Delta-(d-2)/2)\geqslant 0$ in $d$ space-time dimensions.}, we distinguish the following cases for $\Delta_{F^2}\geqslant 1$: 
\begin{itemize}
\item[(I)] $1\leqslant \Delta_{F^2}\leqslant 2$ ($-3\leqslant \gamma_{F^2}\leqslant -2$):  
$I_{\td,{\tD_{F^2}},{\tD_{F^2}}}$ is finite for noninteger $\gamma_{F^2}$ and
IR divergent for the exceptional values $\gamma_{F^2}=-2, -3$ as $\epsilon \rightarrow 0$. The pole in $\epsilon$ occurs in the $\eps$-expansion of the ratio 
$\Gamma\big( {2+v\eps+\gamma_{F^2}}\big)/\Gamma \big(2+\f{(u+v)\eps+\gamma_{F^2}}{2}\big)^2 $ in eq. \eqref{eq:Cuv} according to eqs. \eqref{eq:case11} ($\gamma_{F^2}=-2$) and 
\eqref{eq:case12} ($\gamma_{F^2}=-3$).
\item[(II)]  $2<\Delta_{F^2}< 4$ ($-2<\gamma_{F^2}< 0$): $I_{\td,{\tD_{F^2}},{\tD_{F^2}}}$ is  finite according to the $\eps$-expansion in eq. \eqref{eq:case2}. 
\item[(III)] $\Delta_{F^2}=4$ ($\gamma_{F^2}=0$): $\gamma_{F^2}=0$ is exceptional. 
In the $(u,v)$ scheme the UV divergence of $I_{\td,{\tD_{F^2}},{\tD_{F^2}}}$ in eq. \eqref{eq:Iuv} arises as $\epsilon \rightarrow 0$ from the $\eps$-expansion of the ratio $\Gamma\big( \f{(u-v)\eps }{2}\big)^2 /
 \Gamma \big((u-v)\eps \big)$ 
in eq. \eqref{eq:Cuv} for $\gamma_{F^2}=0$, according to eq. \eqref{eq:case3}. 
In this case, eq. \eqref{eq:Cuv} reads:
\bea\label{appCC0}
 C_{4+2u\eps,2+\f{(u+v)\eps}{2},2+\f{(u+v)\eps }{2}} &=&
 \f{ 
\Gamma\big( {2+v\eps}\big)  
\Gamma\big( \f{(u-v)\eps}{2}\big)^2 
}
{(4\pi)^{2+u\eps}\Gamma \big(2+\f{(u+v)\eps}{2}\big)^2
 \Gamma \big((u-v)\eps\big)}
 \nn\\
  &&\hspace{-0.8truecm}=
\f{1}{16\pi^2}\left(\f{4}{(u-v)\eps}-\f{u}{u-v}4\log(4\pi)+O(\eps) \right)
 \eea

\item[(IV)] $\Delta_{F^2}> 4$  ($\gamma_{F^2}> 0$): We distinguish three cases:
\item[a)] $\gamma_{F^2}$ even integer: $\gamma_{F^2}= 2, 4,\cdots$ are exceptional.  
In the $(u,v)$ scheme the divergence as $\epsilon \rightarrow 0$ arises in the UV from the $\eps$-expansion of the ratio $\Gamma\big( \f{(u-v)\eps -\gamma_{F^2}}{2}\big)^2$ $/\Gamma \big( (u-v)\eps -\gamma_{F^2}\big) $ according to eq. \eqref{eq:case41}.
Eq. \eqref{eq:Cuv} for $\gamma_{F^2}=2n$, $n=1,2,\cdots$ now reads:
\bea\label{appCCexc}
 C_{4+2u\eps, 2+\f{(u+v)\eps}{2}+n,2+\f{(u+v)\eps}{2}+n} 
 &=&
 \f{ 
\Gamma\big( {2+v\eps+2n}\big)  
\Gamma\big( \f{(u-v)\eps}{2}-n\big)^2 
}
{(4\pi)^{2+u\eps}\Gamma \big(2+\f{(u+v)\eps}{2}+n\big)^2
 \Gamma \big((u-v)\eps -2n\big)}\nn\\
 &&\hspace{-0.8truecm}=\f{1}{16\pi^2} \f{4  }{(u-v)\eps} 
 \f{ 
\Gamma\big( {2+2n}\big)}
{\Gamma \big(2+n\big)^2 }
\f{(-2n)!}{(-n)!^2}
\left(1+O(\eps) \right) 
  \eea
\item[b)] $\gamma_{F^2}$ odd integer: For $\gamma_{F^2}= 1, 3,\cdots$ $I_{\td,{\tD_{F^2}},{\tD_{F^2}}}$ is finite and vanishes as $\epsilon \rightarrow 0$, since the $\eps$-expansion in 
eq. \eqref{eq:case42} shows that the ratio 
$\Gamma\big( \f{(u-v)\eps -\gamma_{F^2}}{2}\big)^2 /\Gamma \big( (u-v)\eps -\gamma_{F^2}\big) $ is $O(\eps)$.
\item[c)] $\gamma_{F^2}$ noninteger: $I_{\td,{\tD_{F^2}},{\tD_{F^2}}}$ is finite as $\epsilon \rightarrow 0$, according to eq. \eqref{eq:case5}.
\end{itemize}
By summarizing, for $-3 \leqslant\gamma_{F^2}<0$ the rhs of the LET is finite as $\eps\to 0$, except for $\gamma_{F^2}=-2, -3$, where it is IR divergent as $\epsilon \rightarrow 0$ (cases (I) and (II)). 
For $\gamma_{F^2}=0$, the rhs of the LET is UV divergent as $\epsilon \rightarrow 0$ (case (III)). 
For $\gamma_{F^2}>0$, the rhs of the LET is finite as $\eps \to 0$ (case (IV)c)), with the exception of $\gamma_{F^2}=2,4,\cdots$ where it is UV divergent as $\eps \to 0$ (case (IV)a)).
Moreover, it vanishes for $\gamma_{F^2}=1,3,\cdots$ as $\eps\to 0$ (case (IV)b)).

\subsection{$\eps$-expansion of ratios of $\Gamma$ functions}

We provide the $\eps$-expansion of the ratios of the $\Gamma$ functions in eq. \eqref{eq:Cuv}, which are necessary for the analysis above, on the basis of the following properties. 
$\Gamma(z)$ has simple poles with residue $(-1)^n/n!$ at $z=-n$, with $n$ a nonnegative integer, and  
satisfies the relation $z\Gamma(z)=\Gamma(z+1)$. For $\eps >0$ infinitesimal, it admits the expansion $\Gamma(\eps)= 1/\eps +\Gamma'(1)+O(\eps)$.
To establish (I) for $\gamma_{F^2}=-2$, we employ:
 \bea\label{eq:case11}
 \f{\Gamma\big( {2+v\eps+\gamma_{F^2}}\big)}{\Gamma \big(2+\f{(u+v)\eps +\gamma_{F^2}}{2}\big)^2 }&=&\f{\Gamma(v\eps)}{\Gamma \big(1+\f{(u+v)\eps}{2}\big)^2 }\nn\\
 &=&\Gamma(v\eps)(1+O(\eps))\nn\\
 &=&\f{1}{v\eps}+\Gamma'(1)+O(\eps)
 \eea
To establish (I) for $\gamma_{F^2}=-3$, we employ the following $\eps$-expansion for $\gamma_{F^2}=-(2n+1)$ with $n=1,2,\cdots$:
 \bea\label{eq:case12}
 &&\f{\Gamma\big( {2+v\eps+\gamma_{F^2}}\big)}{\Gamma \big(2+\f{(u+v)\eps +\gamma_{F^2}}{2}\big)^2 }=\f{\Gamma(1-2n+v\eps)}{\Gamma \big(2-\f{2n+1}{2}+\f{(u+v)\eps}{2}\big)^2 }\nn\\
 &&  \hspace{0.5cm}= \f{\Gamma(v\eps)}{(1-2n+v\eps)(2-2n+v\eps)\cdots (-1+v\eps)  }{\Gamma \big(2-\f{2n+1}{2}+\f{(u+v)\eps}{2}\big)^{-2} }  \nn\\
&& \hspace{0.5cm}= \f{1}{v\eps}\f{1}{(1-2n)(2-2n)\cdots (-1)   } \Gamma\big(2-\f{2n+1}{2}\big)^{-2}  (1+O(\eps))
 \eea
 For $\gamma_{F^2}=-2n$ with $n\geqslant 2$, no pole in $\eps$ occurs:
 \bea\label{eq:case13}
&& \f{\Gamma\big( {2+v\eps+\gamma_{F^2}}\big)}{\Gamma \big(2+\f{(u+v)\eps +\gamma_{F^2}}{2}\big)^2 }=\f{\Gamma(2-2n+v\eps)}{\Gamma \big(2-n+\f{(u+v)\eps}{2}\big)^2 }\nn\\
 &&  \hspace{0.5cm}= \f{\Gamma(v\eps) }{\Gamma \big(\f{(u+v)\eps}{2}\big)^2 } 
 \f{(2-n+\f{(u+v)\eps}{2})^2 (1-n+\f{(u+v)\eps}{2})^2\cdots (-1+\f{(u+v)\eps}{2})^2}{(2-2n+v\eps) (1-2n+v\eps)\cdots (-1+v\eps)} 
 \nn\\
 && \hspace{0.5cm}= \f{\f{1}{v\eps}+\Gamma'(1)+O(\eps) }{(\f{2}{(u+v)\eps}+\Gamma'(1)+O(\eps))^2 }
 \f{(2-n+\f{(u+v)\eps}{2})^2 (1-n+\f{(u+v)\eps}{2})^2\cdots (-1+\f{(u+v)\eps}{2})^2}{(2-2n+v\eps) (1-2n+v\eps)\cdots (-1+v\eps)} \nn\\
 && \hspace{0.5cm}= \eps \f{(u+v)^2}{4v}\f{(2-n)!^2 }{ (2-2n)! }   (1+O(\eps))
 \eea
 To establish (II) for $ -2<\gamma_{F^2}<0$, we observe that the ratio:
 \be\label{eq:case2}
  \f{ 
\Gamma\big( {2+v\eps+\gamma_{F^2}}\big)  
\Gamma\big( \f{(u-v)\eps -\gamma_{F^2}}{2}\big)^2 
}
{\Gamma \big(2+\f{(u+v)\eps +\gamma_{F^2}}{2}\big)^2
 \Gamma \big((u-v)\eps -\gamma_{F^2}\big)}
=
 \f{ 
\Gamma\big( {2+\gamma_{F^2}}\big)  
\Gamma\big( \f{-\gamma_{F^2}}{2}\big)^2 
}
{\Gamma \big(2 +\f{\gamma_{F^2}}{2}\big)^2
 \Gamma \big( -\gamma_{F^2}\big)} (1+O(\eps))
 \ee
 is manifestly finite, since $2+\gamma_{F^2}>0$ and $-\gamma_{F^2}>0$. \par
 To establish (III) for $\gamma_{F^2}=0$, we employ:
\bea\label{eq:case3}
 \f{ \Gamma\big( \f{(u-v)\eps}{2}\big)^2 }{ \Gamma \big((u-v)\eps \big)}&=&
 \left( \f{2}{(u-v)\eps}+\Gamma'(1) +O(\eps) \right)^2
 \left( \f{1}{(u-v)\eps}+\Gamma'(1) +O(\eps)   \right)^{-1}\nn\\
 &=& \f{4}{(u-v)\eps}+O(\eps)
\eea 
To establish (IV)a, we employ the $\eps$-expansion for $\gamma_{F^2}=2n$ with $n=1,2,\cdots$:
\bea\label{eq:case41}
&&\f{\Gamma\big( \f{(u-v)\eps -\gamma_{F^2}}{2}\big)^2 }{\Gamma \big((u-v)\eps -\gamma_{F^2}\big) }=\f{\Gamma\big( \f{(u-v)\eps}{2}-n\big)^2 }{\Gamma \big((u-v)\eps -2n\big) }\nn\\
&& \hspace{0.5cm}= \f{\Gamma\big( \f{(u-v)\eps}{2}\big)^2 }{\Gamma \big((u-v)\eps\big) }
\f{(-2n+ (u-v)\eps)(-2n+1+ (u-v)\eps)\cdots(-1+ (u-v)\eps)}{(-n+\f{(u-v)\eps}{2})^2(-n+1+\f{(u-v)\eps}{2})^2\cdots(-1+\f{(u-v)\eps}{2})^2 }\nn\\
&& \hspace{0.5cm}= \f{4}{(u-v)\eps}\left(\f{(-2n)!}{(-n)!^2}+O(\eps)\right)
\eea
To establish (IV)b, we employ the $\eps$-expansion for $\gamma_{F^2}=2n+1$ with $n=0,1,\cdots$:
\bea\label{eq:case42}
&&\f{\Gamma\big( \f{(u-v)\eps -\gamma_{F^2}}{2}\big)^2 }{\Gamma \big((u-v)\eps -\gamma_{F^2}\big) }=\f{\Gamma\big( \f{(u-v)\eps}{2}-n-\f{1}{2}\big)^2 }{\Gamma \big((u-v)\eps -2n-1\big) }\nn\\
&& \hspace{0.5cm}= \f{\Gamma\big(-n-\f{1}{2}\big)^2 }{\Gamma \big((u-v)\eps\big)}(-2n-1+(u-v)\eps)(-2n+(u-v)\eps)\cdots(-1+(u-v)\eps)\nn\\
&& \hspace{0.7cm} (1+O(\eps))\nn\\
&& \hspace{0.5cm}= \eps (u-v)\Gamma\big(-n-\f{1}{2}\big)^2(-2n-1)!(1+O(\eps))
\eea
Finally, to establish (IV)c we observe that the limit $\eps \rightarrow 0$ for noninteger $\gamma_{F^2}>0$:
\bea\label{eq:case5}
 \f{ \Gamma\big( {2+v\eps+\gamma_{F^2}}\big)  
\Gamma\big( \f{(u-v)\eps -\gamma_{F^2}}{2}\big)^2 
}
{\Gamma \big(2+\f{(u+v)\eps +\gamma_{F^2}}{2}\big)^2
 \Gamma \big((u-v)\eps -\gamma_{F^2}\big)}&=& \f{ \Gamma\big( {2+\gamma_{F^2}}\big)  
\Gamma\big( \f{-\gamma_{F^2}}{2}\big)^2 
}
{\Gamma \big(2 +\f{\gamma_{F^2}}{2}\big)^2
 \Gamma \big( -\gamma_{F^2}\big)}(1+O(\eps))~~~~~
 \eea
is finite.

\section{Callan-Symanzik equation in $d=4$ dimensions} \label{CS0}

In a massless QCD-like theory the CS equation \cite{C,S} in $d=4$ dimensions for connected 2-point correlators $G^{(2)}\equiv\braket{O(z) O(0)}$ of a multiplicatively renormalizable gauge-invariant scalar operator $O$ with canonical dimension $D$ expresses the independence of the bare correlator $G^{(2)}_0\equiv\braket{O(z) O(0)}_0$:
\be\label{cs0_1}
G^{(2)}_0(z, \Lambda, g(\Lambda))=Z_O^{-2}(\f{\Lambda}{\mu},g(\mu))G^{(2)}(z, \mu, g(\mu))
\ee
 from the renormalization scale $\mu$: 
 \be\label{cs0_2}
\mu \f{d} {d\mu}G^{(2)}_0\bigg|_{\Lambda, g(\Lambda)}=0
 \ee
where  $\Lambda$ in eq. \eqref{cs0_1} is the UV cutoff of the bare theory in some regularization. Inserting eq. \eqref{cs0_1} into eq. \eqref{cs0_2} we obtain the CS equation \cite{C,S,Zub,Pes}:
\be
\label{CS2_eps0}
\left(\mu\f{\partial}{\partial \mu} + \beta(g)\f{\partial}{\partial g} + 2\gamma_{O}(g)\right) G^{(2)}(z, \mu, g(\mu)) = 0
\ee
with the beta function:
\be\label{beta_eps0}
\beta(g)=\f{dg}{d\log\mu}\bigg|_{\Lambda, g(\Lambda)}
\ee
and the anomalous dimension of $O$:
\be
\gamma_{O}(g)=-\f{d\log Z_O}{d\log\mu}\bigg|_{\Lambda, g(\Lambda)}
\ee
Since the theory is massless at each order in perturbation theory and $O$ has canonical dimension $D$, we may set:
\be
G^{(2)}(z, \mu, g(\mu)) =\f{1}{z^{2D}} 	\bar G^{(2)}(z\mu, g(\mu)) 
\ee
in terms of the dimensionless correlator $ \bar G^{(2)} $, which also satisfies eq. \eqref{CS2_eps0}. Moreover:
\be\label{CS2_eps0_dl}
\left(z\cdot \f{\partial}{\partial z} + \beta(g)\f{\partial}{\partial g} + 2\gamma_{O}(g)\right) \bar G^{(2)}(z\mu, g(\mu)) = 0
\ee
since $ \bar G^{(2)} $ depends on $z$ only via the product $z\mu$. Hence:
\be\label{CS2_eps0_dl0}
\left(z\cdot \f{\partial}{\partial z} + \beta(g)\f{\partial}{\partial g} + 2 D+2\gamma_{O}(g)\right) G^{(2)}(z, \mu, g(\mu)) = 0
\ee
The  solution of the above equation may be written as:
\bea\label{pert2_eps0}
 G^{(2)}(z, \mu, g(\mu)) &=& \frac{1}{z^{2D}}\,\bar G^{(2)}(z\mu, g(\mu))\nn\\
 &=&\frac{1}{z^{2D}}\, \mathcal{G}_{2}^{(O)}(g(z))\, Z^{(O)2}(g(z), g(\mu))
\eea
 in terms of the function $\mathcal{G}_2^{(O)}$ of the RGI coupling $g(z)\equiv g(z\mu, g(\mu))$, which solves: 
 \be\label{gz4}
-\f{d g(z)}{d \log|z|}=\beta(g(z))
\ee
 with the initial condition $g(1,g(\mu))=g(\mu)$ and the renormalized multiplicative factor $Z^{(O)}(g(z), g(\mu))$:
\begin{equation} \label{def_eps0}
Z^{(O)}(g(z), g(\mu)) = \exp \int_{g(\mu)} ^{g(z)}      \frac{ \gamma_O (g) } {\beta(g)} dg 
\end{equation}
that solves:
\be
\gamma_O (g(\mu))=-\f{d \log Z^{(O)}}{d \log\mu}
\ee
Besides, the RG invariance of $g(z)$ implies:
\bea\label{betar_eps01}
0&=&\f{dg(z)}{d\log\mu}\nn\\
&=&\f{\partial g(z)}{\partial \log\mu}+\f{\partial g(z)}{\partial g(\mu)}\f{dg(\mu)}{d\log\mu}\nn\\
&=&\f{d g(z)}{d \log|z|}+\f{\partial g(z)}{\partial g(\mu)}\f{dg(\mu)}{d\log\mu}\nn\\
&=&-\beta(g(z))+\f{\partial g(z)}{\partial g(\mu)}\beta(g(\mu))
\eea
where we have employed eq. \eqref{gz4} and the fact that $g(z)$ depends on $z$ only via the product $z\mu$. It follows:
\bea\label{betar_eps0}
\f{\partial g(z)}{\partial g(\mu)} &=&
\f{\beta(g(z))}{\beta(g(\mu))}
\eea

\section{IR and UV asymptotics of 2-point correlators} \label{D}

In order to clarify under which conditions a massless QCD-like theory is either exactly or asymptotically conformal in the IR/UV if the beta function admits an isolated zero in the IR/UV, we work out the leading and subleading asymptotics of a renormalized 2-point correlator $\braket{O(z)O(0)}$ of a multiplicatively renormalizable operator $O$ and of $F^2$.  

\subsection{Universality of the coefficients of the beta function} \label{DRG1}

\subsubsection{$g_*=0$}  \label{DRG10}

We employ the following shorthand notation for $\alpha = a/(4\pi)$, with $a=g^2$, that may be taken as the natural coupling in a YM theory. Correspondingly:
\be
\beta(a) =\f{da}{d\log\mu}=\f{d a}{d g}\beta(g)=2g\beta(g)
\ee
with:
\bea \label{beta00}
\beta(g)=-\beta_0 g^3 - \beta_1 g^5 -\beta_2 g^7 +  \cdots
\eea
that has a trivial zero $g_*=0$. In a neighborhood of $g_*=0$, we consider the (formal analytic\footnote{We do not assume that formal power series are convergent.}) reparametrization of the coupling, i.e., the change of renormalization scheme:
\be\label{RGC0}
g'=g(1+a_1g^2+a_2g^4+\cdots)
\ee
that guarantees that $g_*=0$ remains a zero of $\beta'(g')=\beta(g'(g)))$ and only odd powers of $g$ enter the expansion of the latter, according to the natural reparametrization that arises as an arbitrary power series in $a$ or $\alpha$ that reduces to the identity at $a=\alpha=0$. \par
Hence, the expansion of $\beta'(g')$ has the very same structure as the expansion of $\beta(g)$ in eq. \eqref{beta00}, but in general with new coefficients:
\be
\beta'(g')=-\beta_0^\prime g^{\prime 3} - \beta_1^\prime g^{\prime 5} - \beta_2^\prime g^{\prime 7}  +\cdots
\ee
It may be expressed in terms of $g$ by means of eq. \eqref{RGC0}:
\bea\label{new0}
\beta(g'(g))&=&-\beta_0^\prime g^3(1+a_1g^2+a_2g^4+\cdots)^3 - \beta_1^\prime g^5(1+a_1g^2+a_2g^4+\cdots)^5~~~~~~~\nn\\
&&~~ -\beta_2^\prime g^7+\cdots \nn\\
&=&-\beta_0^\prime g^3-(\beta_1^\prime +3 a_1 \beta_0^\prime)g^5 -(\beta_2'+5\beta_1'a_1+3\beta_0'a_2+3\beta_0'a_1^2)g^7
+\cdots
\eea
that must agree with:
\bea\label{old0}
\beta'(g')&=&\f{\partial g'(g)}{\partial g}\beta(g)\nn\\
&=&(1+3a_1g^2+5a_2g^4+\cdots)(-\beta_0 g^3-\beta_1 g^5-\beta_2 g^7+\cdots)\nn\\
&=&-\beta_0 g^3-(\beta_1 +3 a_1 \beta_0)g^5 -(\beta_2+3\beta_1a_1+5\beta_0a_2)g^7
+\cdots
\eea
Equating eqs. \eqref{new0} and \eqref{old0} implies that $\beta_0$ and $\beta_1$ are independent of the renormalization scheme, i.e., $\beta_0'=\beta_0$ and $\beta_1'=\beta_1$.\par
Moreover, if $\beta_0=\beta_1=0$, $\beta_2$ is scheme independent. More generally, the first nonvanishing coefficient is scheme independent.

\subsubsection{$g_*>0$} \label{DRG2}

Given:
\bea \label{beta_0}
\beta(g)=\beta^*_0 (g-g_*) +\beta^*_1 (g-g_*)^2 +\beta_2^{*} (g-g_*)^3 +\cdots
\eea
in a neighborhood of the nontrivial zero, $g_*>0$, we consider the most general (formal analytic) reparametrization of the coupling, $g\to g'$ and $g_*\to g_*'$, that preserves the existence of the zero of the beta function:
\be\label{RGC}
g'-g_*'=(g-g_*)(1+b_1(g-g_*)+b_2(g-g_*)^2 +\cdots)
\ee
The expansion of $\beta'(g')$ has the very same structure as the expansion of $\beta(g)$, but in general with new coefficients:
\be
\beta'(g')=\beta_0^{*\prime} (g'-g_*') +\beta_1^{*\prime} (g'-g_*')^2 +\beta_2^{*\prime} (g'-g_*')^3 +\cdots
\ee
It may be expressed in terms of $g$ and $g_*$ by means of eq. \eqref{RGC}:
\bea\label{new}
\beta'(g'(g))&=&\beta_0^{*\prime}(g-g_*)(1+b_1(g-g_*)+b_2(g-g_*)^2 +\cdots)+\beta_1^{*\prime}
(g-g_*)^2 +\cdots\nn\\
&=&\beta_0^{*\prime}(g-g_*)+(\beta_1^{*\prime}+\beta_0^{*\prime}b_1 )(g-g_*)^2 +\cdots
\eea
that must agree with:
\bea\label{old}
\beta'(g')&=&\f{\partial g'(g)}{\partial g}\beta(g)\nn\\
&=&(1+2b_1(g-g_*)+3b_2(g-g_*)^2+\cdots) (\beta_0^{*} (g-g_*) +\beta_1^{*} (g-g_*)^2 +\cdots)\nn\\
&=&\beta_0^{*} (g-g_*) + (\beta_1^{*}+2\beta_0^{*}b_1)(g-g_*)^2 +\cdots
\eea
Equating eqs. \eqref{new} and \eqref{old} implies that $\beta_0^*$ is scheme independent, i.e., $\beta_0^{*\prime}=\beta^*_0$.
Interestingly, the same conclusion is reached by means of eq. \eqref{gF22} that relates $\beta_{0}^*$ to the anomalous dimension of $F^2$ -- a physical quantity -- by the relation:
\be\label{bg0}
 \beta_0^*=\f{\partial\beta(g)}{\partial g}\bigg|_{g=g_*}
=\gamma_{F^2}(g_*)
\ee
Moreover, by assuming that the first coefficients vanish, so that for some positive integer $n$:
\bea
\beta(g)&=&\beta_n^{*} (g-g_*)^{n+1} + \beta_{n+1}^{*} (g-g_*)^{n+2}+\cdots
\eea
the new beta function:
\bea
\beta'(g'(g))&=&\beta_n^{*\prime} (g'-g_*')^{n+1} + \beta_{n+1}^{*\prime} (g'-g_*')^{n+2}+\cdots\nn\\
&=& \beta_n^{*\prime} (g-g_*)^{n+1} (1+b_1(g-g_*)+\cdots)^{n+1} + \beta_{n+1}^{*\prime} (g-g_*)^{n+2}+\cdots\nn\\
&=&\beta_n^{*\prime} (g-g_*)^{n+1}  + (\beta_{n+1}^{*\prime}+(n+1)\beta_n^{*\prime}b_1)
 (g-g_*)^{n+2}+\cdots
\eea
must agree with:
\bea\label{oldn}
\beta'(g')&=&\f{\partial g'(g)}{\partial g}\beta(g)\nn\\
&=&(1+2b_1(g-g_*)+\cdots)(\beta_n^{*} (g-g_*)^{n+1} + \beta_{n+1}^{*} (g-g_*)^{n+2}+\cdots)\nn\\
&=&\beta_n^{*} (g-g_*)^{n+1} +(\beta_{n+1}^{*}+2\beta_n^{*} b_1)(g-g_*)^{n+2}+\cdots
\eea
that implies that the first nonzero coefficient is scheme independent, i.e., $\beta_n^{*\prime}=\beta_n^{*}$. 

\subsection{Asymptotics of the running coupling}

\subsubsection{$g_*=0$ and the AF case with $\beta_0\neq 0$} \label{D.1AF}

For $\beta_0\neq 0$ the theory is AF in the UV or IR for $\beta_0>0$ or $\beta_0<0$ respectively. 
Evaluating asymptotically the integral of the inverse of the beta function in eq. \eqref{beta00}, we obtain: 
\bea\label{intUV}
\int_{g(\mu)}^{g(z)}\f{dg}
{-\beta_{0} g^3 - \beta_1 g^5    +\cdots } = -\int_{\mu^{-1}}^{|z|}d\log |z|
\eea
where both the length scales, $|z|=\sqrt{z^2}$ and $\mu^{-1}$, are assumed to be close to zero (UV) or infinity (IR) in order for $g(z)$ and $g(\mu)$ to stay in a neighborhood of $g_*=0$. 
It follows:
\bea\label{exp-grun}
\frac{1}{g^2(z)}-\f{1}{g^2(\mu)}=-2\beta_0\log|z\mu|-2\f{\beta_1}{\beta_0}\log\left (\f{g(z)}{g(\mu)}\right)+\cdots
\eea 
or, equivalently:
\bea \label{rgi_exp_uv}
&&\f{1}{g^2(z)}+2\beta_0\log |z|+2\f{\beta_1}{\beta_0}\log {g(z)}+C+\cdots \nonumber \\
&&\hspace{0.5truecm}= \f{1}{g^2(\mu)}-2\beta_0\log \mu +2\f{\beta_1}{\beta_0}\log {g(\mu)}+C+\cdots
\eea
where, for later convenience, we have introduced the arbitrary integration constant $C$.
It arises from the indefinite version of the integral in eq. \eqref{intUV} and is scheme dependent. 
Solving iteratively eq. \eqref{exp-grun}, we obtain to order $g^4(\mu)$:
\bea
\label{g2loop_large}
\f{g^2(\mu)}{g^2(z)}& =&  1 - g^2(\mu)2\beta_0\log|z\mu| +g^2(\mu)\f{\beta_1}{\beta_0}
\log\bigg(\f{g^2(\mu)}{g^2(z)} \bigg)+\cdots\nn\\
&=& 1 - g^2(\mu)2\beta_0\log|z\mu|-g^4(\mu)2\beta_1\log|z\mu|+\cdots
\eea
where in the second equality we have employed: 
\be
\log\bigg(\f{g^2(\mu)}{g^2(z)} \bigg)=\log(1- g^2(\mu)2\beta_0\log|z\mu|+\cdots)=-g^2(\mu)2\beta_0\log|z\mu|+\cdots
\ee
Expanding eq. \eqref{g2loop_large} to order $g^4(\mu)$, for $\log|z\mu|$ of order one, yields:
\be
\label{g2loop}
g^2(z) =  g^2(\mu)\left(1 + g^2(\mu)2\beta_0\log|z\mu| + g^4(\mu)(2\beta_1\log|z\mu| 
 + 4\beta_0^2\log^2|z\mu|)+\cdots\right)
\ee
which implies $g(z)=0$ if $g(\mu)=0$ for some -- and thus every -- $\mu\neq 0$. \par
For $g(\mu) \neq g_*=0$ eq. (\ref{rgi_exp_uv}) implies that both sides are independent of $|z|$ and $\mu^{-1}$, i.e., renormalization-group invariant (RGI), though scheme dependent. 
Thus, there exists an RGI mass scale, $ \Lrgi$, which characterizes the physics of the theory, such that: 
\bea\label{C0}
\Lrgi &=& \mu \exp\left (-\f{1}{2\beta_0g^2(\mu)} \right) (g(\mu))^{-\f{\beta_1}{\beta_0^2}}\exp\left( -\f{C}{2\beta_0} +\cdots \right) \nonumber \\
&=& |z|^{-1} \exp\left (-\f{1}{2\beta_0g^2(z)} \right) (g(z))^{-\f{\beta_1}{\beta_0^2}}\exp\left( -\f{C}{2\beta_0} +\cdots \right)
\eea
where the dots stand for a scheme-dependent series in even powers of $g(\mu)$ and $g(z)$ respectively. 
Hence, solving eq. \eqref{C0} in terms of $g(z)$ asymptotically as $|z|\to +\infty/0^+$ in the IR/UV, we get:
\bea \label{alfa2}
g^2(z) &\sim& \dfrac{1}{-2\beta_0 \log({|z|\Lrgi })} 
\bigg(1+\frac{\beta_1}{2\beta_0^2}\frac{\log(-2\beta_0\log({|z|\Lrgi }))}{\log({|z|\Lrgi })}\nn\\
&& -\f{C}{2\beta_0\log  ({|z|\Lrgi }  )  } 
+ \cdots\bigg)
\eea
The IR/UV asymptotics in eq. \eqref{alfa2} is selfconsistently realized if:
\bea\label{AFcases}
\beta_0 < 0~~{\textrm{for}}~|z|\to +\infty~~\textrm{(IR)}\nn\\
\beta_0 > 0~~{\textrm{for}}~|z|\to 0^+~~\textrm{(UV)}
\eea
A well known realization of the latter case is the QCD asymptotic freedom in the UV where, incidentally, the $\overline{MS}$ scheme may be defined by the choice $C=(\beta_1/\beta_0)\log\beta_0$ that cancels in eq. \eqref{alfa2} the last term against the contribution $\frac{\beta_1}{2\beta_0^2}\f{\log2\beta_0}{\log({|z|\Lrgi } ) }$ in the first line.\par
It follows the universal -- i.e., renormalization-scheme independent -- IR/UV asymptotics of the running coupling as $ |z|\to +\infty/0^+$:
\be \label{alfa}
g^2(z) \sim \dfrac{1}{-2\beta_0 \log({|z|\Lrgi })} 
\left(
1+\frac{\beta_1}{2\beta_0^2}\frac{\log(-2\beta_0\log({|z|\Lrgi } ))}{\log({|z|\Lrgi })} 
\right)
\ee 
Indeed, we easily verify that in eq. \eqref{alfa} we may change the normalization of $\Lrgi$, which amounts to a change of $C$, without changing the asymptotics as $ |z|\to +\infty/0^+$.

\subsubsection{$g_*=0$ and the AF case with $\beta_0=0$} \label{D.1AF2}
 
For $\beta_0=0$ the beta function in a neighborhood of $g_*=0$ reads:
\be\label{UEbeta}
\beta(g)=-\beta_1g^5-\beta_2g^7-\beta_3g^9+\cdots
\ee
For example, eq. \eqref{UEbeta} with $\beta_1<0$ is realized in massless QCD at the upper edge of the conformal window, where the theory is asymptotically free in the IR. The integral of the inverse of the beta function now reads:
\bea\label{intIRAF}
\int_{g(\mu)}^{g(z)}\f{dg}
{-\beta_{1} g^5 - \beta_2 g^7 -\beta_3g^9   +\cdots } = -\int_{\mu^{-1}}^{|z|}d\log |z|
\eea
where for $\beta_1<0\,(\beta_1>0)$ both the length scales, $|z|$ and $\mu^{-1}$, are assumed to be close to infinity (zero) in order for $g(z)$ and $g(\mu)$ to stay in a neighborhood of $g_*=0$. It follows:
\bea\label{exp-grun-IRAF}
&&\f{1}{g^4(z)}-\f{1}{g^4(\mu)}-\f{2\beta_2}{\beta_1}\left(
\frac{1}{g^2(z)}-\f{1}{g^2(\mu)}\right) \nonumber \\
&&=-4\beta_1\log|z\mu| +4\left(\f{\beta_2^2}{\beta_1^2}-\f{\beta_3}{\beta_1}\right)\log\left (\f{g(z)}{g(\mu)}\right)
+\cdots
\eea 
or equivalently:
\bea \label{rgi_exp_IRAF}
&& \frac{1}{g^4(z)}- \f{2\beta_2}{\beta_1}  \f{1}{g^2(z)}+4\beta_1\log |z|-4 \left(\f{\beta_2^2}{\beta_1^2}-\f{\beta_3}{\beta_1}\right)
\log {g(z)}+C+\cdots \nonumber \\
&&\hspace{0.5truecm}=  \frac{1}{g^4(\mu)}- \f{2\beta_2}{\beta_1}  \f{1}{g^2(\mu)}-4\beta_1\log \mu-4 \left(\f{\beta_2^2}{\beta_1^2}-\f{\beta_3}{\beta_1}\right)
\log {g(\mu)}+C+\cdots 
\eea
where again we have introduced the arbitrary integration constant $C$.

Solving iteratively eq. \eqref{exp-grun-IRAF}, we obtain to order $g^6(\mu)$:
\bea\label{gIRAFlarge}
\f{g^4(\mu)}{g^4(z)}&=&
1-g^4(\mu)4\beta_1\log|z\mu| +g^2(\mu)\f{2\beta_2}{\beta_1}\big(\sqrt{1-g^4(\mu)4\beta_1\log|z\mu|+\cdots}-1\big)\nn\\
&&-g^4(\mu)\left(\f{\beta_2^2}{\beta_1^2}-\f{\beta_3}{\beta_1}\right)
\log\left(1-g^4(\mu)4\beta_1\log|z\mu|+\cdots\right)+\cdots\nn\\
&=&
1-g^4(\mu)4\beta_1\log|z\mu| +g^2(\mu)\f{2\beta_2}{\beta_1}\big(-g^4(\mu)2\beta_1\log|z\mu|+\cdots\big)\nn\\
&&-g^4(\mu)\left(\f{\beta_2^2}{\beta_1^2}-\f{\beta_3}{\beta_1}\right)
\left(-g^4(\mu)4\beta_1\log|z\mu|+\cdots\right)+\cdots\nn\\
&=&1-g^4(\mu)4\beta_1\log|z\mu| -g^6(\mu)4\beta_2\log|z\mu|+\cdots
\eea
 Eq. \eqref{gIRAFlarge}, expanded to order $g^6(\mu)$
 for  $\log|z\mu|$ of order one, yields:  
\bea
\label{gIRAFsmall}
g^4(z) &=&  g^4(\mu)\big(1 + g^4(\mu)4\beta_1\log|z\mu| + g^6(\mu)4\beta_2\log|z\mu| 
+\cdots\big)
\eea
which implies $g(z)=0$ if $g(\mu)=0$ for some -- and thus every -- $\mu\neq 0$. 
Analogously to eq. \eqref{rgi_exp_uv}, for $g(\mu) \neq g_*=0$ eq. (\ref{rgi_exp_IRAF}) implies
the existence of the RGI mass scale $ \Lrgi$: 
\bea\label{C0_IRAF}
\Lrgi &=&\mu \exp\left (-\f{1}{4\beta_1g^4(\mu)} + \f{\beta_2}{2\beta_1^2g^2(\mu) }\right) 
(g(\mu))^{\f{\beta_2^2}{\beta_1^3}-\f{\beta_3}{\beta_1^2}}
\exp\left( -\f{C}{4\beta_1} +\cdots \right)
\nn\\
&&\hspace{-1.0truecm}= |z|^{-1} \exp\left (-\f{1}{4\beta_1g^4(z)} + \f{\beta_2}{2\beta_1^2g^2(z) }\right) 
(g(z))^{\f{\beta_2^2}{\beta_1^3}-\f{\beta_3}{\beta_1^2}}
\exp\left( -\f{C}{4\beta_1} +\cdots \right)
\eea
where the dots stand for a scheme-dependent series in even powers of $g(\mu)$ and $g(z)$ respectively. 
Hence, solving eq. \eqref{C0_IRAF} in terms of $g(z)$ asymptotically as $ |z|\to +\infty/0^+$ in the IR/UV, we get:
\bea \label{alfa2_IRAF}
g^4(z) &\sim& \f{1}{-4\beta_1\log(|z| \Lrgi)}\bigg( 1+\f{\beta_2}{2\beta_1^2}
\f{\sqrt{-4\beta_1}}{\sqrt{\log(|z| \Lrgi)} }
-\f{1}{4\beta_1} \left(\f{\beta_2^2}{\beta_1^2}-\f{\beta_3}{\beta_1}\right)\nn\\
&&\hspace{-1.5truecm}
\f{\log(-4\beta_1\log(|z| \Lrgi)  ) }{\log(|z| \Lrgi) }
-\f{C}{4\beta_1\log(|z| \Lrgi) } 
-\f{\beta_2^2}{\beta_1^3\log(|z| \Lrgi)}
+ \cdots\bigg)
\eea
The corresponding universal asymptotics reads:
\bea\label{alfa2_IRas}
g^4(z) &\sim& \f{1}{-4\beta_1\log(|z| \Lrgi)}
\eea
The IR/UV asymptotics in eq. \eqref{alfa2_IRas} is selfconsistently realized if:
\bea\label{AFcases2}
\beta_1 < 0~~{\textrm{for}}~|z|\to +\infty~~\textrm{(IR)}\nn\\
\beta_1 > 0~~{\textrm{for}}~|z|\to 0^+~~\textrm{(UV)}
\eea

\subsubsection{$g_*> 0$ and $\gamma_{F^2}\neq 0$}

For $\beta_0^*\neq 0$, evaluating the integral of the inverse of the beta function in eq. \eqref{beta_0}, we get:
\bea\label{int}
&&\int_{g(\mu)}^{g(z)}\f{dg}
{g-g_* +\f{\beta_1^*}{\beta_0^*} (g-g_*)^2    +\cdots } = -\beta_0^*\int_{\mu^{-1}}^{|z|}d\log |z|
\eea 
where both scales, $\mu^{-1}$ and $|z|$, are close to infinity (IR zero) or zero (UV zero) for $g(z),g(\mu)\lesssim g_*$, and the IR (UV) zero $g_*$ is reached as
$ |z|\to +\infty/0^+$. 
Eq. \eqref{int} implies:
\bea\label{rgi_exp_new}
&&\log\left(\f{g_*-g(z)}{g_*-g(\mu)}\right) =
-{\beta_0^*} \log |z\mu|  +
\f{\beta_1^* }{\beta_0^*}(g(z)-g(\mu))+\cdots
\eea
where the dots stand for powers of $g_*-g(z)$ and $g_*-g(\mu)$ higher than one.
Equivalently:
\bea\label{rgi_exp}
&&\log(g_*-g(z)) +{\beta_0^*} \log |z|  +\f{\beta_1^* }{\beta_0^*}(g_*-g(z))+C_*+\cdots \nn\\
&&\hspace{0.5truecm}= \log(g_*-g(\mu)){-\beta_0^*}\log\mu +\f{\beta_1^* }{\beta_0^*}(g_*-g(\mu))+C_*+\cdots
\eea
where, as in the AF case, the arbitrary integration constant $C_*$ arises from the indefinite version of the integral in the lhs of eq. \eqref{int} and is scheme dependent. \par
Solving iteratively eq. \eqref{rgi_exp_new}, we obtain:
\bea\label{D7}
g_*-g(z)&=&(g_*-g(\mu))|z\mu|^{-\beta_0^*}\exp \left(\f{\beta_1^* }{\beta_0^*} (g(z)-g(\mu))+\cdots\right )\nn\\
&=&
(g_*-g(\mu))|z\mu|^{-\beta_0^*}\exp \left(
\f{\beta_1^* }{\beta_0^*} (g_*-g(\mu))\left( 1-  |z\mu|^{-\beta_0^*}  \right)+\cdots
\right )
\eea
whose perturbative expansion to order $(g_*-g(\mu))^2$ is:
\be \label{LL}
g_*-g(z)=(g_*-g(\mu))|z\mu|^{-\beta_0^*}\bigg(1+\f{\beta_1^* }{\beta_0^*} (g_*-g(\mu))\left( 1-  
|z\mu|^{-\beta_0^*}  \right)
+\cdots\bigg)
\ee
It implies $g(z)=g_*$ if $g(\mu)=g_*$ for some -- and thus every -- $\mu\neq 0$. \par
For $g(\mu) \neq g_*$ eq. \eqref{rgi_exp} implies the existence
of the RGI -- though scheme-dependent -- scale:
\bea\label{CI}
\Lrgi&=&\mu \left(g_* - g(\mu)\right)^{-\f{1}{\beta_0^*}}  \exp \left(-\f{\beta_1^*}{\beta_0^{*2}}(g_*-g(\mu))-\f{C_*}{\beta_0^*} +\cdots\right )
\nn\\
&=& |z|^{-1} \left(g_* - g(z)\right)^{-\f{1}{\beta_0^*}}  \exp \left(-\f{\beta_1^*}{\beta_0^{*2}}(g_*-g(z))-\f{C_*}{\beta_0^*} +\cdots\right )
\eea
since both sides are independent of $\mu^{-1}$ and $z$.
Solving eq. \eqref{CI} in terms of $g_*-g(z)$ asymptotically as $ |z|\to +\infty/0^+$ in the IR/UV, we get:
\be\label{int_rgi}
g_*-g(z) \sim  \left (|z|\Lrgi \right)^{-\beta_0^*}
\exp \left(- \f{\beta_1^* }{\beta_0^*}\left (|z|\Lrgi \right)^{-\beta_0^*}
-C_*+\cdots\right )
\ee
The corresponding universal asymptotics reads:
\be\label{int_rgi_uni}
g_*-g(z) \sim  \left (|z|\Lrgi \right)^{-\beta_0^*} \exp \left(-C_*\right)
\ee
The IR/UV asymptotics in eq. \eqref{int_rgi_uni} is selfconsistently realized, somehow in analogy with the AF case, if:
\bea\label{D8}
\beta_0^* > 0~~{\textrm{for}}~|z|\to +\infty~~\textrm{(IR)}\nn\\
\beta_0^* < 0~~{\textrm{for}}~|z|\to 0^+~~\textrm{(UV)}
\eea
The IR case ($\beta_0^*>0$) may be realized by the RG flow that connects an AF gauge theory in the UV to $g_*$. For example, QCD in the conformal window. \par
The UV case ($\beta_0^*<0$) may be realized by the RG flow that connects a gauge theory AF in the IR to $g_*$. For example, QCD above the conformal window, where AF in the UV is lost, by admitting the existence of the interacting zero $g_*>0$ according to the hypothesis of the asymptotic safety. \par
Since in a massless QCD-like theory AF in the UV the only free parameter is $\Lambda_{\scriptscriptstyle{UV}}$, in the corresponding conformal window $\Lambda_{\scriptscriptstyle{IR}}$ must be proportional to $\Lambda_{\scriptscriptstyle{UV}}$, with a proportionality constant that is determined by the exact RG flow.\par

\subsubsection{$g_*> 0$ and $\gamma_{F^2}= 0$ with $\beta_1^* \neq 0$}
\label{appDn1}

Firstly, we assume that $\beta_1^* \neq 0$, which now is universal because $\beta_0^* = 0$. Then, eq. \eqref{beta_0} reads:
\bea\label{IRexp_beta_zero_b1}
\beta(g)&=&\beta_1^* (g-g_*)^2+\beta_2^* (g-g_*)^3+\cdots
\eea 
and the integral of its inverse:
\bea\label{int_bla}
&&\int_{g(\mu)}^{g(z)}\f{dg}
{\beta_1^* (g-g_*)^2+\beta_2^* (g-g_*)^3    +\cdots } = -\int_{\mu^{-1}}^{|z|}d\log |z|
\eea 
yields:
\be\label{one}
\f{1}{g(z)-g_*}  -\f{1}{g(\mu)-g_*}={\beta_1^*}\log{|z\mu|}+\f{\beta_2^*}{\beta_1^*}\log\left(\f{g_*-g(\mu)}{g_*-g(z)}\right)+\cdots
\ee
or equivalently for $g(z), g(\mu) \lesssim g_*$ analogously to eq. \eqref{rgi_exp}:
\bea\label{onee}
&&-\f{1}{g_*-g(z )}-{\beta_1^*}\log|z| +\f{\beta_2^*}{\beta_1^*}\log\left({g_*-g(z)}\right)    +C_*+\cdots \nn\\
&&\hspace{1.0truecm}= -\f{1}{g_*-g(\mu)}+{\beta_1^*}\log{\mu}+\f{\beta_2^*}{\beta_1^*}\log\left({g_*-g(\mu)}\right)
+C_*+\cdots
\eea
where the dots in eqs. \eqref{one}  and \eqref{onee} stand for positive powers of $g_*-g(z)$ and $g_*-g(\mu)$. 
Solving iteratively eq. \eqref{one}, we obtain to order $(g_*-g(\mu))^2$:
\bea\label{D13_rgi}
\f{g_*-g(\mu)}{g_*-g(z)}&=&
1-(g_*-g(\mu))\beta_1^*\log{|z\mu|}-(g_*-g(\mu))\f{\beta_2^*}{\beta_1^*}\nn\\
&&\log\big(1-(g_*-g(\mu))\beta_1^*\log  |z\mu|+\cdots \big)+\cdots\nn\\
&=&
1-(g_*-g(\mu))\beta_1^*\log{|z\mu|}+(g_*-g(\mu))^2\beta_2^*\log{|z\mu|}+\cdots
\eea
Its inverse, expanded to order $(g_*-g(\mu))^2$ for $\log|z\mu|$ of order one yields:
\bea\label{D13_rgi_small}
g_*-g(z)&=&(g_*-g(\mu))\big( 1+(g_*-g(\mu))\beta_1^*\log|z\mu|- (g_*-g(\mu))^2\nn\\
&&\big(\beta_2^*\log|z\mu|-\beta_1^{*2} \log^2|z\mu|\big) +\cdots\big)
\eea
which implies $g(z)=g_*$ if $g(\mu)=g_*$ for some -- and thus every -- $\mu\neq 0$. \par
For $g(\mu) \neq g_*$ eq. \eqref{onee} implies the existence of the RGI -- though scheme-dependent -- scale:
\bea \label{RGII}
\Lrgi&=&\mu \exp{\left(\f{1}{-\beta_1^*(g_*-g(\mu))}\right)}(g_*-g(\mu))^\f{\beta_2^*}{\beta_1^{*2}}
\exp{\left(
\f{C_*}{\beta_1^*}+\cdots \right)}
\nn\\
&=&
|z|^{-1}\exp{\left(\f{1}{-\beta_1^*(g_*-g(z))}\right)}(g_*-g(z))^\f{\beta_2^*}{\beta_1^{*2}}
\exp{\left(
\f{C_*}{\beta_1^*}+\cdots \right)}
\eea
Solving eq. \eqref{RGII} asymptotically in terms of $g_*-g(z)$ as $|z|\to+\infty /0^+$ in the IR/UV yields:
\bea\label{D17}
g_*-g(z)&\sim& \f{1}{-\beta_1^*\log(|z|\Lrgi)}\bigg(
1-\f{\beta_2^*}{\beta_1^{*2}}
\f{\log(-\beta_1^*  \log(|z|\Lrgi))}{\log(|z|\Lrgi)}\nn\\
&&+\f{C_*}{\beta_1^*\log(|z|\Lrgi) }+\cdots\bigg)
\eea
The corresponding universal asymptotics is:
\bea\label{D17_as}
g_*-g(z)&\sim& \f{1}{-\beta_1^*\log(|z|\Lrgi)}
\eea
The IR/UV asymptotics is selfconsistently realized if:
\bea\label{D15}
\beta_1^*< 0~~{\textrm{for}}~|z|\to +\infty~~\textrm{(IR)}\nn\\
\beta_1^* > 0~~{\textrm{for}}~|z|\to 0^+~~\textrm{(UV)}
\eea

\subsubsection{$g_*> 0$ and $\gamma_{F^2}= 0$ with $\beta_1^*= 0$}
\label{appDn2}
Remarkably, the LET for $O=F^2$ implies that if $\beta_0^*=0$ also $\beta_1^*=0$. Then eq. \eqref{IRexp_beta_zero_b1} reads:
\bea\label{IRexp_beta_zero}
\beta(g)&=&\beta_2^* (g-g_*)^3+\beta_3^* (g-g_*)^4+\beta_4^* (g-g_*)^5+\beta_5^* (g-g_*)^6+\cdots
\eea
with $\beta_2^*$ universal.
It can be viewed as the analog of the AF case, with the replacements $-\beta_0\to  \beta_2^*$ and $g\to g-g_*$, with $\beta_0$ in eq. \eqref{beta00}. 
By replacing the beta function in eq. \eqref{int_bla} with eq. \eqref{IRexp_beta_zero}, the integral in eq. \eqref{int_bla} now yields:
\bea\label{twob}
&&\f{1}{(g(z )-g_*)^2}  -\f{1}{(g(\mu)-g_*)^2} -\f{2\beta_3^*}{\beta_2^*}\left(\f{1}{g(z )-g_*}  -\f{1}{g(\mu)-g_*}\right)\nn\\
&&
\hspace{0.5truecm}
={2\beta_2^*}\log{|z\mu|}
+2\left(\f{\beta_3^{*2}}{\beta_2^{*2}}-\f{\beta_4^*}{\beta_2^*} 
\right)
\log\left(\f{g_*-g(z)}{g_*-g(\mu)} \right)+\cdots
\eea
or equivalently for $g(z),g(\mu) \lesssim g_*$:
\bea\label{twoo}
&&\f{1}{(g_*-g(z ))^2} 
+\f{2\beta_3^*}{\beta_2^*(g_*-g(z ))}
-{2\beta_2^*}\log{|z|}
-2\left(\f{\beta_3^{*2}}{\beta_2^{*2}}-\f{\beta_4^*}{\beta_2^*} 
\right)
\log\left({g_*-g(z)} \right)\nn\\
&&~~~+C_*+\cdots =
 \f{1}{(g_*-g(\mu))^2} 
 +\f{2\beta_3^*}{\beta_2^*(g_*-g(\mu ))}
 +{2\beta_2^*}\log{\mu}\nn\\
&&\hspace{3.0truecm} -2\left(\f{\beta_3^{*2}}{\beta_2^{*2}}-\f{\beta_4^*}{\beta_2^*} 
\right)
\log\left({g_*-g(\mu)} \right)
+C_*+\cdots
\eea
Solving eq. \eqref{twob} iteratively, we obtain to order $(g_*-g(\mu))^3$:
\bea\label{D16_rgi}
&&\hspace{-1.0truecm}\f{(g_*-g(\mu))^2}{(g_*-g(z))^2}=
1+(g_*-g(\mu))^2 2\beta_2^*\log|z\mu|\nn\\
&&\hspace{-0.6truecm}-(g_*-g(\mu))\f{2\beta_3^*}{\beta_2^*}
\bigg(\sqrt{1+(g_*-g(\mu))^2 2\beta_2^*\log|z\mu|+\cdots}-1\bigg)\nn\\
&&\hspace{-0.6truecm}-(g_*-g(\mu))^2
\bigg(\f{\beta_3^{*2}}{\beta_2^{*2}}-\f{\beta_4^*}{\beta_2^*}\bigg)
\log\big(1+(g_*-g(\mu))^2 2\beta_2^*\log|z\mu|+\cdots\big)+\cdots\nn\\
&&\hspace{-0.6truecm}=
1+(g_*-g(\mu))^2 2\beta_2^*\log|z\mu|-(g_*-g(\mu))^3 2\beta_3^*\log|z\mu|+\cdots
\eea
Eq. \eqref{D16_rgi} expanded to order $(g_*-g(\mu))^3$, for $\log|z\mu| $ of order one, yields: 
\bea\label{D16_rgi_small}
(g_*-g(z))^2&=&(g_*-g(\mu))^2 \big( 1-(g_*-g(\mu))^2 2\beta_2^*\log|z\mu|\nn\\
&&+(g_*-g(\mu))^32\beta_3^*\log|z\mu|
+\cdots\big)
\eea
which implies $g(z)=g_*$ if $g(\mu)=g_*$ for some -- and thus every -- $\mu\neq 0$. \par
For $g(\mu) \neq g_*$ eq. \eqref{twoo} implies the existence of the RGI scale:
\bea \label{RGIII}
\Lrgi&=&\mu\exp{\bigg(\f{1}{2\beta_2^*(g_*-g(\mu))^2}+\f{\beta_3^{*}}{\beta_2^{*2}(g_*-g(\mu))}\bigg)}
(g_*-g(\mu))^{ -\f{\beta_3^{*2}}{\beta_2^{*3}}+\f{\beta_4^*}{ \beta_2^{*2}}}\nn\\
&&~~\exp{\bigg(\f{C_*}{2\beta_2^*}+\cdots \bigg)}
\nn\\
&=&|z|^{-1}\exp{\bigg(\f{1}{2\beta_2^*(g_*-g(z))^2}+\f{\beta_3^{*}}{\beta_2^{*2}(g_*-g(z))}\bigg)}
(g_*-g(z))^{ -\f{\beta_3^{*2}}{\beta_2^{*3}}+\f{\beta_4^*}{ \beta_2^{*2}}}\nn\\
&&~~\exp{\bigg(\f{C_*}{2\beta_2^*}+\cdots \bigg)}
\eea
Solving eq. \eqref{RGIII} asymptotically in terms of $g_*-g(z)$ as $|z|\to+\infty /0^+$ in the IR/UV yields:
\bea\label{D24}
&&\hspace{-0.5truecm}(g_*-g(z))^2\sim\f{1}{2\beta_2^*\log(|z|\Lrgi)}\bigg( 1+
\f{\beta_3^{*}}{\beta_2^{*2}}\f{\sqrt{2\beta_2^* }}
{\sqrt{\log(|z|\Lrgi)}}\nn\\
&&-\left(-\f{\beta_3^{*2}}{\beta_2^{*3}}+\f{\beta_4^*}{ \beta_2^{*2}}\right)
\f{\log(2\beta_2^*\log(|z|\Lrgi)  ) }{2\log(|z|\Lrgi)}
+\f{C_*}{2\beta_2^*\log(|z|\Lrgi)}\nn\\
&&+\f{2\beta_3^{*2}}{\beta_2^{*3}\log(|z|\Lrgi)}
+\cdots\bigg)
\eea
The corresponding universal asymptotics is:
\bea\label{D24_as}
&&\hspace{-0.5truecm}(g_*-g(z))^2\sim\f{1}{2\beta_2^*\log(|z|\Lrgi)}
\eea
The IR/UV asymptotics in eqs. \eqref{D24} and \eqref{D24_as} is selfconsistently realized if:
\bea
\label{D22eq}
\beta_2^* > 0~~{\textrm{for}}~|z|\to +\infty~~\textrm{(IR)}\nn\\
\beta_2^* < 0~~{\textrm{for}}~|z|\to 0^+~~\textrm{(UV)}
\eea
Indeed, these conditions are analogous to the ones for $\beta_0$ in order to realize the asymptotic freedom in the UV ($-\beta_0<0$) or IR ($-\beta_0>0$).

\subsubsection{$g_*> 0$ and $\gamma_{F^2}= 0$ with $\beta_1^*= \cdots=\beta_{n-1}^*=0$}
 
Then eq. \eqref{IRexp_beta_zero_b1} reads:
\bea\label{IRexp_beta_zero_n}
\beta(g)&=&\beta_n^* (g-g_*)^{n+1}+\beta_{n+1}^* (g-g_*)^{n+2}+\cdots
\eea
with $\beta_n^*$ universal.
 The integral in eq. \eqref{int_bla} now yields:
\bea\label{twob_n}
&&\f{1}{(g(z )-g_*)^n}  -\f{1}{(g(\mu)-g_*)^n} -\f{n\beta_{n+1}^*}{(n-1)\beta_n^*}\left(\f{1}{(g(z )-g_*)^{n-1}}  -\f{1}{(g(\mu)-g_*)^{n-1}}\right)
+\cdots \nn\\
&&
={n\beta_n^*}\log{|z\mu|} -n\alpha_2
\log\left(\f{g_*-g(z)}{g_*-g(\mu)} \right)+\cdots
\eea
 where the dots stand for subleading contributions and the coefficient $\alpha_{2}$ depends on the $\beta_i^*$. Equivalently: 
\bea\label{twoo_n}
&&\f{1}{(g(z)-g_*)^n} -\f{n\beta_{n+1}^*}{(n-1)\beta_n^*}\f{1}{(g(z)-g_*)^{n-1}}
+\cdots -{n\beta_n^*}\log{|z|}+n\alpha_2
\log\left({g_*-g(z)} \right)\nn\\
&&+C_*+\cdots =
\f{1}{(g(\mu)-g_*)^n} -\f{n\beta_{n+1}^*}{(n-1)\beta_n^*}\f{1}{(g(\mu)-g_*)^{n-1}}
+\cdots 
  +{n\beta_n^*}\log{\mu} \nn\\
&&+n\alpha_2
\log\left({g_*-g(\mu)} \right)
+C_*+\cdots
\eea
Solving eq. \eqref{twob_n} iteratively, we obtain to order $(g(\mu)-g_*)^{n+1}$:
\be\label{D16_rgi_n}
\left(\f{g_*-g(\mu)}{g_*-g(z)}\right)^n=
1+(g(\mu)-g_*)^n n\beta_n^*\log|z\mu|+(g(\mu)-g_*)^{n+1} n\beta_{n+1}^*\log|z\mu|+\cdots
\ee
Eq. \eqref{D16_rgi_n} expanded to order $(g(\mu)-g_*)^{n+1}$, for $\log|z\mu| $ of order one, yields: 
\bea\label{D16_rgi_small_n}
(g_*-g(z))^n&=&(g_*-g(\mu))^n \big( 1-(g(\mu)-g_*)^n n\beta_n^*\log|z\mu|\nn\\
&&-(g(\mu)-g_*)^{n+1}n\beta_{n+1}^*\log|z\mu|+\cdots\big)
\eea
which implies $g(z)=g_*$ if $g(\mu)=g_*$ for some -- and thus every -- $\mu\neq 0$. \par
For $g(\mu) \neq g_*$ eq. \eqref{twoo_n} implies the existence of the RGI scale:
\bea \label{RGIII_n}
\Lrgi&=&\mu\exp{\bigg(\f{1}{n\beta_n^*(g(\mu)-g_*)^n}-\f{\beta_{n+1}^{*}}{(n-1)\beta_n^{*2}(g(\mu)-g_*)^{n-1}}+\cdots\bigg)}\nn\\
&&~~\big(g_*-g(\mu)\big)^{\f{\alpha_2}{\beta_n^*}}
\exp{\bigg(\f{C_*}{n\beta_n^*}+\cdots \bigg)}
\nn\\
&=&|z|^{-1}\exp{\bigg(\f{1}{n\beta_n^*(g(z)-g_*)^n}-\f{\beta_{n+1}^{*}}{(n-1)\beta_n^{*2}(g(z)-g_*)^{n-1}}+\cdots \bigg)}
\nn\\
&&~~\big(g_*-g(z)\big)^{\f{\alpha_2}{\beta_n^*}}
\exp{\bigg(\f{C_*}{n\beta_n^*}+\cdots \bigg)}
\eea
Solving eq. \eqref{RGIII_n} asymptotically in terms of $g_*-g(z)$ as $|z|\to+\infty /0^+$ in the IR/UV, for $g(z),g(\mu) \lesssim g_*$ yields:
\bea\label{D24_n}
(g_*-g(z))^n&\sim&\f{1}{(-1)^n n\beta_n^*\log(|z|\Lrgi)}
\bigg( 1+
\f{n\beta_{n+1}^{*}}{(n-1)\beta_n^{*}}
\left(\f{1}
{(-1)^n n\beta_n^*\log(|z|\Lrgi)}\right)^{\f{1}{n}}\nn\\
&&
-\alpha_2
\f{\log((-1)^n n\beta_n^*\log(|z|\Lrgi)  ) }{n\beta_n^*\log(|z|\Lrgi)}
+\f{C_*}{n\beta_n^*\log(|z|\Lrgi)}\nn\\
&&
+\left(\f{n\beta_{n+1}^{*}}{(n-1)\beta_n^{*}} \right)^2
\left(\f{1}
{(-1)^n n\beta_n^*\log(|z|\Lrgi)}\right)^{\f{2}{n}}
+\cdots\bigg)
\eea
The corresponding universal asymptotics is:
\bea\label{D24_as_n}
&&\hspace{-0.5truecm}(g_*-g(z))^n\sim\f{(-1)^n}{n\beta_n^*\log(|z|\Lrgi)}
\eea
The IR/UV asymptotics in eqs. \eqref{D24_n} and \eqref{D24_as_n} is selfconsistently realized if:
\bea
\label{D22eq_n}
n\,\textrm{even}:\,\beta_n^* > 0~~{\textrm{for}}~|z|\to +\infty~~\textrm{(IR)}\nn\\
\hspace{2.0truecm}\beta_n^* < 0~~{\textrm{for}}~|z|\to 0^+~~\textrm{(UV)}\nn\\
n\,\textrm{odd}:\,\beta_n^* < 0~~{\textrm{for}}~|z|\to +\infty~~\textrm{(IR)}\nn\\
\hspace{2.0truecm}\beta_n^* > 0~~{\textrm{for}}~|z|\to 0^+~~\textrm{(UV)}
\eea
These conditions are analogous to the ones for $\beta_0$ in order to realize the asymptotic freedom in the UV ($-\beta_0<0$) or IR ($-\beta_0>0$) and agree with the results in appendix \ref{appDn2} for $n=2$ and $\alpha_2=\f{\beta_4^*}{\beta_2^*}-\f{\beta_3^{*2}}{\beta_2^{*2}}$.

\subsection{Asymptotics of $\braket{O(z)O(0)}$}\label{D.2AF}

\subsubsection{$g_*=0$ and the AF case with $\beta_0\neq 0$}

Perturbatively\footnote{In the present paper the convention about the sign of the coefficients $\gamma^{(O)}_{i}$ agrees with \cite{MBM,MBN,BB}, but is opposite to the standard one \cite{MB00}.}:
\bea \label{a}
\gamma_{O}(g)= -\gamma^{(O)}_{0} g^2 - \gamma^{(O)}_{1} g^4+ \cdots
\eea 
where, in general, only the first coefficient, $\gamma^{(O)}_{0} $, is universal, i.e., renormalization-scheme independent. 
Then, $Z^{(O)}(g(z), g(\mu))$ in eq. \eqref{def_eps0} can be straightforwardly evaluated asymptotically in the UV ($\beta_0>0$) or IR ($\beta_0<0$) by means of eq. (\ref{a}) if $g(\mu)\neq 0$ for some $\mu\neq 0$: 
\bea \label{zeta00}
&&Z^{(O)}(g(z), g(\mu)) \sim \left(\frac{g(z)}{g(\mu)}\right)^{\frac{\gamma_0^{(O)}}{\beta_0}} \exp \bigg( \frac{\gamma_1^{(O)}  \beta_0 - \gamma_{0}^{(O)} \beta_1}{2 \beta_0^2} (g^2(z)-g^2(\mu))+\cdots \bigg)\nn\\
&&\sim \left(\frac{1}{-g^2(\mu)2\beta_0 \log({|z|\Lrgi })}
\left(
1+\frac{\beta_1}{2\beta_0^2}\frac{\log(-\beta_0\log({|z|\Lrgi } ))}{\log({|z|\Lrgi })} 
\right)\right)^{\frac{\gamma_0^{(O)}}{2\beta_0}} Z^{(O)'}(g(\mu))\nn\\
\eea
where in the second equality we have employed eq. \eqref{alfa} for $g(z)$ in terms of $\Lrgi$ and:
\be\label{ZOF}
 Z^{(O)'}(g(\mu))=\exp \bigg(- \frac{\gamma_1^{(O)}  \beta_0 - \gamma_{0}^{(O)} \beta_1}{2 \beta_0^2} g^2(\mu)+\cdots \bigg)
 \ee
 is the limit as $g(z)\to 0$ of the subleading exponential in eq. \eqref{zeta00}.
 By means of eq. \eqref{zeta00}, the asymptotics of the 2-point correlator in the UV ($\beta_0>0$) \cite{MBM,MBN,BB} or IR ($\beta_0<0$) is: 
\bea
\label{CSS0}
 \langle O(z) O(0) \rangle &\sim& \f{\mathcal{G}^{(O)}_2(0)}{z^{2D}} \left( \f{g(z)}{g(\mu)} \right)^{\f{2 \gamma^{(O)}_{0}}{\beta_0}} 
\exp \bigg( \frac{\gamma_1^{(O)}  \beta_0 - \gamma_{0}^{(O)} \beta_1}{ \beta_0^2} (g^2(z)-g^2(\mu))+\cdots \bigg)
\nn\\
 &&\hspace{-2.0truecm}\sim \f{\mathcal{G}^{(O)}_2(0)}{z^{2D}}
 \left(\frac{1}{-g^2(\mu)2\beta_0 \log({|z|\Lrgi })}
\left(
1+\frac{\beta_1}{2\beta_0^2}\frac{\log(-\beta_0\log({|z|\Lrgi } ))}{\log({|z|\Lrgi })} 
\right)\right)^{\f{ \gamma^{(O)}_{0}}{\beta_0}} \nn\\
&&\hspace{-1.0truecm}Z^{(O)'2}(g(\mu))
 \eea
where the logarithmic dependence on $|z|$ prevents the correlator from being asymptotically conformal. 
We observe that the $g(\mu)$-dependent normalization may be singular if we further take the limit $g(\mu)\to 0^+$ as $\mu\to 0^+/+\infty$ (IR/UV).\par
Inserting in the first line of eq. \eqref{zeta00} the perturbative expansion for $g(z)$ in terms of $g(\mu)$ in eq. \eqref{g2loop}, we now obtain:
\bea \label{zeta00_small}
Z^{(O)}(g(z), g(\mu)) &=& 1+g^2(\mu) \gamma_0^{(O)}\log|z\mu|+g^4(\mu)\bigg(\gamma_1^{(O)}\log|z\mu| \nn\\
&&+\big(\gamma_0^{(O)}\beta_0+\f{\gamma_0^{(O)2}}{2} \big)\log^2|z\mu|
\bigg)+
\cdots
\eea
where, as an aside, we observe that the dependence on the coefficients $\beta_i$ of the beta function starts to order $g^4(\mu)$. Eq. \eqref{zeta00_small} shows that $Z^{(O)}=1$ if $g(\mu)=0$ for any $\mu$. It follows:
\be\label{CSS0_ex}
\langle O(z) O(0) \rangle =\f{\mathcal{G}^{(O)}_2(0)}{z^{2D}}
\ee
Hence, the correlator is exactly conformal and free, if $g(\mu)=0$ for any $\mu$ according to eq. 
\eqref{CSS0_ex}, but is not asymptotically conformal if $g(\mu)\neq 0$ for some $\mu\neq 0$ according to eq. \eqref{CSS0}.

\subsubsection{$g_*=0$ and the AF case with $\beta_0=0$}

For $\beta_0=0$:
\bea \label{zeta00_IRAF}
Z^{(O)}(g(z), g(\mu))& \sim& 
\exp\bigg(-\f{\gamma_0^{(O)}}{2\beta_1} \left(\f{1}{g^2(z)}-\f{1}{g^2(\mu)}\right)\bigg)
\left(\frac{g(z)}{g(\mu)}\right)^{\frac{\gamma_1^{(O)}\beta_1-\gamma_0^{(O)}\beta_2    }{\beta_1^2}  } \nn\\
&&\exp {\bigg(  O(g^2(z)-g^2(\mu))\bigg)}\nn\\
&\sim& 
\exp\bigg(
-\f{\gamma_0^{(O)}}{2\beta_1} 
\sqrt{-4\beta_1\log(|z|\Lrgi)}\bigg)
\exp\bigg(\f{\gamma_0^{(O)}}{2\beta_1g^2(\mu)} \bigg)
\nn\\
&&
\bigg(\f{1}{-4\beta_1 g^4(\mu)\log(|z|\Lrgi)}  \bigg)^{\frac{\gamma_1^{(O)}\beta_1-\gamma_0^{(O)}\beta_2    }{4\beta_1^2}  }
Z^{(O)'}(g(\mu))
\eea
where, assuming $g(\mu)\neq 0$ for some $\mu\neq 0$, we have employed  in the second equality eq. \eqref{alfa2_IRas} for $g(z)$ in terms of $\Lrgi$ and
$Z^{(O)'}(g(\mu))=\exp \big(O(g^2(\mu)) \big)$ is the limit $g(z)\to 0$ of the last subleading exponential in the first equality in eq. \eqref{zeta00_IRAF}.
Hence, the asymptotics of the 2-point correlator in the IR ($\beta_1<0$) or UV ($\beta_1>0$) is:
\bea
\label{CSS0_IRAF}
\langle O(z) O(0) \rangle &\sim& \f{\mathcal{G}^{(O)}_2(0)}{z^{2D}}
\exp\bigg(-\f{\gamma_0^{(O)}}{\beta_1} \left(\f{1}{g^2(z)}-\f{1}{g^2(\mu)}\right)\bigg)
\left(\frac{g(z)}{g(\mu)}\right)^{2\frac{\gamma_1^{(O)}\beta_1-\gamma_0^{(O)}\beta_2    }{\beta_1^2}  } \nn\\
&&\exp {\bigg(  O(g^2(z)-g^2(\mu))\bigg)}\nn\\
 & \sim& \f{\mathcal{G}^{(O)}_2(0)}{z^{2D}}
 \exp\bigg(
-\f{\gamma_0^{(O)}}{\beta_1} 
\sqrt{-4\beta_1\log(|z|\Lrgi)}\bigg)
\exp\bigg(\f{\gamma_0^{(O)}}{\beta_1g^2(\mu)} \bigg)
\nn\\
&&\hspace{0.8truecm}
\bigg(\f{1}{-4\beta_1 g^4(\mu)\log(|z|\Lrgi)}  \bigg)^{\frac{\gamma_1^{(O)}\beta_1-\gamma_0^{(O)}\beta_2    }{2\beta_1^2}  }
Z^{(O)'2}(g(\mu))
 \eea
where, analogously to eq. \eqref{CSS0}, the logarithmic dependence on $|z|$ prevents the correlator from being asymptotically conformal and, again, the $g(\mu)$-dependent normalization may be singular by further taking the limit $g(\mu)\to 0^+$ as $\mu\to 0^+/+\infty$ in the IR/UV.
Inserting in the first equality of eq. \eqref{zeta00_IRAF} the perturbative expansion for $g(z)$ in terms of $g(\mu)$ in eq. \eqref{gIRAFsmall}, we now obtain:
\bea \label{zeta00_IRAF_small}
Z^{(O)}(g(z), g(\mu)) &= &
1+g^2(\mu)\gamma_0^{(O)}\log|z\mu|+g^4(\mu)\big( \gamma_1^{(O)}\log|z\mu|
\nn\\&&
+\f{\gamma_0^{(O)2}}{2}\log^2|z\mu|\big)+\cdots
\eea
where the dependence on the beta function coefficients $\beta_i$ only occurs to orders higher than $g^4(\mu)$.
Eq. \eqref{zeta00_IRAF_small} shows that $Z^{(O)}=1$ if $g(\mu)=0$ for any $\mu$. Correspondingly, we obtain the exactly conformal and free correlator:
\be\label{CSS0_ex2}
\langle O(z) O(0) \rangle =\f{\mathcal{G}^{(O)}_2(0)}{z^{2D}}
\ee
as in eq. \eqref{CSS0_ex}.

\subsubsection{$g_*>0$ and $F^2$ irrelevant in the IR/UV} \label{D.2.1}

Analogously to the beta function, the most general perturbative expansion of the anomalous dimension of a multiplicatively renormalizable operator $O$ in a neighborhood of $g_*>0$ reads:
\bea\label{Tgamma}
\gamma_O(g) &=&\gamma_O(g_*)+ \gamma_O^\prime(g_*)(g-g_*)+ \f{1}{2}\gamma_O^{\prime\prime}(g_*)(g-g_*)^2+\cdots\nonumber\\
&\equiv& \gamma_0^{(O)*}+ \gamma_1^{(O)*}(g-g_*)+ \gamma_2^{(O)*}(g-g_*)^2+\cdots
\eea
where only the first coefficient $\gamma_0^{(O)*}$, i.e., the anomalous dimension of $O$ at the zero of the beta function $\gamma_O(g_*)$, is renormalization-scheme independent.
For $\beta_0^*\neq 0$ and generic values of $\gamma_{0}^{(O)*}$ and $\gamma_{1}^{(O)*}$ with $g(\mu)\neq g_*$, eq. \eqref{def_eps0} yields:
\bea\label{ZO_IR}
Z^{(O)}(g(z), g(\mu))&\sim& \exp \int_{g(\mu)}^{g(z)}
\f{\gamma_0^{(O)*} + \gamma_1^{(O)*}(g-g_*)  + \cdots   }{\beta_0^*(g-g_*) +  \beta_1^*  (g-g_*)^2  + \cdots   }\,dg\nonumber\\
&\sim& \left( \f{g_*-g(z)}{g_*-g(\mu)}\right )^{\f{\gamma_0^{(O)*}}{\beta_0^*}}\exp \left ( 
\f{\gamma_1^{(O)*} \beta_0^* -  \gamma_0^{(O)*} \beta_1^* }{\beta_0^{*2}}
(g(z)-g(\mu))+\cdots     \right ) \nonumber\\
&\sim& (|z|\Lrgi)^{-\gamma_0^{(O)*}}\left( \f{1}{g_*-g(\mu)}\right )^{\f{\gamma_0^{(O)*}}{\beta_0^*}} Z^{(O)'}(g_*-g(\mu))
\eea
where, assuming $g(\mu)\neq g_*$ for some $\mu\neq 0$, we have employed in the last equality eq. \eqref{int_rgi_uni} for $g(z)$ in terms of $\Lrgi$ and:
\be\label{ZO_IR_mu}
Z^{(O)'}(g_*-g(\mu))=\exp \left ( 
\f{\gamma_1^{(O)*} \beta_0^* -  \gamma_0^{(O)*} \beta_1^* }{\beta_0^{*2}}
(g_*-g(\mu))+\cdots     \right ) 
\ee
Hence, by means of eq. \eqref{ZO_IR} the IR ($\beta_0^*>0$) or UV ($\beta_0^*<0$) asymptotics of the correlator in eq. \eqref{pert2_eps0} is:
\be
\label{CSS0_IR}
 \langle O(z) O(0) \rangle \sim\f{\mathcal{G}^{(O)}_2(g_*)}{z^{2D}} 
 (|z|\Lrgi)^{-2\gamma_0^{(O)*}}\left( \f{1}{g_*-g(\mu)}\right )^{\f{2\gamma_0^{(O)*}}{\beta_0^*}} Z^{(O)'2}(g_*-g(\mu))
\ee
 Eq. \eqref{CSS0_IR} shows that if $g(\mu)\neq g_*$ for some $\mu\neq 0,+\infty$ the correlator is asymptotically conformal with anomalous dimension $\gamma_{O}(g_*)=\gamma_0^{(O)*}$.
Inserting in the first equality of eq. \eqref{ZO_IR} the perturbative expansion for $g(z)$ in terms of $g(\mu)$ in eq. \eqref{LL}, we now obtain:
\bea \label{ZO_IR_small}
Z^{(O)}(g(z), g(\mu)) &= &|z\mu|^{-\gamma_0^{(O)*}}\bigg( 1+( g_*-g(\mu)) \f{\gamma_1^{(O)*}}{\beta_0^*} \big( 1-|z\mu|^{-\beta_0^*}\big)+\cdots
\bigg)
\eea
where the dependence on the beta function coefficients $\beta_i$ starts to order $g_*-g(\mu)$. Eq. \eqref{ZO_IR_small}
shows that
 $Z^{(O)}=|z\mu|^{-\gamma_0^{(O)*}}$ if $g(\mu)=g_*$ for any $\mu$, so that we obtain the exactly conformal  correlator:
\be\label{CSS0_IR_ex}
\langle O(z) O(0) \rangle =\f{\mathcal{G}^{(O)}_2(g_*)}{z^{2D}}|z\mu|^{-2\gamma_0^{(O)*}}
\ee
We observe that the normalization of the correlators in eqs. \eqref{CSS0_IR} and \eqref{CSS0_IR_ex} may not coincide. In fact, the normalization in eq. \eqref{CSS0_IR} may be singular in the limit $g(\mu)\to g_*$ as $\mu\to 0^+/+\infty$ in the IR/UV.

\subsubsection{$g_*>0$ and $F^2$ marginal in the IR/UV with $\beta_1^*\neq 0$ }\label{D.2.2} 
 
For $\beta_0^*=\gamma_{F^2}(g_*)=0$ and nonvanishing $\beta_1^*$ eq. \eqref{ZO_IR} now reads:
\bea\label{ZO_IR_one}
Z^{(O)}(g(z), g(\mu))&\sim& \exp \int_{g(\mu)}^{g(z)}
\f{\gamma_0^{(O)*} + \gamma_1^{(O)*}(g-g_*) + \gamma_2^{(O)*}(g-g_*)^2 +\cdots   }{ \beta_1^*  (g-g_*)^2 + \beta_2^*  (g-g_*)^3  +\cdots }\,dg\nonumber\\
&&\hspace{-3.2truecm}\sim 
\exp \left ( 
-\f{\gamma_0^{(O)*}}{\beta_1^* }\left(\f{1}{g(z)-g_* } -\f{1}{g(\mu)-g_*} \right) \right)
\left( \f{g(z)-g_*}{g(\mu)-g_*}\right )^{
\f{ \gamma_1^{(O)*}\beta_1^*- \gamma_0^{(O)*}\beta_2^*  }{\beta_1^{*2}}}
\nn\\
&&\hspace{-2.8truecm}\exp\big (O (g(z)-g(\mu)) \big)
\nn\\
&&\hspace{-3.2truecm}\sim (|z|\Lrgi)^{-\gamma_0^{(O)*}}
\exp \left ( 
-\f{\gamma_0^{(O)*}}{\beta_1^*(g_*-g(\mu)) }\right)
\nn\\
&&\hspace{-3.2truecm}\left( \f{1}{-(g_*-g(\mu))\beta_1^*\log(|z|\Lrgi)}
\right)^{
\f{ \gamma_1^{(O)*}\beta_1^*- \gamma_0^{(O)*}\beta_2^*  }{\beta_1^{*2}}}
Z^{(O)'}(g_*-g(\mu))
\eea
where, assuming $g(\mu)\neq g_*$ for some $\mu\neq 0$, we have employed in the last equality eq. \eqref{D17_as} for $g_*-g(z)$ in terms of $\Lrgi$ and $Z^{(O)'}(g_*-g(\mu))=\exp{(O(g_*-g(\mu)))}$ is the last subleading exponential in the second equality for $g(z)=g_*$. 

Hence, by means of eq. \eqref{ZO_IR_one} the IR ($\beta_1^*<0$) or UV ($\beta_1^*>0$) asymptotics of the correlator in eq. \eqref{pert2_eps0} is:
\bea
\label{D23}
 \langle O(z) O(0) \rangle &\sim&\f{\mathcal{G}^{(O)}_2(g_*)}{z^{2D}} 
 \exp \left ( 
-\f{2\gamma_0^{(O)*}}{\beta_1^* }\left(\f{1}{g(z)-g_* } -\f{1}{g(\mu)-g_*} \right) \right)
\nn\\
&&\left( \f{g(z)-g_*}{g(\mu)-g_*}\right )^{2
\f{ \gamma_1^{(O)*}\beta_1^*- \gamma_0^{(O)*}\beta_2^*  }{\beta_1^{*2}}}
 Z^{(O)'2}(g_*-g(\mu))\nn\\
 &&\hspace{-3.0truecm}\sim\f{\mathcal{G}^{(O)}_2(g_*)}{z^{2D}}  
  (|z|\Lrgi)^{-2\gamma_0^{(O)*}}
\exp \left ( 
-\f{2\gamma_0^{(O)*}}{\beta_1^*(g_*-g(\mu)) }\right)\nn\\
&&\hspace{-2.8truecm}\left( \f{1}{-(g_*-g(\mu))\beta_1^*\log(|z|\Lrgi)}
\right)^{2
\f{ \gamma_1^{(O)*}\beta_1^*- \gamma_0^{(O)*}\beta_2^*  }{\beta_1^{*2}}}
Z^{(O)'2}(g_*-g(\mu))
 \eea
As in the AF case, also for $g_*>0$ the correlator in eq. \eqref{D23} is not asymptotically conformal if $g(\mu)\neq g_*$ for some $\mu\neq 0$, due to the logarithmic dependence on $|z|$. Moreover, as in all the other cases, the $g(\mu)$-dependent normalization may be singular if we further take the limit $g(\mu)\to g_*$ as $\mu\to 0^+/+\infty$ in the IR/UV.\par
Inserting in the second equality of eq. \eqref{ZO_IR_one} the perturbative expansion for $g_*-g(z)$ in terms of $g_*-g(\mu)$ in eq. \eqref{D13_rgi_small}, we obtain:
 \be\label{Zmarg_small} 
 Z^{(O)}(g(z), g(\mu))=|z\mu|^{-\gamma_0^{(O)*}}\big( 1+(g_*-g(\mu))\gamma_1^{(O)*}\log|z\mu| +\cdots\big)
 \ee
 where the dependence on the beta function coefficients $\beta_i^*$ occurs to orders higher than $g_*-g(\mu)$. Eq. \eqref{Zmarg_small} 
shows that
 $Z^{(O)}=|z\mu|^{-\gamma_0^{(O)*}}$ if $g(\mu)=g_*$ for any $\mu$, so that we obtain:
\be\label{D23_ex}
\langle O(z) O(0) \rangle =\f{\mathcal{G}^{(O)}_2(g_*)}{z^{2D}}|z\mu|^{-2\gamma_0^{(O)*}}
\ee
Hence, the correlator is exactly conformal if $g(\mu)=g_*$ for any $\mu$ according to eq. 
\eqref{D23_ex}, but is not asymptotically conformal if $g(\mu)\neq g_*$ for some $\mu\neq 0$ according to eq. \eqref{D23}.

 \subsubsection{$g_*>0$ and $F^2$ marginal in the IR/UV with $\beta_1^*= 0$ }\label{D.2.2b} 

For $\gamma_{F^2}=\beta_0^*=0$ the LET further implies $\beta_1^*=0$, according to eq. \eqref{eq:fs0}. Then, for generic $\gamma_O(g)$  eq. \eqref{ZO_IR} reads:
\bea\label{ZO_IR_two}
Z^{(O)}(g(z), g(\mu))&\sim& \exp \int_{g(\mu)}^{g(z)}
\f{\gamma_0^{(O)*} + \gamma_1^{(O)*}(g-g_*)+ \gamma_2^{(O)*}(g-g_*)^2  +\cdots  }{ \beta_2^*  (g-g_*)^3  + \beta_3^*  (g-g_*)^4 + \beta_4^*  (g-g_*)^5 +\cdots }\,dg\nonumber\\
&&\hspace{-2.2truecm}\sim 
\exp \bigg ( 
-\f{\gamma_0^{(O)*}}{2\beta_2^* }\left(\f{1}{(g(z)-g_*)^2 } -\f{1}{(g(\mu)-g_*)^2} \right) \bigg)
\nn\\
&&\hspace{-1.8truecm}
\exp \bigg ( 
\f{\gamma_0^{(O)*} \beta_3^* -\gamma_1^{(O)*}\beta_2^{*}}{\beta_2^{*2} }
\left(\f{1}{g(z)-g_* } -\f{1}{g(\mu)-g_*} \right) \bigg)\nn\\
&&\hspace{-1.8truecm}
\left( \f{g(z)-g_*}{g(\mu)-g_*}\right )^{\f{\gamma_2^{(O)*}\beta_2^*- \gamma_1^{(O)*}\beta_3^*  +\gamma_0^{(O)*} \alpha_3}{\beta_2^{*2}}}
 \exp \bigg ( O(g(z)-g(\mu)) \bigg)
\nn\\
&&\hspace{-2.2truecm}\sim (|z|\Lrgi)^{-\gamma_0^{(O)*}}
\exp \bigg ( -
\f{\gamma_0^{(O)*} \beta_3^* -\gamma_1^{(O)*}\beta_2^{*}}{\beta_2^{*2} }
\sqrt{2\beta_2^*\log(|z|\Lrgi) }\bigg)\nn\\
&&\hspace{-1.8truecm}
\left( \f{1}{2\beta_2^*\log(|z|\Lrgi) }\right )^{\f{1}{2}\f{\gamma_2^{(O)*}\beta_2^*- \gamma_1^{(O)*}\beta_3^*  +\gamma_0^{(O)*} \alpha_3}{\beta_2^{*2}}}\nn\\
&&\hspace{-1.8truecm}\exp \bigg ( 
-\f{\gamma_0^{(O)*}}{2\beta_2^* }\left( -\f{1}{(g(\mu)-g_*)^2} \right) \bigg)
\exp \bigg ( 
\f{\gamma_0^{(O)*} \beta_3^* -\gamma_1^{(O)*}\beta_2^{*}}{\beta_2^{*2} }
\left( -\f{1}{g(\mu)-g_*} \right) \bigg)\nn\\
&&\hspace{-1.8truecm}
\left( \f{1}{g_*-g(\mu)}\right )^{\f{\gamma_2^{(O)*}\beta_2^*- \gamma_1^{(O)*}\beta_3^*  +\gamma_0^{(O)*} \alpha_3}{\beta_2^{*2}}} Z^{(O)'}(g_*-g(\mu))
\eea
where, assuming $g(\mu)\neq g_*$ for some $\mu\neq 0$, we have employed in the last equality eq. \eqref{D24_as} for $g_*-g(z)$ in terms of $\Lrgi$,  $\alpha_3=\f{\beta_3^{*2}-\beta_4^*\beta_2^*}{\beta_2^{*}}$ and
$Z^{(O)'}(g_*-g(\mu))=\exp{(O(g_*-g(\mu)))}$ is the last subleading exponential in the second equality for $g(z)=g_*$.
Hence, by means of eq. \eqref{ZO_IR_two} the IR ($\beta_2^*>0$) or UV ($\beta_2^*<0$) asymptotics of the correlator in eq. \eqref{pert2_eps0} is:
\bea
\label{OOnew}
 \langle O(z) O(0) \rangle &\sim&\f{\mathcal{G}^{(O)}_2(g^*)}{z^{2D}} 
  (|z|\Lrgi)^{-2\gamma_0^{(O)*}}\nn\\
  &&\hspace{-1.6truecm} 
\exp \bigg ( -2
\f{\gamma_0^{(O)*} \beta_3^* -\gamma_1^{(O)*}\beta_2^{*}}{\beta_2^{*2} }
\sqrt{2\beta_2^*\log(|z|\Lrgi) }\bigg)
\nn\\&&\hspace{-1.6truecm}
\left( \f{1}{2\beta_2^*\log(|z|\Lrgi) }\right )^{\f{\gamma_2^{(O)*}\beta_2^*- \gamma_1^{(O)*}\beta_3^*  +\gamma_0^{(O)*} \alpha_3}{\beta_2^{*2}}}\nn\\
&&\hspace{-1.6truecm}\exp \bigg ( 
-\f{\gamma_0^{(O)*}}{\beta_2^* }\left( -\f{1}{(g(\mu)-g_*)^2} \right) \bigg)
\exp \bigg ( 2
\f{\gamma_0^{(O)*} \beta_3^* -\gamma_1^{(O)*}\beta_2^{*}}{\beta_2^{*2} }
\left( -\f{1}{g(\mu)-g_*} \right) \bigg)\nn\\
&&\hspace{-1.6truecm}
\left( \f{1}{g_*-g(\mu)}\right )^{2\f{\gamma_2^{(O)*}\beta_2^*- \gamma_1^{(O)*}\beta_3^*  +\gamma_0^{(O)*} \alpha_3}{\beta_2^{*2}}} Z^{(O)'2}(g_*-g(\mu))
\eea
As for $\beta_1^*\neq 0$ in eq. \eqref{D23}, for $g(\mu)\neq g_*$ and marginal $F^2$ with $\beta_1^*= 0$, the correlator in eq. \eqref{OOnew}
is not asymptotically conformal due to the logarithmic dependence on $|z|$. 
As in all the other cases, the $g(\mu)$-dependent normalization may be singular if we further take the limit $g(\mu)\to g_*$ as $\mu\to 0^+/+\infty$ (IR/UV).
Inserting in the second equality of eq. \eqref{ZO_IR_two} the perturbative expansion for $g_*-g(z)$ in terms of $g_*-g(\mu)$ in eq. \eqref{D16_rgi_small}, we obtain:
 \be\label{Zmarg_small2}
 Z^{(O)}(g(z), g(\mu))=|z\mu|^{-\gamma_0^{(O)*}}\big( 1+(g_*-g(\mu))\gamma_1^{(O)*}\log|z\mu| +\cdots\big)
 \ee
 which coincides to this order with the analogous expansion in eq. \eqref{Zmarg_small}. It again shows that
 $Z^{(O)}=|z\mu|^{-\gamma_0^{(O)*}}$ if $g(\mu)=g_*$ for any $\mu$, so that we obtain:
\be\label{OOnew_ex}
\langle O(z) O(0) \rangle =\f{\mathcal{G}^{(O)}_2(g_*)}{z^{2D}}|z\mu|^{-2\gamma_0^{(O)*}}
\ee
Hence, the correlator is exactly conformal if $g(\mu)=g_*$ for any $\mu$ according to eq. 
\eqref{OOnew_ex}, but is not asymptotically conformal if $g(\mu)\neq g_*$ for some $\mu\neq 0$ according to eq. \eqref{OOnew}.

 \subsubsection{$g_*>0$ and $F^2$ marginal in the IR/UV with $\beta_1^*= \cdots=\beta_{n-1}^*=0$ }\label{D.2.2b_n} 
 Eq. \eqref{ZO_IR} reads:
\bea\label{ZO_IR_two_n}
Z^{(O)}(g(z), g(\mu))&\sim& \exp \int_{g(\mu)}^{g(z)}
\f{\gamma_0^{(O)*} + \gamma_1^{(O)*}(g-g_*)+ \gamma_2^{(O)*}(g-g_*)^2  +\cdots  }{ \beta_n^*  (g-g_*)^{n+1}  + \beta_{n+1}^*  (g-g_*)^{n+2} + \cdots }\,dg\nonumber\\
&&\hspace{-2.truecm}\sim 
\exp \bigg ( 
-\f{\gamma_0^{(O)*}}{n\beta_n^* }\left(\f{1}{(g(z)-g_*)^n } -\f{1}{(g(\mu)-g_*)^n} \right) \bigg)
\nn\\
&&\hspace{-1.5truecm}
\exp \bigg ( 
\f{\gamma_0^{(O)*} \beta_{n+1}^* -\gamma_1^{(O)*}\beta_n^{*}}{(n-1)\beta_n^{*2} }
\left(\f{1}{(g(z)-g_*)^{n-1} } -\f{1}{(g(\mu)-g_*)^{n-1}} \right) +\cdots \bigg)\nn\\
&&\hspace{-1.5truecm}
\left( \f{g(z)-g_*}{g(\mu)-g_*}\right )^{\f{\alpha_2}{\beta_n^*}}
 \exp \bigg ( O(g(z)-g(\mu)) \bigg)
\nn\\
&&\hspace{-2.truecm}\sim (|z|\Lrgi)^{-\gamma_0^{(O)*}}
\exp \bigg ( 
\f{\gamma_0^{(O)*} \beta_{n+1}^* -\gamma_1^{(O)*}\beta_n^{*}}{(n-1)\beta_n^{*2} }
(n\beta_n^*\log(|z|\Lrgi) )^{\f{n-1}{n}}+\cdots\bigg)\nn\\
&&\hspace{-1.5truecm}
\left( \f{1}{n\beta_n^*\log(|z|\Lrgi) }\right )^{\f{\alpha_2}{n\beta_n^*}}\nn\\
&&\hspace{-1.5truecm}
\exp \bigg ( 
\f{\gamma_0^{(O)*}}{n\beta_n^* } \f{1}{(g(\mu)-g_*)^n} \bigg)
\exp \bigg ( -
\f{\gamma_0^{(O)*} \beta_{n+1}^* -\gamma_1^{(O)*}\beta_n^{*}}{(n-1)\beta_n^{*2} }
\f{1}{(g(\mu)-g_*)^{n-1}} \bigg ) \nn\\
&&\hspace{-1.5truecm}
\left( \f{1}{g(\mu)-g_*}\right )^{\f{\alpha_2}{\beta_n^*}}Z^{(O)'}(g_*-g(\mu))
\eea
where, assuming $g(\mu)\neq g_*$ for some $\mu\neq 0$, we have employed in the last equality eq. \eqref{D24_as_n} for $g(z)-g_*$ in terms of $\Lrgi$ and
$Z^{(O)'}(g_*-g(\mu))=\exp{(O(g_*-g(\mu)))}$ is the last subleading exponential in the second equality for $g(z)=g_*$.
Hence, by means of eq. \eqref{ZO_IR_two_n} the IR ($\beta_n^*>0 (<0)$ for $n$ even (odd)) or UV ($\beta_n^*<0 (>0)$ for $n$ even (odd)) asymptotics of the correlator in eq. \eqref{pert2_eps0} is:
\bea
\label{OOnew_n}
 \langle O(z) O(0) \rangle &\sim&\f{\mathcal{G}^{(O)}_2(g^*)}{z^{2D}} 
  (|z|\Lrgi)^{-2\gamma_0^{(O)*}}
  \nn\\&&\hspace{-2.5truecm}
\exp \bigg ( 2
\f{\gamma_0^{(O)*} \beta_{n+1}^* -\gamma_1^{(O)*}\beta_n^{*}}{(n-1)\beta_n^{*2} }
(n\beta_n^*\log(|z|\Lrgi) )^{\f{n-1}{n}}+\cdots\bigg)
\left( \f{1}{n\beta_n^*\log(|z|\Lrgi) }\right )^{\f{2\alpha_2}{n\beta_n^*}}\nn\\
&&\hspace{-2.5truecm}
\exp \bigg ( 
\f{2\gamma_0^{(O)*}}{n\beta_n^* }\f{1}{(g(\mu)-g_*)^n} \bigg)
\exp \bigg ( -2
\f{\gamma_0^{(O)*} \beta_{n+1}^* -\gamma_1^{(O)*}\beta_n^{*}}{(n-1)\beta_n^{*2} }
\f{1}{(g(\mu)-g_*)^{n-1}}+\cdots \bigg )
\nn\\
&&\hspace{-2.5truecm}
\left( \f{1}{g(\mu)-g_*}\right )^{\f{2\alpha_2}{\beta_n^*}}Z^{(O)'2}(g_*-g(\mu))
  \eea
As for $\beta_1^*\neq 0$ in eq. \eqref{D23} and $\beta_2^*\neq 0$ ($n=2$) in eq. \eqref{OOnew}, for $g(\mu)\neq g_*$ and marginal $F^2$ with $\beta_1^*= \cdots=\beta_{n-1}^*=0$ for any $n\geq 2$, the correlator in eq. \eqref{OOnew_n}
is not asymptotically conformal due to the logarithmic dependence on $|z|$. 
As in all the other cases, the $g(\mu)$-dependent normalization may be singular if we further take the limit $g(\mu)\to g_*$ as $\mu\to 0^+/+\infty$ (IR/UV).

Inserting in the second equality of eq. \eqref{ZO_IR_two_n} the perturbative expansion for $g_*-g(z)$ in terms of $g_*-g(\mu)$ in eq. \eqref{D16_rgi_small_n}, we obtain:
 \be\label{Zmarg_small2_n}
 Z^{(O)}(g(z), g(\mu))=|z\mu|^{-\gamma_0^{(O)*}}\big( 1+(g_*-g(\mu))\gamma_1^{(O)*}\log|z\mu| +\cdots\big)
 \ee
 which coincides to this order with the analogous expansion in eqs. \eqref{Zmarg_small} and \eqref{Zmarg_small2}. It again shows that
 $Z^{(O)}=|z\mu|^{-\gamma_0^{(O)*}}$ if $g(\mu)=g_*$ for any $\mu$, so that we obtain:
\be\label{OOnew_ex_n}
\langle O(z) O(0) \rangle =\f{\mathcal{G}^{(O)}_2(g_*)}{z^{2D}}|z\mu|^{-2\gamma_0^{(O)*}}
\ee
Hence, the correlator is exactly conformal if $g(\mu)=g_*$ for any $\mu$ according to eq. 
\eqref{OOnew_ex_n}, but is not asymptotically conformal if $g(\mu)\neq g_*$ for some $\mu\neq 0$ according to eq. \eqref{OOnew_n}.

\subsection{Asymptotics of $\braket{F^2(z)F^2(0)}$}\label{D.3}

$Z^{(F^2)}$ may be computed in a closed form in terms of the beta function. Indeed, eq. \eqref{gF22} relates the anomalous dimension of $F^2$ to the beta function, so that $Z^{(F^2)}$ in eq. \eqref{def_eps0} admits a closed form in terms of the beta function:
\bea \label{z}
Z^{(F^2)}(g(z), g(\mu)) &=& \exp \int_{g(\mu)} ^{g(z)}      \frac{ \gamma_{F^2} (g) } {\beta(g)} dg \nonumber \\
                                                  &=& \exp \int_{g(\mu)} ^{g(z)}      \frac{\f{\partial}{\partial g}\left(\f{\beta(g)}{g}\right) } {\frac{\beta(g)}{g}} dg  \nonumber \\
                                                  &=& \f{ \beta(g(z)) }{g(z) }  \f{g(\mu) } { \beta(g(\mu))}
\eea
Correspondingly, the solution of the CS equation admits the form:
\be
\label{CS_2ptF2}
 \braket{F^2(z) F^2(0)} = \f{\mathcal{G}^{(F^2)}_2(g(z))}{z^{8}}
\left (\f{ \beta(g(z)) }{g(z) }\right )^2 \left ( \f{g(\mu) } { \beta(g(\mu))} \right)^2
 \ee
 
\subsubsection{$g_*=0$ and the AF case}

The IR/UV asymptotics of $\braket{F^2(z)F^2(0)}$ in the AF case 
is obtained by expanding $\f{\beta(g(z))}{g(z)}$  in eq. 
\eqref{CS_2ptF2} around $g_*=0$, and assuming $g(\mu)\neq 0$ for some $\mu\neq 0$. For $\beta_0\neq 0$ eq. \eqref{beta00} yields: 
\bea
\f{ \beta(g(z)) }{g(z) } &=& -\beta_0g^2(z)\bigg(1+\f{\beta_1}{\beta_0}g^2(z)+\cdots\bigg)
\eea
It follows the IR ($\beta_0<0$) or UV ($\beta_0>0$) asymptotics of the 2-point correlator by means of eq. \eqref{alfa}: 
\bea
&&\braket{ F^2(z) F^2(0)} \sim \f{\mathcal{G}^{( F^2)}_2(0)}{z^{8}}g^4(z)\beta_0^2
\left( \f{g(\mu) } { \beta(g(\mu))} \right)^2\nn\\
&&\hspace{1.5truecm}\sim\f{\mathcal{G}^{( F^2)}_2(0)}{z^{8}}
\left(\f{1}{-2\beta_0\log(|z|\Lrgi)}
\left(
1+\frac{\beta_1}{2\beta_0^2}\frac{\log(-\beta_0\log({|z|\Lrgi } ))}{\log({|z|\Lrgi })} 
\right)
\right)^2\nn\\
&&\hspace{1.8truecm}\left ( \f{\beta_0g(\mu) } { \beta(g(\mu))} \right)^2
\eea
which agrees with eq. \eqref{CSS0} for $O=F^2$, $D=4$, $\gamma_0^{(F^2)}=2\beta_0$ and:
\bea
\f{g(\mu) } { \beta(g(\mu)) }&=& \f{1}{-\beta_0g^2(\mu)}\bigg(1-\f{\beta_1}{\beta_0}g^2(\mu)+\cdots\bigg)
\eea
that implies:
\bea
\left(\f{\beta_0g(\mu) } { \beta(g(\mu)) }\right)^2&=& \f{1}{g^4(\mu)}\bigg(1-2\f{\beta_1}{\beta_0}g^2(\mu)+\cdots\bigg)\nn\\
&=& \f{1}{g^4(\mu)}Z^{(F^2)'2}(g(\mu))
\eea
with $Z^{(F^2)'}(g(\mu))$ in eq. \eqref{ZOF} for $O=F^2$ and $\gamma_1^{(F^2)}=4\beta_1$.

By means of the perturbative expansion of $g(z)$ in terms of $g(\mu)$ in eq. \eqref{g2loop},
we obtain:
\bea
Z^{(F^2)}(g(z), g(\mu)) &=&\left(\f{ \beta(g(z)) }{g(z) }\right)\left(\f{g(\mu) } { \beta(g(\mu)) }\right)=\f{g^2(z)}{g^2(\mu)}
\bigg(1+\f{\beta_1}{\beta_0}(g^2(z)-g^2(\mu))+\cdots\bigg)\nn\\
&&\hspace{-1.5truecm}=1+g^2(\mu)2\beta_0\log|z\mu|+g^4(\mu)(4\beta_1\log|z\mu|+4\beta_0^2\log^2|z\mu|)+\cdots
\eea
which agrees with eq. \eqref{zeta00_small} for $O=F^2$, $\gamma_0^{(F^2)}=2\beta_0$ and $\gamma_1^{(F^2)}=4\beta_1$.
Hence, $Z^{(F^2)}=1$ if $g(\mu)=0$ for any $\mu$. Correspondingly, the correlator is conformal and free:
\be\label{freeFF}
\braket{ F^2(z) F^2(0)} = \f{\mathcal{G}^{( F^2)}_2(0)}{z^{8}}
\ee
that, of course, agrees with eq. \eqref{CSS0_ex} for $O=F^2$. 

For $\beta_0=0$ eq. \eqref{beta00} yields: 
\bea
\f{ \beta(g(z)) }{g(z) } &=& -\beta_1g^4(z)\bigg(1+\f{\beta_2}{\beta_1}g^2(z)+\cdots\bigg)
\eea
and by means of eq. \eqref{alfa2_IRAF} the asymptotics of the 2-point correlator now reads:
\bea
\braket{ F^2(z) F^2(0)} &\sim &\f{\mathcal{G}^{( F^2)}_2(0)}{z^{8}}g^8(z)\beta_1^2
\left( \f{g(\mu) } { \beta(g(\mu))} \right)^2\nn\\
&\sim&\f{\mathcal{G}^{( F^2)}_2(0)}{z^{8}}
\left(\f{1}{-4\beta_1\log(|z|\Lrgi)}\right)^2
\left ( \f{\beta_1g(\mu) } { \beta(g(\mu))} \right)^2
\eea
which agrees with eq. \eqref{CSS0_IRAF} for $O=F^2$, with $\gamma_0^{(F^2)}=2\beta_0=0$, $\gamma_1^{(F^2)}=4\beta_1$ and:
\bea
\left(\f{\beta_1g(\mu) } { \beta(g(\mu)) }\right)^2&=& \f{1}{g^8(\mu)}\bigg(1-2\f{\beta_2}{\beta_1}g^2(\mu)+\cdots\bigg)\nn\\
&=& \f{1}{g^8(\mu)}Z^{(F^2)'2}(g(\mu))
\eea
 with $Z^{(F^2)'}(g(\mu))$ in eq. \eqref{CSS0_IRAF} for $O=F^2$.
By means of the perturbative expansion of $g(z)$ in terms of $g(\mu)$ in eq. \eqref{gIRAFsmall},
we obtain:
\bea
Z^{(F^2)}(g(z), g(\mu)) &=&\left(\f{ \beta(g(z)) }{g(z) }\right)\left(\f{g(\mu) } { \beta(g(\mu)) }\right)=\f{g^4(z)}{g^4(\mu)}
\bigg(1+\f{\beta_2}{\beta_1}(g^2(z)-g^2(\mu))+\cdots\bigg)\nn\\
&=&1+g^4(\mu)4\beta_1\log|z\mu|+g^6(\mu)6\beta_2\log|z\mu|+\cdots
\eea
which agrees with eq. \eqref{zeta00_IRAF_small} for $O=F^2$, $\gamma_0^{(F^2)}=0$, since $\beta_0=0$ and $\gamma_1^{(F^2)}=4\beta_1$.
Hence, $Z^{(F^2)}= 1$ if $g(\mu)=0$ for any $\mu$. Correspondingly, the correlator is the free one in eq. \eqref{freeFF}.

\subsubsection{$g_*>0$ and $F^2$ irrelevant in the IR/UV } 
  
The asymptotics of $\braket{F^2(z)F^2(0)}$ for $g(\mu)\neq g_*$ for some $\mu\neq 0$
is now obtained 
by expanding $\f{\beta(g(z))}{g(z)}$ around $g_*>0$. 
Then, eq. \eqref{beta_0} yields: 
\bea\label{ratiobg_IR}
\f{ \beta(g(z)) }{g(z) } &=& 
\f{\beta_0^*(g(z)-g_*) +\beta_1^*(g(z)-g_*)^2 +\cdots }
{g_*+(g(z)-g_*)}\nonumber\\
&=&-\f{\beta_0^*}{g_*}(g_*-g(z))\left ( 
1-
\left (\f{\beta_1^*}{\beta_0^*}-\f{1}{g_*}  \right) (g_*-g(z)) +\cdots  \right )
\eea
By means of eq. \eqref{int_rgi_uni} the asymptotics of the correlator in eq. \eqref{CS_2ptF2}
now reads:
\bea\label{2pt_trace_IR}
\braket{ F^2(z) F^2(0)} &\sim &\f{\mathcal{G}^{( F^2)}_2(g_*)}{z^{8}}(g_*-g(z))^2\f{\beta_0^{*2}}{g_*^2}\left(\f{g(\mu) }{ \beta(g(\mu)) }\right)^2\nn\\
&\sim& \f{\mathcal{G}^{( F^2)}_2(g_*)}{z^{8}}
(|z|\Lrgi)^{-2\beta_0^*} 
\left(\f{\beta_0^*g(\mu) }{ g_*\beta(g(\mu)) }\right)^2
\eea
which agrees with eq. \eqref{CSS0_IR} for $O=F^2$, since by eq. \eqref{bg0} $\gamma_{F^2}(g_*)=\gamma_0^{(F^2)*}=\beta_0^*$ and:
\bea
\left(\f{\beta_0^*g(\mu) }{ g_*\beta(g(\mu)) }\right)^2
&=&\bigg( \f{1}{g_*-g(\mu)}\bigg)^2\bigg(1+2\left (\f{\beta_1^*}{\beta_0^*}-\f{1}{g_*}  \right)
(g_*-g(\mu))+\cdots\bigg)\nn\\
&=& 
\bigg( \f{1}{g_*-g(\mu)}\bigg)^2
Z^{(F^2)'2}(g_*-g(\mu))
\eea
with $Z^{(F^2)'}(g_*-g(\mu))$ in eq. \eqref{ZO_IR_mu} for $O=F^2$ and:
\bea\label{relOF} 
\gamma_0^{(F^2)*}&=&\gamma_{F^2}(g_*)=\beta_0^*\nn\\
 \gamma_1^{(F^2)*}&=&\gamma_{F^2}^\prime(g_*)=2\beta_1^*-\f{\beta_0^*}{g_*}
 \eea
By means of the perturbative expansion of $g_*-g(z)$ in terms of $g_*-g(\mu)$ in eq. \eqref{LL},
we obtain:
\bea\label{ZFirrD}
Z^{(F^2)}(g(z), g(\mu)) &=&\left(\f{ \beta(g(z)) }{g(z) }\right)\left(\f{g(\mu) } { \beta(g(\mu)) }\right)\nn\\
&&\hspace{-2.0truecm}=\left(\f{g_*-g(z)}{g_*-g(\mu)}\right)
\bigg(1+\left (\f{\beta_1^*}{\beta_0^*}-\f{1}{g_*}  \right) (g(z)-g(\mu))+\cdots\bigg)\nn\\
&&\hspace{-2.0truecm}=|z\mu|^{-\beta_0^*}
\bigg(1+(g_*-g(\mu))\f{\beta_1^*}{\beta_0^*}(1-|z\mu|^{-\beta_0^*})+\cdots\bigg)\nn\\
&&\hspace{-2.0truecm}~~~\bigg(1+(g_*-g(\mu))\left(\f{\beta_1^*}{\beta_0^*}-\f{1}{g_*}\right)(1-|z\mu|^{-\beta_0^*})+\cdots\bigg)\nn\\
&&\hspace{-2.0truecm}=|z\mu|^{-\beta_0^*}\bigg(1+(g_*-g(\mu))\left(\f{2\beta_1^*}{\beta_0^*}-\f{1}{g_*}\right)(1-|z\mu|^{-\beta_0^*})+\cdots\bigg)\nn\\
&&\hspace{-2.0truecm}=|z\mu|^{-\gamma_0^{(F^2)*}}\bigg(1+(g_*-g(\mu))\f{\gamma_1^{(F^2)*}}{\beta_0^*}
(1-|z\mu|^{-\gamma_0^{(F^2)*}})+\cdots\bigg)
\eea
where in the last equality we have employed eq. \eqref{relOF}.
Eq. \eqref{ZFirrD}
agrees with eq. \eqref{ZO_IR_small} for $O=F^2$ and shows that $Z^{(F^2)}=|z\mu|^{-\gamma_0^{(F^2)*}}$ if $g(\mu)=g_*$ for any $\mu$. Correspondingly, we obtain the exactly conformal correlator:
\be
\braket{ F^2(z) F^2(0)} = \f{\mathcal{G}^{( F^2)}_2(g_*)}{z^{8}}|z\mu|^{-2\gamma_0^{(F^2)*}}
\ee
in agreement with eq. \eqref{CSS0_IR_ex} for $O=F^2$.\par

\subsubsection{$g_*>0$ and $F^2$ marginal in the IR/UV}  

 For $\beta_0^*= 0$ and $\beta_1^*\neq 0$ 
 the expansion in eq. \eqref{ratiobg_IR} 
 becomes: 
   \bea
   \label{ratiobg_IR3_2}
\f{ \beta(g(z)) }{g(z) } &=& 
\f{\beta_1^*(g(z)-g_*)^2 +\beta_2^*(g(z)-g_*)^3 +\beta_3^*(g(z)-g_*)^4\cdots }
{g_*+(g(z)-g_*)}\nonumber\\
&=&\f{\beta_1^*}{g_*}(g_*-g(z))^2\left ( 
1-
\left (\f{\beta_2^*}{\beta_1^*}-\f{1}{g_*}  \right) (g_*-g(z)) +\cdots  \right )
\eea
 Hence, the corresponding asymptotics of the correlator in eq. \eqref{CS_2ptF2} is:
\bea\label{2pt_trace_IR3_2}
\braket{ F^2(z) F^2(0)} &\sim &\f{\mathcal{G}^{( F^2)}_2(g_*)}{z^{8}}
(g_*-g(z))^4
\f{\beta_1^{*2}}{g_*^2}\left(\f{g(\mu) }{ \beta(g(\mu)) }\right)^2\nn\\
&\sim &\f{\mathcal{G}^{( F^2)}_2(g_*)}{z^{8}}
\left(\f{1}{-\beta_1^*\log(|z|\Lrgi)}\right)^4
\left(\f{\beta_1^*g(\mu) }{ g_*\beta(g(\mu)) }\right)^2
\eea
where in the second equality we have employed eq. \eqref{D17_as}.
Eq. \eqref{2pt_trace_IR3_2} reproduces the asymptotics in eq. 
 \eqref{D23} for $O=F^2$, since for $\gamma_0^{(F^2)*}=\beta_0^*=0$ eq. \eqref{relOF} implies $\gamma_1^{(F^2)*}=2\beta_1^*$ and:
 \bea
\left(\f{\beta_1^*g(\mu) }{ g_*\beta(g(\mu)) }\right)^2
&=&
\bigg( \f{1}{g_*-g(\mu)}\bigg)^2\bigg(1+2\left (\f{\beta_2^*}{\beta_1^*}-\f{1}{g_*}  \right)
(g_*-g(\mu))+\cdots\bigg)\nn\\
&=& 
\bigg( \f{1}{g_*-g(\mu)}\bigg)^2
Z^{(F^2)'2}(g_*-g(\mu))
\eea
with $Z^{(F^2)'}(g_*-g(\mu))$ in eq. \eqref{D23} for $O=F^2$. \par
By means of the perturbative expansion of $g_*-g(z)$ in terms of $g_*-g(\mu)$ in eq. \eqref{D13_rgi_small}, we obtain:
\bea\label{ZFirrDmar}
Z^{(F^2)}(g(z), g(\mu)) &=&\left(\f{ \beta(g(z)) }{g(z) }\right)\left(\f{g(\mu) } { \beta(g(\mu)) }\right)
\nn\\
&=& 
\left(\f{g_*-g(z)}{g_*-g(\mu)}\right)^2
\bigg(1+\left (\f{\beta_2^*}{\beta_1^*}-\f{1}{g_*}  \right) (g(z)-g(\mu))+\cdots\bigg)\nn\\
&=&1+(g_*-g(\mu))2\beta_1^*\log|z\mu|+(g_*-g(\mu))^2\nn\\
&&
\bigg(-\beta_1^*\left(\f{3\beta_2^*}{\beta_1^*}-\f{1}{g_*}\right)\log|z\mu| +3\beta_1^{*2}\log^2|z\mu|
\bigg)+\cdots
\eea
It agrees with eq. \eqref{Zmarg_small} for $O=F^2$ by means of eq. \eqref{relOF} with $\gamma_0^{(F^2)*} =0$.
It follows that if $g(\mu)=g_*$ for any $\mu$ then $Z^{(F^2)}=1$ in eq. \eqref{ZFirrDmar}, which yields the free correlator:
\be
\braket{ F^2(z) F^2(0)} = \f{\mathcal{G}^{( F^2)}_2(g_*)}{z^{8}}
\ee
in agreement with eq. \eqref{D23_ex} for $O=F^2$ and $F^2$ marginal.\par
 
 For $\beta_0^*= 0$, the LET implies $\beta_1^*=0$, and 
 the expansion in eq. \eqref{ratiobg_IR3_2} 
 becomes: 
   \bea
   \label{ratiobg_IR3}
\f{ \beta(g(z)) }{g(z) } &=& 
\f{\beta_2^*(g(z)-g_*)^3 +\beta_3^*(g(z)-g_*)^4\cdots }
{g_*+(g(z)-g_*)}\nonumber\\
&=&\f{\beta_2^*}{g_*}(g(z)-g_*)^3\left ( 
1+
\left (\f{\beta_3^*}{\beta_2^*}-\f{1}{g_*}  \right) (g(z)-g_*) +\cdots  \right )
\eea
Hence, the corresponding asymptotics of the correlator in eq. \eqref{CS_2ptF2} is:
\bea\label{2pt_trace_IR3}
\braket{ F^2(z) F^2(0)} &\sim &\f{\mathcal{G}^{( F^2)}_2(g_*)}{z^{8}}
(g_*-g(z))^6\f{\beta_2^{*2}}{g_*^2}\left(\f{g(\mu) }{ \beta(g(\mu)) }\right)^2\nn\\
&\sim& \f{\mathcal{G}^{( F^2)}_2(g_*)}{z^{8}}\bigg(\f{1}{2\beta_2^*\log(|z|\Lrgi)} \bigg)^3
\left(\f{\beta_2^*g(\mu) }{ g_*\beta(g(\mu)) }\right)^2
\eea
where in the second equality we have employed eq. \eqref{D24_as}.
Eq. \eqref{2pt_trace_IR3} agrees with the asymptotics in eq. 
 \eqref{OOnew} for $O=F^2$ by employing eq. \eqref{relOF} with
 $\gamma_0^{(F^2)*}=\gamma_1^{(F^2)*}=0$ and
  $\f{\gamma_2^{(F^2)*}}{\beta_2^*}=3\f{\gamma_{F^2}^{\prime\prime}(g_*)}{\beta^{\prime\prime\prime}(g_*)}=3$, given that $\gamma_{F^2}^{\prime\prime}(g_*)= \beta^{\prime\prime\prime}(g_*)$ for $\beta(g_*)=\beta^{\prime}(g_*)=\beta^{\prime\prime}(g_*)=0$ and:  
   \bea
\left(\f{\beta_2^*g(\mu) }{ g_*\beta(g(\mu)) }\right)^2
&=&
\bigg( \f{1}{g_*-g(\mu)}\bigg)^6\bigg(1+2\left (\f{\beta_3^*}{\beta_2^*}-\f{1}{g_*}  \right)
(g_*-g(\mu))+\cdots\bigg)\nn\\
&=& 
\bigg( \f{1}{g_*-g(\mu)}\bigg)^6
Z^{(F^2)'2}(g_*-g(\mu))
\eea
with $Z^{(F^2)'}(g_*-g(\mu))$ in eq. \eqref{OOnew} for $O=F^2$.\par
By means of the perturbative expansion of $g_*-g(z)$ in terms of $g_*-g(\mu)$ in eq. \eqref{D16_rgi_small}, we obtain:
\bea\label{ZFirrDmar2}
Z^{(F^2)}(g(z), g(\mu)) &=&\left(\f{ \beta(g(z)) }{g(z) }\right)\left(\f{g(\mu) } { \beta(g(\mu)) }\right)
\nn\\
&=& 
\left(\f{g_*-g(z)}{g_*-g(\mu)}\right)^3
\bigg(1+\left (\f{\beta_3^*}{\beta_2^*}-\f{1}{g_*}  \right) (g(z)-g(\mu))+\cdots\bigg)\nn\\
&=&1-(g_*-g(\mu))^23\beta_2^*\log|z\mu|+(g_*-g(\mu))^3\nn\\
&&
\bigg(\beta_2^*\left(\f{4\beta_3^*}{\beta_2^*}-\f{1}{g_*}\right)\log|z\mu|\bigg)
+\cdots
\eea
It agrees with eq. \eqref{Zmarg_small2} for $O=F^2$ by means of eq. \eqref{relOF} with $\gamma_0^{(F^2)*} =\gamma_1^{(F^2)*}=0$.
It follows that, if $g(\mu)=g_*$ for any $\mu$, $Z^{(F^2)}=1$ in eq. \eqref{ZFirrDmar2}, which yields the free correlator:
\be
\braket{ F^2(z) F^2(0)} = \f{\mathcal{G}^{( F^2)}_2(g_*)}{z^{8}}
\ee
in agreement with eq. \eqref{OOnew_ex} for $O=F^2$ and $F^2$ marginal.
Generalizing the latter analysis to the case in which $\beta_1^*=\cdots=\beta_{n-1}^*=0$ for any $n\geq 2$,  the expansion in eq. \eqref{ratiobg_IR3_2} 
becomes: 
   \bea
   \label{ratiobg_IR3_n}
\f{ \beta(g(z)) }{g(z) } &=& 
\f{\beta_n^*(g(z)-g_*)^{n+1} +\beta_{n+1}^*(g(z)-g_*)^{n+2}\cdots }
{g_*+(g(z)-g_*)}\nonumber\\
&=&\f{\beta_n^*}{g_*}(g(z)-g_*)^{n+1}\left ( 
1+
\left (\f{\beta_{n+1}^*}{\beta_n^*}-\f{1}{g_*}  \right) (g(z)-g_*) +\cdots  \right )
\eea
Hence, the corresponding asymptotics of the correlator in eq. \eqref{CS_2ptF2} is:
\bea\label{2pt_trace_IR3_n}
\braket{ F^2(z) F^2(0)} &\sim &\f{\mathcal{G}^{( F^2)}_2(g_*)}{z^{8}}
(g(z)-g_*)^{2(n+1)}\f{\beta_n^{*2}}{g_*^2}\left(\f{g(\mu) }{ \beta(g(\mu)) }\right)^2\nn\\
&\sim& \f{\mathcal{G}^{( F^2)}_2(g_*)}{z^{8}}
\bigg(\f{1}{n\beta_n^*\log(|z|\Lrgi)} \bigg)^{\f{2(n+1)}{n}}
\left(\f{\beta_n^*g(\mu) }{ g_*\beta(g(\mu)) }\right)^2
\eea
where in the second equality we have employed eq. \eqref{D24_as_n}.
Eq. \eqref{2pt_trace_IR3_n} agrees with the asymptotics in eq. 
 \eqref{OOnew_n} for $O=F^2$ with 
 $\alpha_2=(n+1)\beta_n^*$
 and:
 \bea\label{gen_n}
 &&\gamma_0^{(F^2)*}=\gamma_1^{(F^2)*}=\cdots =\gamma_{n-1}^{(F^2)*}=0\nn\\
&& \gamma_n^{(F^2)*}=(n+1)\beta_n^*\nn\\
 && \gamma_{n+1}^{(F^2)*}=(n+2)\beta_{n+1}^*-\f{\beta_n^*}{g_*}\nn\\
 && \gamma_{n+2}^{(F^2)*}=(n+3)\beta_{n+2}^*-\f{\beta_{n+1}^*}{g_*}+\f{\beta_n^*}{g_*^2}\nn\\
 &&\cdots
 \eea
 obtained by employing eq. \eqref{gF22} and the expansion in eq. \eqref{IRexp_beta_zero_n} for $\beta(g)$ with:
 \be
 \gamma_{F^2}(g)=\gamma_n^{(F^2)*}(g-g_*)^n+\gamma_{n+1}^{(F^2)*}(g-g_*)^{n+1}+\gamma_{n+2}^{(F^2)*}(g-g_*)^{n+2}+\cdots
 \ee
 Moreover:  
   \bea
\left(\f{\beta_n^*g(\mu) }{ g_*\beta(g(\mu)) }\right)^2
&=&
\bigg( \f{1}{g_*-g(\mu)}\bigg)^{2(n+1)}\bigg(1+2\left (\f{\beta_{n+1}^*}{\beta_n^*}-\f{1}{g_*}  \right)
(g_*-g(\mu))+\cdots\bigg)\nn\\
&=& 
\bigg( \f{1}{g_*-g(\mu)}\bigg)^{2(n+1)}
Z^{(F^2)'2}(g_*-g(\mu))
\eea
with $Z^{(F^2)'}(g_*-g(\mu))$ in eq. \eqref{OOnew_n} for $O=F^2$.\par
By means of the perturbative expansion of $g_*-g(z)$ in terms of $g_*-g(\mu)$ in eq. \eqref{D16_rgi_small_n}, we obtain:
\bea\label{ZFirrDmar2_n}
&&Z^{(F^2)}(g(z), g(\mu)) =\left(\f{ \beta(g(z)) }{g(z) }\right)\left(\f{g(\mu) } { \beta(g(\mu)) }\right)
\nn\\
&&= 
\left(\f{g_*-g(z)}{g_*-g(\mu)}\right)^{n+1}
\bigg(1+\left (\f{\beta_{n+1}^*}{\beta_n^*}-\f{1}{g_*}  \right) (g(z)-g(\mu))+\cdots\bigg)\nn\\
&&=1-(g(\mu)-g_*)^n (n+1)\beta_n^*\log|z\mu|-(g(\mu)-g_*)^{n+1} 
\bigg( (n+2)\beta_{n+1}^*-\f{\beta_n^*}{g_*}\bigg) \nn\\
&&~~~~\log|z\mu|
+\cdots\nn\\
&&=1-(g(\mu)-g_*)^n \gamma_n^{(F^2)*}
\log|z\mu|-(g(\mu)-g_*)^{n+1}\gamma_{n+1}^{(F^2)*}
\log|z\mu|
+\cdots
\eea
where in the last equality we employed eq. \eqref{gen_n}.
It follows that, if $g(\mu)=g_*$ for any $\mu$, $Z^{(F^2)}=1$ in eq. \eqref{ZFirrDmar2_n}, which yields the free correlator:
\be
\braket{ F^2(z) F^2(0)} = \f{\mathcal{G}^{( F^2)}_2(g_*)}{z^{8}}
\ee
in agreement with eq. \eqref{OOnew_ex_n} for $O=F^2$ and $F^2$ marginal.

\section{Callan-Symanzik equation in $\td=4-2\eps$ dimensions} \label{CS1}
The CS equation in $\td=4-2\eps$ dimensions reads:
\be
\label{CS2_eps}
\left(\mu\f{\partial}{\partial \mu} + \beta(g,\eps)\f{\partial}{\partial g} + 2\gamma_{O}(g)\right) G^{(2)}(z, \mu, g(\mu)) = 0
\ee
where:
\be
\beta(g,\eps)=\f{dg}{d\log\mu}=-\eps g+\beta(g)
\ee
The solution of eq. \eqref{CS2_eps}, analogously to eq. \eqref{pert2_eps0} in $d=4$ dimensions, reads:
\bea\label{pert2_eps}
 G^{(2)}(z, \mu, g(\mu)) &=& \frac{1}{z^{2\tilde D}}\,\bar G^{(2)}(z\mu, g(\mu))\nn\\
 &=&\frac{1}{z^{2\tilde D}}\, \mathcal{G}_{2}^{(O)}(\tg(z))\, Z^{(O)2}(\tg(z), g(\mu))
\eea
where $\tilde D$ is the canonical dimension of $O$ in $\td$ dimensions, the renormalized multiplicative factor $Z^{(O)}(\tg(z), g(\mu))$ is:
\begin{equation} \label{def_eps}
Z^{(O)}(\tg(z), g(\mu)) = \exp \int_{g(\mu)} ^{\tg(z)}      \frac{ \gamma_O (g) } {\beta(g,\eps)} dg 
\end{equation}
that solves:
\be
\gamma_O (g(\mu))=-\f{d \log Z^{(O)}}{d \log\mu}
\ee
and
$\mathcal{G}_2^{(O)}$ is a function of the RGI coupling $\tg(z)\equiv \tg(z\mu, g(\mu))$ in $\td$ dimensions, which solves:
 \be\label{gz4_eps}
-\f{d \tg(z)}{d \log |z|}=\beta(\tg(z),\eps)
\ee
with the initial condition $\tg(1,g(\mu))=g(\mu)$.
Besides, the RG invariance of $\tg(z)$ implies:
\bea\label{betar_eps1}
0&=&\f{d\tg(z)}{d\log\mu}\nn\\
&=&\f{\partial \tg(z)}{\partial \log\mu}+\f{\partial \tg(z)}{\partial g(\mu)}\f{dg(\mu)}{d\log\mu}\nn\\
&=&\f{d \tg(z)}{d \log|z|}+\f{\partial \tg(z)}{\partial g(\mu)}\f{dg(\mu)}{d\log\mu}\nn\\
&=&-\beta(\tg(z),\eps)+\f{\partial \tg(z)}{\partial g(\mu)}\beta(g(\mu),\eps)
\eea
where we have employed eq. \eqref{gz4_eps} and the fact that $\tg(z)$ depends on $z$ only via the product $z\mu$. It follows:
\bea\label{betar_eps}
\f{\partial \tg(z)}{\partial g(\mu)} &=&
\f{\beta(\tg(z),\eps)}{\beta(g(\mu),\eps)}
\eea

\subsection{Asymptotics of the running coupling}
The running coupling satisfies: 
\bea\label{int_eps}
&&\int_{g(\mu)}^{\tg(z)}\f{dg}
{\beta(g,\eps)} = -\int_{\mu^{-1}}^{|z|}d\log |z|
\eea 
 If $\beta(g)$ has a zero at some $g_*>0$ in $d=4$, then $\beta(g,\eps)$ has a (physical or unphysical) zero at some $\tg_*=g_*+\delta g_*(\eps)$ in $\td=4-2\eps$, which is therefore solution of:
\bea\label{b_eps}
0&=&\beta(\tg_*,\eps)\nn\\
&=&-\eps\tg_*+\beta(\tg_*)\nn\\
&=&-\eps\tg_*+\beta_0^*(\tg_*-g_*)+\beta_1^*(\tg_*-g_*)^2+\cdots
\eea
where in the last line 
we have employed the expansion of $\beta(g)$ around $g_*$:
\be\label{exxx}
\beta(g)=\beta_0^*(g-g_*)+\beta_1^*(g-g_*)^2+\cdots
\ee
evaluated at $g=\tg_*$. Since we are interested in the leading-order $\eps$ dependence of $\tg(z)-g(\mu)$ for $g(\mu)=g_*$, it is sufficient to include in eq. \eqref{int_eps} only the first nonzero universal contribution in eq. \eqref{exxx}. Therefore:
\bea\label{ntrunc}
&&\int_{g(\mu)}^{\tg(z)}\f{dg}
{-\eps g + \beta_n^*(g-g_*)^{n+1}} = -\log |z\mu|
\eea
where $\beta_n^*$ is the first nonzero coefficient in eq. \eqref{exxx}.

\subsubsection{$g_*>0$ and $\gamma_{F^2}\neq 0$}

For $\beta_0^*=\gamma_{F^2}\neq 0$ the operator $F^2$ is relevant or irrelevant at $g_*$ in $d=4$ dimensions. Eq. \eqref{ntrunc} reads:
\bea\label{ntrunc_0}
&&\int_{g(\mu)}^{\tg(z)}\f{dg}
{-\eps g + \beta_0^*(g-g_*)} = -\log |z\mu|
\eea
and  for $x=g-g_*$:
\bea\label{ntrunc_00}
&&\int_{g(\mu)-g_*}^{\tg(z)-g_*}\f{dx}
{ (\beta_0^*-\eps)\big(x-\f{\eps g_*}{\beta_0^*-\eps}\big)} = -\log |z\mu|
\eea
where:
\be\label{S1}
\tg_*-g_*=\f{\eps g_*}{\beta_0^*-\eps}=\f{\eps g_*}{\beta_0^*}+O(\eps^2)
\ee
is the root of the polynomial in the denominator. Hence, we obtain:
\be
\f{1}{\beta_0^*-\eps}\log\f{\big|\tg(z)-g_*-\f{\eps g_*}{\beta_0^*-\eps}\big|}{\big|g(\mu)-g_*-\f{\eps g_*}{\beta_0^*-\eps}\big|}=-\log|z\mu|
\ee
and for $g(\mu)=g_*$:
\be\label{SOL}
\bigg|\tg(z)-g_*-\f{\eps g_*}{\beta_0^*-\eps}\bigg|=\f{\eps g_*}{|\beta_0^*-\eps|}|z\mu|^{-(\beta_0^*-\eps)}
\ee
Employing eq. \eqref{S1} we rewrite the lhs of eq. \eqref{SOL} to obtain:
\be\label{SOL2}
\big|\tg(z)-\tg_*\big|=\f{\eps g_*}{|\beta_0^*-\eps|}|z\mu|^{-(\beta_0^*-\eps)}
\ee
Moreover, by eq. \eqref{S1} for small $\eps$:
\bea
&&\tg_*>g_*~~\textrm{for}~~\beta_0^*>0\nn\\
&&\tg_*<g_*~~\textrm{for}~~\beta_0^*<0
\eea
Besides, $\beta(g_*)=0$ implies $\beta(g_*,\eps)<0$, so that, if $\tg_*> g_*$, $\tg_*$ is an IR zero of $\beta(g,\eps)$ and, if $\tg_*< g_*$, $\tg_*$ is an UV zero of $\beta(g,\eps)$. Since we are interested in the UV-IR flow around $g_*$, we should select the solutions of eq. \eqref{SOL2} with $\tg(z)$ in the interval between $g_*$ and $\tg_*$, i.e.:
\bea\label{COND}
&&g_*<\tg(z)<\tg_*~~\textrm{for}~~\beta_0^*>0\nn\\
&&\tg_*<\tg(z)<g_*~~\textrm{for}~~\beta_0^*<0
\eea
Then, eq. \eqref{SOL2} for both signs of $\beta_0^*$ yields:
\be\label{notm_eps_ex}
\tg(z)-g(\mu)\big|_{g(\mu)=g_*}= \tg(z)-g_*=\f{\eps g_*}{\beta_0^*-\eps}\big(1-|z\mu|^{-(\beta_0^*-\eps)}\big)
\ee
whose expansion in $\eps$ reads:
 \bea\label{notm_eps}
 &&\tg(z)-g(\mu)\big|_{g(\mu)=g_*}= \tg(z)-g_*\nn\\
 &&=\f{\eps g_*}{\beta_0^*}
 \bigg(1-|z\mu|^{-\beta_0^*}-\f{\eps}{\beta_0^*}|z\mu|^{-\beta_0^*}(1+\beta_0^*\log|z\mu|)
+\cdots
 \bigg)
 \eea
 where the dots stand for $O(\eps)$ contributions depending on $\beta_i^*$ with $i\geq 1$.

 \subsubsection{$g_*>0$ and $\gamma_{F^2}= 0$ with $\beta_1^*\neq 0$}
\label{app:b1pos}

For $\beta_0^*=\gamma_{F^2}= 0$ the operator $F^2$ is marginal at $g_*$ in $d=4$ dimensions. 
For $\beta_1^*\neq 0$, eq. \eqref{ntrunc} reads:
\bea\label{int_eps_m}
&&\int_{g(\mu)}^{\tg(z)}\f{dg}
{-\eps g+\beta_1^*(g-g_*)^2} = -\log |z\mu|
\eea 
and equivalently:
\bea\label{int_eps_m2}
&&\int_{g(\mu)-g_*}^{\tg(z)-g_*}\f{dx}
{x^2-\f{\eps}{\beta_1^*}x-  \f{\eps g_*}{\beta_1^*}} = -\beta_1^*\log |z\mu|
\eea 
with $x=g-g_*$.
The discriminant of the quadratic polynomial in the denominator: 
\bea
\Delta&=&\bigg(\f{\eps}{\beta_1^*}\bigg)^2+\f{4\eps g_*}{\beta_1^*}\nn\\
&=&\f{4\eps g_*}{\beta_1^*}\bigg(1+\f{\eps}{4g_*\beta_1^*} \bigg)
\eea
satisfies:
\bea
&&\Delta >0~ {\mbox {for}}~\beta_1^*>0\nn\\
&&\Delta <0~ {\mbox {for}} ~\beta_1^*<0
\eea
For example, for $\beta_1^*>0$ the quadratic polynomial has two real distinct roots:
\be\label{roots}
(\tg_*-g_*)_\pm=\f{\eps}{2\beta_1^*}\pm \sqrt{\f{\eps g_*}{\beta_1^* }\bigg(1+\f{\eps}{4g_*\beta_1^*} \bigg)}
\ee
By employing the decomposition:
\be
\f{1}{ax^2+bx+c} = \f{1}{a(x-p)(x-q)}=\f{A}{x-p}+\f{B}{x-q}
\ee
with roots $p,q$ and:
\be
A=\f{1}{a(p-q)}, ~~B=-A=\f{1}{a(q-p)}
\ee
we obtain:
\be\label{gen}
\int_{x_0}^{x_1} \f{dx}{ax^2+bx+c}=\f{1}{a(p-q)} \int_{x_0}^{x_1} dx\bigg( \f{1}{x-p}-\f{1}{x-q}\bigg)=\f{1}{a(p-q)}\log \f{|x-p|}{|x-q|}\bigg|_{x_0}^{x_1}
\ee
Hence, by employing eq. \eqref{roots} and applying eq. \eqref{gen}, eq. \eqref{int_eps_m2} yields:
\bea\label{int_eps_m2_ex}
\f{1}{2\sqrt{\f{\eps g_*}{\beta_1^*}\big(1+\f{\eps}{4g_*\beta_1^*} \big)
 }} \log\f{\big|x-\f{\eps}{2\beta_1^*}-\sqrt{\f{\eps g_*}{\beta_1^*}\big(1+\f{\eps}{4g_*\beta_1^*} \big)}\big|}
 {\big|x-\f{\eps}{2\beta_1^*}+\sqrt{\f{\eps g_*}{\beta_1^*}\big(1+\f{\eps}{4g_*\beta_1^*} \big)}\big|}
 \Bigg|_{g(\mu)-g_*}^{\tg(z)-g_*}
 = -\beta_1^*\log |z\mu|
\eea
or equivalently:
\bea\label{complete_m}
&& \f{\big|\tg(z)-g_*-\f{\eps}{2\beta_1^*}-\sqrt{\f{\eps g_*}{\beta_1^*}\big(1+\f{\eps}{4g_*\beta_1^*} \big)}\big|}
 {\big|\tg(z)-g_*-\f{\eps}{2\beta_1^*}+\sqrt{\f{\eps g_*}{\beta_1^*}\big(1+\f{\eps}{4g_*\beta_1^*} \big)}\big|}
 \f{\big|g(\mu)-g_*-\f{\eps}{2\beta_1^*}+\sqrt{\f{\eps g_*}{\beta_1^*}\big(1+\f{\eps}{4g_*\beta_1^*} \big)}\big|}
 {\big|g(\mu)-g_*-\f{\eps}{2\beta_1^*}-\sqrt{\f{\eps g_*}{\beta_1^*}\big(1+\f{\eps}{4g_*\beta_1^*} \big)}\big|}\nn\\
 &&=\exp{\bigg(-2\beta_1^*\sqrt{\f{\eps g_*}{\beta_1^*}\big(1+\f{\eps}{4g_*\beta_1^*} \big)}\log |z\mu|\bigg)}
\eea
We set $g(\mu)=g_*$, expand both sides of eq. \eqref{complete_m} in powers of $\eps$ and show that $\tg(z)-g_*$ is of order $\eps$, so that the terms $O(\sqrt{\eps})$ should be treated as leading order in the $\eps$ expansion of the lhs. For the latter we thus obtain:
\bea\label{eccoNLO}
&&\mbox{lhs}\eqref{complete_m}=
\f{\sqrt{\f{\eps g_*}{\beta_1^*} \big(1+\f{\eps}{4g_*\beta_1^*} \big) }+\f{\eps}{2\beta_1^*}+g_*-\tg(z)}
{\sqrt{\f{\eps g_*}{\beta_1^*}\big(1+\f{\eps}{4g_*\beta_1^*} \big)}-\f{\eps}{2\beta_1^*}+\tg(z)-g_*}\,\,
\f{\sqrt{\f{\eps g_*}{\beta_1^*}\big(1+\f{\eps}{4g_*\beta_1^*} \big)}-\f{\eps}{2\beta_1^*}}
{\sqrt{\f{\eps g_*}{\beta_1^*}\big(1+\f{\eps}{4g_*\beta_1^*} \big)}+\f{\eps}{2\beta_1^*}}\nn\\
&&=\f{\sqrt{\f{\eps g_*}{\beta_1^*}}\big(1+\f{\eps}{8g_*\beta_1^*}+\cdots \big)      
+\f{\eps}{2\beta_1^*}+g_*-\tg(z)}
{\sqrt{\f{\eps g_*}{\beta_1^*}}\big(1+\f{\eps}{8g_*\beta_1^*}+\cdots \big) 
-\f{\eps}{2\beta_1^*}+\tg(z)-g_*}\,\,
\f{\sqrt{\f{\eps g_*}{\beta_1^*}}\big(1+\f{\eps}{8g_*\beta_1^*}+\cdots \big) -\f{\eps}{2\beta_1^*}}
{\sqrt{\f{\eps g_*}{\beta_1^*}}\big(1+\f{\eps}{8g_*\beta_1^*}+\cdots \big) +\f{\eps}{2\beta_1^*}}\nn\\
&&=\bigg(1+\sqrt{\f{\eps}{g_*\beta_1^*}}+2(g_*-\tg(z))\sqrt{\f{\beta_1^*}{\eps g_*}}+\f{\eps}{2g_*\beta_1^*}+\f{2(g_*-\tg(z))}{g_*}\nn\\
&&~~~+2(g_*-\tg(z))^2\f{\beta_1^*}{\eps g_*}+\cdots\bigg)\bigg(1-\sqrt{\f{\eps}{g_*\beta_1^*}}+\f{\eps}{2g_*\beta_1^*}+\cdots\bigg)
\nn\\
&&=1+2(g_*-\tg(z))\sqrt{\f{\beta_1^*}{\eps g_*}}+2(g_*-\tg(z))^2\f{\beta_1^*}{\eps g_*}+
\cdots
\eea
where the dots  stand for $O(\eps^2)$ terms in the second equality and $O(\eps\sqrt{\eps})$ terms in the following equalities. 
Hence, eq. \eqref{complete_m} reads:
\bea
&&1+2(g_*-\tg(z))\sqrt{\f{\beta_1^*}{\eps g_*}}+2(g_*-\tg(z))^2\f{\beta_1^*}{\eps g_*}+
\cdots\nn\\
&&=
1-2\sqrt{{\eps g_*\beta_1^*}}\log|z\mu|+2\eps g_*\beta_1^*\log^2|z\mu|+\cdots
\eea
where the dots stand for $O(\eps\sqrt{\eps})$ terms. 
By solving iteratively for $\tg(z)-g_*$, it follows: 
\be\label{m_eps}
\tg(z)-g(\mu)|_{g(\mu)=g_*}=\tg(z)-g_*=\eps g_*\log|z\mu|+\cdots
\ee
where the dots stand for $O(\eps^2)$ terms that can also be generated by the higher-order coefficients $\beta_i^*$ with $i\geq 2$.

\subsection{Asymptotics of $Z^{(O)}(\tg(z), g(\mu))$}

$Z^{(O)}(\tg(z), g(\mu))$ in eq. \eqref{def_eps} may be computed asymptotically in a neighborhood of $g_*$. Specifically, we are interested in the leading-order corrections in $\eps$ to the $d=4$ solution. Hence, we may limit ourselves to:
\bea\label{ZO_IR_eps}
&&Z^{(O)}(\tg(z), g(\mu))\sim \exp \int_{g(\mu)}^{\tg(z)}
\f{\gamma_0^{(O)*}  }{-\eps g +\beta_n^*(g-g_*)^{n+1}   }\,dg
\eea
where in the expansion of the anomalous dimension $\gamma_O(g)$:
\be
\gamma_O(g)=\gamma_0^{(O)*}+\gamma_1^{(O)*}(g-g_*)+\gamma_2^{(O)*}(g-g_*)^2+\cdots
\ee
we have assumed $\gamma_0^{(O)*}\neq0$.

\subsubsection{$g_*>0$ and $\gamma_{F^2}\neq 0$}

For $\beta_0^*\neq 0$, we obtain:
\bea
Z^{(O)}(\tg(z), g(\mu))&\sim& \exp \int_{g(\mu)}^{\tg(z)}
\f{\gamma_0^{(O)*}  }{-\eps g +\beta_0^*(g-g_*)   }\,dg\nn\\
&=&\exp \int_{g(\mu)-g_*}^{\tg(z)-g_*}
\f{\gamma_0^{(O)*}  }{(\beta_0^*-\eps) \big(x-\f{\eps g_*}{\beta_0^*-\eps}\big)  }\,dx\nn\\
&=&\exp\bigg(\f{\gamma_0^{(O)*}}{\beta_0^*-\eps} \log \f{\big|\tg(z)-g_*-\f{\eps g_*}{\beta_0^*-\eps}\big|}{\big|g(\mu)-g_*-\f{\eps g_*}{\beta_0^*-\eps}\big|}\bigg)
\eea
that for $g(\mu)=g_*$, by employing eq. \eqref{S1}, yields:
\bea\label{ZO_IR_small_eps}
Z^{(O)}(\tg(z), g(\mu))\big|_{g(\mu)=g_*}&\sim& 
\exp\bigg(\f{\gamma_0^{(O)*}}{\beta_0^*-\eps} \log \f{\big|\tg(z)-\tg_*\big|\big|\beta_0^*-\eps\big|}{\eps g_*}\bigg)\nn\\
&=&
\bigg(\f{\big|\tg(z)-\tg_*\big|\big|\beta_0^*-\eps\big|}{\eps g_*}\bigg)^\f{\gamma_0^{(O)*}}{\beta_0^*-\eps}
\eea
We finally employ the conditions in eq. \eqref{COND} to obtain:
\bea\label{Z_notm_eps}
Z^{(O)}(\tg(z), g(\mu))\big|_{g(\mu)=g_*}&\sim& 
\bigg(\f{(\tg_*-\tg(z))(\beta_0^*-\eps)}{\eps g_*}\bigg)^\f{\gamma_0^{(O)*}}{\beta_0^*-\eps}\nn\\
&\sim&\bigg(1+\f{(g_*-\tg(z))(\beta_0^*-\eps)}{\eps g_*}\bigg)^\f{\gamma_0^{(O)*}}{\beta_0^*-\eps}\nn\\
&\sim&\bigg(1-\big(1-|z\mu|^{-(\beta_0^*-\eps)}\big)\bigg)^\f{\gamma_0^{(O)*}}{\beta_0^*-\eps}\nn\\
&\sim&|z\mu|^{-\gamma_0^{(O)*}}(1+O(\eps))
\eea
where in the last but one equality we have employed eq. \eqref{notm_eps_ex}.  
Eq. \eqref{Z_notm_eps} reproduces the result in eq. \eqref{ZO_IR_small} in $d=4$ dimensions for $g(\mu)=g_*$, so that
the perturbative solution $Z^{(O)}(\tg(z), g(\mu))$ in $\td=4-2\eps$ dimensions admits the expansion:
\bea
Z^{(O)}(\tg(z), g(\mu))\big|_{g(\mu)=g_*} &= & Z^{(O)}(g(z), g(\mu)) \big |_{g(\mu)=g_*} +O(\eps)
\eea
where $Z^{(O)}(g(z), g(\mu)) \big |_{g(\mu)=g_*} $ is given by eq. \eqref{ZO_IR_small} for $g(\mu)=g_*$ and the $O(\eps)$ contributions depend on $\gamma_i^{(O)*}$ and $\beta_i^*$ with $i\geq 1$.

\subsubsection{$g_*>0$ and $\gamma_{F^2}= 0$ with $\beta_1^*\neq 0$}

For $\beta_0^*= 0$ and $\beta_1^*>0$ (appendix \ref{app:b1pos}):
\bea
&&Z^{(O)}(\tg(z), g(\mu))\sim \exp \int_{g(\mu)}^{\tg(z)}
\f{\gamma_0^{(O)*}  }{-\eps g +\beta_1^*(g-g_*)^2   }\,dg
\nn\\
&&\sim \exp \f{\gamma_0^{(O)*}  }{\beta_1^*} \int_{g(\mu)-g_*}^{\tg(z)-g_*}\f{dx}{x^2-\f{\eps}{\beta_1^*}x-\f{\eps g_*}{\beta_1^*}}\nn\\
&&\sim\exp\bigg(\f{\gamma_0^{(O)*}}{2\beta_1^*
\sqrt{\f{\eps g_*}{\beta_1^* }\big(1+\f{\eps}{4g_*\beta_1^*} \big)}} 
\log 
\f{\big|x-\f{\eps}{2\beta_1^*}-\sqrt{\f{\eps g_*}{\beta_1^*}\big(1+\f{\eps}{4g_*\beta_1^*} \big)}\big|}
 {\big|x-\f{\eps}{2\beta_1^*}+\sqrt{\f{\eps g_*}{\beta_1^*}\big(1+\f{\eps}{4g_*\beta_1^*} \big)}\big|}
 \Bigg|_{g(\mu)-g_*}^{\tg(z)-g_*}
\bigg)
\eea
that, for $g(\mu)=g_*$ and $\tg(z)-g_*$ of $O(\eps)$ according to eq. \eqref{m_eps}, yields:
\bea\label{ZO_IR_small_eps_2}
&&Z^{(O)}(\tg(z), g(\mu))\big|_{g(\mu)=g_*}
 \sim 
\exp\Bigg(\f{\gamma_0^{(O)*}}{2\beta_1^*
\sqrt{\f{\eps g_*}{\beta_1^* }\big(1+\f{\eps}{4g_*\beta_1^*} \big)}} \nn\\
&&\hspace{1.5truecm}\log 
\bigg(
\f{ \sqrt{\f{\eps g_*}{\beta_1^*}\big(1+\f{\eps}{4g_*\beta_1^*} \big)}+ \f{\eps}{2\beta_1^*}+g_*-\tg(z)}
 {\sqrt{\f{\eps g_*}{\beta_1^*}\big(1+\f{\eps}{4g_*\beta_1^*} \big)}-\f{\eps}{2\beta_1^*}
 +\tg(z)-g_*}\,\,
 \f{\sqrt{\f{\eps g_*}{\beta_1^*}\big(1+\f{\eps}{4g_*\beta_1^*} \big)}-\f{\eps}{2\beta_1^*}}
 {\sqrt{\f{\eps g_*}{\beta_1^*}\big(1+\f{\eps}{4g_*\beta_1^*} \big)}+\f{\eps}{2\beta_1^*}}
\bigg)
 \Bigg)\nn\\ 
 &&\sim 
\exp\Bigg(\f{\gamma_0^{(O)*}}{2
\sqrt{{\eps g_*}{\beta_1^* }\big(1+\f{\eps}{4g_*\beta_1^*} \big)}} 
\log 
\bigg(
1+2(g_*-\tg(z))\sqrt{\f{\beta_1^*}{\eps g_*}}
+2(g_*-\tg(z))^2{\f{\beta_1^*}{\eps g_*}}
\nn\\
&&\hspace{1.5truecm}
+\cdots
\bigg)
 \Bigg)\nn\\
  &&\sim 
\exp\Bigg(\f{\gamma_0^{(O)*}}{2
\sqrt{{\eps g_*}{\beta_1^* }\big(1+\f{\eps}{4g_*\beta_1^*} \big)}} 
\log 
\bigg(
1-2\sqrt{\eps g_*\beta_1^*}\log|z\mu|
+2{\eps g_*\beta_1^*}\log^2|z\mu|
+\cdots
\bigg)
 \Bigg)\nn\\ 
  &&\sim 
\exp\Bigg(\f{\gamma_0^{(O)*}}{2
\sqrt{{\eps g_*}{\beta_1^* }
}}  \bigg(1-\f{\eps}{8g_*\beta_1^*}+O(\eps^2)\bigg)
\bigg(
-2\sqrt{\eps g_*\beta_1^*}\log|z\mu|
+2{\eps g_*\beta_1^*}\log^2|z\mu|
\nn\\
&&\hspace{1.5truecm}
-2{\eps g_*\beta_1^*}\log^2|z\mu|
+\cdots
\bigg)
 \Bigg)\nn\\
&&\sim 
\exp\Bigg(\f{\gamma_0^{(O)*}}{2\sqrt{{\eps g_*\beta_1^*}}}\bigg(1-\f{\eps}{8g_*\beta_1^*}+O(\eps^2)\bigg)
\bigg(-2\sqrt{\eps g_*\beta_1^*}\log|z\mu|+\cdots \bigg)
\Bigg)\nn\\
&&\sim 
\exp\Bigg(-\gamma_0^{(O)*} \log|z\mu|+O(\eps)
\Bigg)\nn\\
&&\sim |z\mu|^{-\gamma_0^{(O)*} } (1+O(\eps))
\eea
where the dots stand for $O(\eps\sqrt{\eps})$ terms and we have employed the expansion in eq. \eqref{eccoNLO} in the second equality and
eq. \eqref{m_eps} in the third equality.
Hence, as for the nonmarginal case, eq. \eqref{ZO_IR_small_eps_2} reproduces the result in eq. \eqref{Zmarg_small} in $d=4$ dimensions for $g(\mu)=g_*$, so that
the perturbative solution $Z^{(O)}(\tg(z), g(\mu))$ in $\td=4-2\eps$ dimensions admits the expansion:
\bea
Z^{(O)}(\tg(z), g(\mu))\big|_{g(\mu)=g_*} &= & Z^{(O)}(g(z), g(\mu)) \big |_{g(\mu)=g_*} +O(\eps)
\eea
where $Z^{(O)}(g(z), g(\mu)) \big |_{g(\mu)=g_*} $ is given by eq. \eqref{Zmarg_small} for $g(\mu)=g_*$, no $O(\sqrt{\eps})$ contributions are present and the $O(\eps)$ contributions depend on  $\gamma_i^{(O)*}$ with $i\geq 0$ and $\beta_j^*$ with $j\geq 1$. 
\section{LET in the $(u,v_{F^2})$ scheme}  \label{E}
We apply to the LET a version of the more general $(u,v_O)$-regularization scheme introduced in \cite{Skenderis} that preserves conformal invariance, but not necessarily gauge invariance. In the $(u,v_O)$ scheme the space-time dimension $d$ and the canonical dimension $\Delta_{O_0}$ of a generic operator $O$ respectively become:
\bea 
\label{eq:DRuv}
d\to \tilde{d}&=&d+2u\eps\nn\\
\Delta_{O_0}\to \tilde{\Delta}_{O_0}&=&\Delta_{O_0} +(u+v_O)\eps
\eea
with arbitrary real parameters $(u,v_O)$, where $v_O$ may be chosen independently for each operator $O$. 
Correspondingly, the CS equation for the $2$-point correlator is deformed as follows:
\be
\label{CS2uv}
\left(z \cdot \f{\partial}{\partial z} + \beta(g,\eps ,u,v_{F^2})  \f{\partial}{\partial g} + 2 \tilde{\Delta}_{O_0}+ 2\gamma_{O}(g)\right) G^{(2)}(z, \mu, g(\mu)) = 0
\ee
where $\beta(g,\eps ,u,v_{F^2})$ is defined in eq. \eqref{eq:betauv}. Thus we employ the $(u,v_{F^2})$ scheme that combines dimensional regularization in $\td=4+2u\eps$ dimensions with a deformation of the canonical dimension $\tD_{F^2_0}$ of the operator $F^2$. Dimensional regularization in $\tilde d= d+2u\epsilon$ dimensions corresponds for $F^2$ to $u=v_{F^2}$ and standard dimensional regularization to $u=v_{F^2}=-1$. \par
Moreover, dimensional regularization in $\tilde d= d+2u\epsilon$ dimensions for $O\neq F^2$ corresponds to a certain choice of $v_O(u)$ dictated by the content of the composite operator $O$ in terms of its elementary fields to order $g^0$ in perturbation theory. 

\subsection{LET for bare correlators in the $(u,v_{F^2})$ scheme} \label{E1}

We adapt the $(u,v_{F^2})$ scheme to the LET. Specifically, the shift $\Delta_{F^2_0}\to\tD_{F^2_0}= \Delta_{F^2_0}+(u+v_{F^2})\eps$ for the bare operator $F_0^2\equiv 2 \Tr  F^2_0$ implies that the canonically normalized YM action that is inserted in the rhs of the LET acquires a dimension proportional to $\eps$:
\bea
 [S]=(v_{F^2}-u)\eps
 \eea
 with:
 \bea
 S&=&\mu^{\td - \tD_{F^2_0}}  \f{1}{2} \int \Tr F_0^2\,d^\td x\nn\\
 &=&\mu^{(u-v_{F^2})\eps}  \f{1}{2}\int  \Tr F_0^2\,d^\td x
 \eea
The bare coupling has now dimension $[g_0]=-\tD_{F^2_0}/2+2= -(u+v_{F^2})\eps/2$ and thus is related to the dimensionless renormalized one by the generalization of eq. \eqref{eq:gg0}:
 \bea
g_0&=&Z_g(g,\eps,u,v_{F^2}) \mu^{-\f{(u+v_{F^2})}{2}\eps} g
\eea
By the substitution $\eps\to -\f{(u+v_{F^2})}{2}\eps$, the beta function in eq. \eqref{eq:beta} now reads:
\bea\label{eq:betauv}
\beta(g,\eps ,u,v_{F^2}) &=& \f{1}{2} (u+v_{F^2}) \eps g +\beta(g)
\eea
which recovers dimensional regularization in eq. \eqref{eq:beta} for $u=v_{F^2}=-1$.
As a consequence, the LET regularized in the $(u,v_{F^2})$ scheme reads for canonically normalized bare operators:
\bea
\label{eq:LETCuv}
&&\sum^{k=n}_{k=1} c_{O_k} \braket{O_1\cdots O_n}_0 +
\f{\partial}{\partial\log {g}_0} \braket{O_1\cdots O_n}_0 \bigg|_{(u,v_{F^2})}\nn\\
&&\hspace{0.5truecm}= \f{1}{2}\mu^{(u-v_{F^2})\eps}\int\, \braket{O_1\cdots O_n F^2(x)}_0 
- \braket{O_1\cdots O_n}_0\braket{F^2(x)}_0\, d^\td x \bigg|_{(u,v_{F^2})} 
\eea
where the residual dependence on the scale $\mu$ appears in the rhs for dimensional reasons. 

\subsection{LET for renormalized correlators in the $(u,v_{F^2})$ scheme} \label{E.2}

In the $(u,v_{F^2})$ scheme, as in dimensional regularization, it is convenient to rewrite eq. \eqref{eq:LETCuv} in terms of renormalized correlators. 
Firstly, we need to generalize eqs. \eqref{eq:gg0rel0}:
 \bea\label{eq:gg0rel_uv}
  \f{\partial\log g}{\partial\log g_0} &=& \left ( 1+ \f{\partial\log Z_g}{\partial\log g}\right )^{-1}\nn\\
  &=&\left ( 1+\f{2\beta(g)}{(u+v_{F^2})\eps g}\right )
  \eea
and \eqref{eq:Zrel0}:
 \bea\label{eq:Zrel_uv}
  \f{\partial\log Z_O^{-2}}{\partial\log g_0}&=&-2\f{\partial\log Z_O}{\partial\log g_0}\nn\\
  &=&-2\f{d\log Z_O}{d\log\mu}\f{d\log\mu}{d\log g}\f{\partial\log g}{\partial\log g_0}\nn\\
  &=&2\gamma_O(g)\left(\f{1}{g}\f{d g}{d\log\mu} \right)^{-1}\left ( 1+ \f{\partial\log Z_g}{\partial\log g}\right )^{-1}\nn\\
  &=&2\gamma_O(g)\left(\f{\beta(g,\eps ,u,v_{F^2})}{g} \right)^{-1}\left ( 1+ \f{2\beta(g)}{(u+v_{F^2})\eps g}\right )\nn\\
  &=&4\f{\gamma_O(g)}{(u+v_{F^2})\eps}\left ( 1+\f{2\beta(g)}{(u+v_{F^2})\eps g}\right )^{-1}\left ( 1+\f{2\beta(g)}{(u+v_{F^2})\eps g}\right )\nn\\
   &=&4\f{\gamma_O(g)}{(u+v_{F^2})\eps}
  \eea
where the $u,v_{F^2}$ dependence in the last line is inherited from the one in eq. \eqref{eq:betauv} and thus is related to the specific redefinition of the dimension of $F_0^2$ and $g_0$. 
Moreover, in the $(u,v_{F^2})$ scheme the renormalization factor $Z_{F^2}$ is related to $Z_g$ by:
 \bea\label{eq:ZFZg}
 Z_{F^2}^{-1}(g,\eps,u,v_{F^2})&=&\left(1+g\f{\partial}{\partial g}\log Z_g(g,\eps,u,v_{F^2})\right)^{-1}\nn\\
 &=&\left ( 1+\f{2\beta(g)}{(u+v_{F^2})\eps g}\right )
 \eea
As for dimensional regularization, we consider the simplest case where no operator mixing occurs under renormalization up to operators that vanish by the equations of motion and we focus on eq. \eqref{eq:LETCuv} with $n=2$ and 
$O_1=O_2= O$, with $O$ a primary conformal operator. Then, eq. \eqref{eq:LETCuv} reads for $z \neq 0$:
\bea
\label{eq:LETn2bare}
&&2c_O \braket{O(z)O(0)}_0 +
\f{\partial}{\partial\log {g}_0} \braket{O(z)O(0)}_0 \bigg|_{(u,v_{F^2})} \nn\\
&&\hspace{0.5truecm}= \f{1}{2}\mu^{(u-v_{F^2})\eps}\int \braket{O(z) O(0) F^2(x)}_0 
- \braket{O(z)O(0)}_0\braket{F^2(x)}_0
 d^\td x \bigg|_{(u,v_{F^2})}
\eea
We assume as well that $F_0^2$ is multiplicatively renormalizable up to operators that vanish by the equations of motion. \par
Hence, the bare 1-, 2- and 3-point correlators are related to the renormalized ones by a multiplicative renormalization up to contact terms. Therefore, up to contact terms, we obtain:
\bea
\label{eq:RvsB}
\braket{F^2(x)}_0&=&Z_{F^2}^{-1}(g,\eps ,u,v_{F^2})\braket{F^2(x)}\nn\\
\braket{O(z)O(0)}_0&=&Z_O^{-2}(g,\eps ,u,v_{F^2})\braket{O(z)O(0)}\nn\\
\braket{O(z)O(0)F^2(x)}_0&=&Z_O^{-2}(g,\eps ,u,v_{F^2})Z_{F^2}^{-1}(g,\eps ,u,v_{F^2})\braket{O(z)O(0)F^2(x)}
\eea
where the renormalization factors $Z_O$ and $Z_{F^2}$ are functions of the renormalized coupling $g$ and the parameters of the regularization, $u$ and $v_{F^2}$, through the beta function. \par
The lhs and rhs of the regularized LET in eq. \eqref{eq:LETn2bare} can now be rewritten in terms of the renormalized 1-, 2- and 3-point correlators by means of eq. \eqref{eq:RvsB}.\par
We rewrite the lhs of eq. \eqref{eq:LETn2bare} without contact terms, since we assume $z\neq 0$:
 \bea
 \label{eq:lhs_gen}
 \textrm{lhs}&=& 2c_O\, Z_O^{-2}\braket{O(z)O(0)} +
g_0\f{\partial}{\partial {g}_0} Z_O^{-2}\braket{O(z)O(0)}
+Z_O^{-2}g_0\f{\partial}{\partial {g}_0} \braket{O(z)O(0)}\bigg|_{(u,v_{F^2})}\nn\\
&=& Z_O^{-2}\braket{O(z)O(0)}\left( 2c_O +\f{\partial\log Z_O^{-2}}{\partial\log g_0}
+\f{\partial\log \braket{O(z)O(0)}}{\partial\log g_0}
\right) \bigg|_{(u,v_{F^2})}\nn\\
&=& Z_O^{-2}\braket{O(z)O(0)}\left( 2c_O + \f{4\gamma_O(g)}{(u+v_{F^2})\eps}+\left ( 1+\f{2\beta(g)}{(u+v_{F^2})\eps g}\right ) \right.\nn\\ 
&&\left. 
\f{\partial\log \braket{O(z)O(0)}}{\partial\log g}\right)\bigg|_{(u,v_{F^2})}
\eea
where in the last line we have employed:
\bea
\frac{\partial}{\partial\log g_0} =\frac{\partial \log g}{\partial\log g_0} \frac{\partial}{\partial\log g}
\eea
and eqs. \eqref{eq:gg0rel_uv} and \eqref{eq:Zrel_uv}.\par
We rewrite the rhs of eq. \eqref{eq:LETn2bare} by means of eq. \eqref{eq:RvsB}:
\bea\label{eq:rhs_gen}
\textrm{rhs}&=&\f{1}{2}\mu^{(u-v_{F^2})\eps}\int \braket{O(z) O(0) F^2(x)}_0 - \braket{O(z)O(0)}_0\braket{F^2(x)}_0 d^\td x \bigg|_{(u,v_{F^2})}\nn\\
&=&\f{1}{2} \mu^{(u-v_{F^2})\eps}Z_O^{-2}Z_{F^2}^{-1}\int \braket{O(z) O(0) F^2(x)}' - \braket{O(z)O(0)}\braket{F^2(x)} d^\td x \bigg|_{(u,v_{F^2})}\nn\\
&&+\f{1}{2} \mu^{(u-v_{F^2})\eps}\int \braket{O(z)O(0)F^2(x)}_{0;\textrm{c.t.}} \, d^\td x \bigg|_{(u,v_{F^2})}
\eea
where the ${}^\prime$ in the second line denotes the correlator at different points.

\subsection{LET for a generic operator $O$ with $F^2$ marginal} \label{E3}

The LET may be rewritten in terms of the conformal correlators in eq. \eqref{eq:CFT23}, where we replace all the dimensions $\Delta_{O}\to \tilde{\Delta}_{O}$ in the $(u,v_{F^2})$ regularization scheme, yielding in the lhs:
\bea
&&g\f{\partial}{\partial {g}} \braket{O(z)O(0)} \bigg|_{(u,v_{F^2})}  =   \f{N_2(g)}{|z|^{2\tD_{O_0}}} |z\mu|^{-2\gamma_O}    \left ( \f{\partial\log \mathcal{G}^{(O)}_2}{\partial\log g}-2g\f{\partial\gamma_O}{\partial g} \log{|z\mu |} \right ) + \cdots  \bigg|_{g \rightarrow g_*} \nn\\
   &&~~~~= \braket{O(z)O(0)}_{\text{conf}} \bigg|_{(u,v_{F^2})}   \left ( \f{\partial\log \mathcal{G}^{(O)}_2}{\partial\log g}-2g\f{\partial\gamma_O}{\partial g} \log{|z\mu |} \right ) +\cdots   \bigg|_{g \rightarrow g_*}
\eea
where:
\bea
\braket{O(z)O(0)}_{\text{conf}} \bigg|_{(u,v_{F^2})}=   \f{N_2(g)}{|z|^{2\tD_{O_0}}} |z\mu|^{-2\gamma_O}     \bigg|_{g \rightarrow g_*}
\eea
The LET follows by equating the lhs and rhs:
 \bea
 \label{eq:LETappE}
&&\braket{O(z)O(0)}_{\text{conf}} \bigg|_{(u,v_{F^2})}\left( 2c_O + \f{4\gamma_O(g)}{(u+v_{F^2})\eps}+\left ( 1+\f{2\beta(g)}{(u+v_{F^2})\eps g}\right )
\left( -2g\f{\partial\gamma_O}{\partial g} \log{|z\mu |} \right.\right.  \nn\\
&&\left.\left.
+ \f{\partial\log  \mathcal{G}^{(O)}_2}{\partial\log g}          \right) + \mbox{finite terms}+ \cdots \right)  \bigg|_{g \rightarrow g_*}=\braket{O(z)O(0)}_{\text{conf}} \bigg|_{(u,v_{F^2})} 
\nn\\
&&Z_{F^2}^{-1}\,
\f{1}{2}  \f{N_3(g)}{N_2(g)}
(2\pi)^{4+2u\eps} |z\mu |^{(u-v_{F^2})\eps -\gamma_{F^2}} 
C_{4+2u\eps, \f{4+(u+v_{F^2})\eps +\gamma_{F^2}}{2},\f{4+(u+v_{F^2})\eps +\gamma_{F^2}}{2}} \nn\\
&&+\f{1}{2}\mu^{(u-v_{F^2})\eps} Z_O^{2}\int \braket{O(z)O(0)F^2(x)}_{0;\textrm{c.t.}} \, d^\td x \bigg|_{(u,v_{F^2})} + \mbox{finite terms}+ \cdots  \bigg|_{g \rightarrow g_*}
\eea
with $C_{4+2u\eps, \f{4+(u+v_{F^2})\eps +\gamma_{F^2}}{2},\f{4+(u+v_{F^2})\eps +\gamma_{F^2}}{2}}$ given by eq. \eqref{eq:Cuv}. The $\eps$ expansion in the rhs of eq. \eqref{eq:LETappE}, without including the contact terms, yields:
\bea\label{eq:rhs23_eps_noct}
&&\braket{O(z)O(0)}_{\text{conf}} \bigg|_{(u,v_{F^2})}Z_{F^2}^{-1}
\f{1}{2}  \f{N_3(g)}{N_2(g)}
(2\pi)^{4+2u\eps}\,  |z\mu|^{(u-v_{F^2})\eps} C_{4+2u\eps,\f{4+(u+v_{F^2})\eps}{2},\f{4+(u+v_{F^2})\eps }{2}}  \bigg|_{g \rightarrow g_*} 
\nn\\
&&=\braket{O(z)O(0)}_{\text{conf}} \bigg|_{(u,v_{F^2})}Z_{F^2}^{-1}\f{1}{2}  \f{N_3(g)}{N_2(g)}
(2\pi)^4 \left(1+2u\eps\log(2\pi)+O(\eps^2)\right) \f{1}{16\pi^2}
\nn\\
&&~~~
\left(\f{4}{(u-v_{F^2})\eps}
-\f{u}{u-v_{F^2}}4\log(4\pi)+O(\eps) \right)
\left(1+(u-v_{F^2})\eps\log|z\mu|+O(\eps^2)\right)\bigg|_{g \rightarrow g_*}\nn\\
&&=\braket{O(z)O(0)}_{\text{conf}} \bigg|_{(u,v_{F^2})} Z_{F^2}^{-1}\f{\pi^2}{2}  \f{N_3(g)}{N_2(g)}
\bigg(\f{4}{(u-v_{F^2})\eps}+4\log|z\mu|+\f{4u}{u-v_{F^2}}\log\pi 
\nn\\&&
~~~+\cdots\bigg)  \bigg|_{g \rightarrow g_*} 
\eea
where we have employed eq. \eqref{appCC0}.
 Hence, eq. \eqref{eq:rhs23_eps_noct} is UV divergent as $\epsilon \rightarrow 0$\footnote{The integral $I_{d,\Delta_{F^2},\Delta_{F^2}}$ is IR finite for $\Delta_{F^2}>d/2$, i.e., for $\Delta_{F^2}>2$ in $d=4$.} -- with a semilocal divergence according to \cite{Skenderis} -- in $d=4$ dimensions due to the short-distance singularities at $x\sim 0$ and $x\sim z$.
Finally, by inserting eq. \eqref{eq:rhs23_eps_noct} into
eq. \eqref{eq:LETappE}, the LET reads: 
\bea\label{eq:LETgamzero}
 &&\braket{O(z)O(0)}_{\text{conf}} \bigg|_{(u,v_{F^2})}\left( 2c_O 
+\f{4\gamma_O(g)}{(u+v_{F^2})\eps}+
\left ( 1+\f{2\beta(g)}{(u+v_{F^2})\eps g}\right ) 
\left(
-2g\f{\partial\gamma_O}{\partial g} \log{|z\mu|} \right.\right. \nn\\
&&\hspace{3.0truecm}\left.\left.+ \f{\partial\log   \mathcal{G}^{(O)}_2 }{\partial\log g}\right)+ \mbox{finite terms} +\cdots
\right) \bigg|_{g \rightarrow g_*} \nn\\
&&\hspace{0.5truecm}=\braket{O(z)O(0)}_{\text{conf}} \bigg|_{(u,v_{F^2})} Z_{F^2}^{-1}
\f{\pi^2}{2}  \f{N_3(g)}{N_2(g)}
\bigg(\f{4}{(u-v_{F^2})\eps}+4\log|z\mu| 
+\f{4u}{u-v_{F^2}}\log\pi 
\nn\\
&&\hspace{0.8truecm}+\cdots\bigg)+\f{1}{2} \mu^{(u-v_{F^2})\eps} Z_O^{2}\int \braket{O(z)O(0)F^2(x)}_{0;\textrm{c.t.}}  \, d^\td x + \mbox{finite terms} +\cdots  \bigg|_{g \rightarrow g_*} \nn\\
\eea
In standard dimensional regularization, i.e., for $u=v_{F^2}=-1$, the integral in the rhs of the LET is not regularized, due to the $\frac{1}{(u-v_{F^2})\epsilon}$ singularity. Besides, the coefficient of the singularity in the rhs as $\epsilon \rightarrow 0$ is sensitive to order $\epsilon$ corrections to the canonical dimension of $F^2$.  \par
We may set $\mbox{finite terms}=0$ provided that we specialize to $u=0$, since then the correlators reduce to their form in $d=4$ dimensions
with only a deformation of the canonical dimension due to $v_{F^2} \neq 0$. Indeed, for $\gamma_{F^2}=0$ the scaling dimension of $F^2$ in the $(0,v_{F^2})$ scheme is $\tD_{F^2}=\tD_{F^2_0}=4+\eps v_{F^2}$.

\subsection{A perturbative check of the LET} \label{E5}

Therefore, we set $u=0$. Perturbatively, $F^2$ is an exactly marginal operator to order $g^0$. Moreover, after setting $\mbox{finite terms}=0$, by matching the finite logarithmic terms in the lhs and rhs of eq. \eqref{eq:LETgamzero}, it turns out that $\f{N_3(g)}{N_2(g)}$ should be of order $g^2$ so that in the rhs of the LET $F^2$ should be consistently taken to be marginal.
Hence, we get the following form of the LET that is consistent up to order $g^2$, where $F^2$ develops a nontrivial anomalous dimension $\gamma_{F^2}(g)=-2 \beta_0 g^2 + \cdots$ (appendix \ref{C}) but the beta function vanishes:
\bea\label{eq:LETgamzeroa}
 &&\braket{O(z)O(0)}\left( 2c_O 
+\f{4\gamma_O(g)}{v_{F^2}\eps}
-2g\f{\partial\gamma_O}{\partial g} \log{|z\mu|} + \f{\partial\log N_2}{\partial\log g}
\right)   \bigg|_{\textrm{Order up to } g^2}\nn\\
&&=\braket{O(z)O(0)}\f{\pi^2}{2}  \f{N_3(g)}{N_2(g)}
\bigg(-\f{4}{v_{F^2}\eps}+4\log|z\mu| \bigg)  \bigg|_{\textrm{Order up to }  g^2} \nn\\
&&\,\,\,\,\,\,\,+\f{1}{2} \mu^{-v_{F^2}\eps} Z_O^{2}\int \braket{O(z)O(0)F^2(x)}_{0;\textrm{c.t.}}  \, d^\td x   \bigg|_{\textrm{Order up to }  g^2}
\eea 
where we have set $\mathcal{G}^{(O)}_2(g)=N_2(g)$ because the theory is conformal up to order $g^2$.\par
It is instructive to analyze both the perturbative selfconsistency of the above regularization of the LET and its compatibility with the available computations in the literature.
Since no divergence arises in the lhs to order $g^0$, it must be $N_3(0)=0$ in the rhs, i.e.:
\bea \label{lo}
\braket{O(z)O(0)F^2(x)} \bigg|_{\textrm{Order }  g^0}
=0
\eea
As a consequence, only a finite contact term (section \ref{2.3}) may contribute in the rhs to compensate for the lhs. Hence, it must hold as $\epsilon \rightarrow 0$:
\bea
F^2(x) O(0) = 2c_O \, \delta^{(4)}(x) O(0) + \cdots
\eea
 Remarkably, the two equations above are satisfied for $O=F^2$ as predicted by the LET, the first one by direct computation and the second one with $c_{F^2}=2$ \cite{BB}.\par
To order $g^2$, the lhs is divergent because of the anomalous dimension $\gamma_O=-\gamma_0 g^2+\cdots$, and the divergence must be matched by the rhs, where it may originate either from a divergence of the integral of the $3$-point correlator or from a divergent contact term.\par
By direct computation for $O=F^2$ \cite{BB,Z1} the contact term proportional to a $\delta^{(4)}$ is finite to order $g^2$ in dimensional regularization. We assume that it is finite in the $(0,v_{F^2})$ scheme as well. \par
Hence, the divergence in the rhs of the LET may only arise from the
integral of the $3$-point correlator. Indeed, though $O$ and $F^2$ have developed an anomalous dimension to order $g^2$, since $\f{N_3(g)}{N_2(g)}$ is necessarily of order $g^2$ according to eq. \eqref{lo}, the corrections of order $g^2$ to the anomalous dimensions of $O$ and $F^2$ only contribute to higher orders in the rhs, so that $F^2$ stays effectively marginal in the evaluation of the rhs to order $g^2$.\par
As a consequence, the form above of the LET remains valid to order $g^2$, as we have anticipated.\par
Besides, the matching to order $g^2$ of the logarithmic terms in the lhs and rhs, which is independent of the regularization, leads to the constraint:
\be\label{eq:log0}
-2g\f{\partial\gamma_O}{\partial g} \log{|z\mu|}=4\f{\pi^2}{2}  \f{N_3(g)}{N_2(g)}\log|z\mu|
\ee
that implies:
\be\label{eq:log0_sim}
g\f{\partial\gamma_O}{\partial g}=-{\pi^2}  \f{N_3(g)}{N_2(g)}
\ee
Remarkably, we verify it for $O=F^2$ by direct computation by means of \cite{BB}. Indeed, $\gamma_{F^2}(g)=-2\beta_0 g^2 + \cdots$ and we read from the OPE of $F^2$ with itself  \cite{BB} that $N_2= \frac{48(N^2-1)}{\pi^4}$ and $N_3=N_2 \frac{4\beta_0}{\pi^2}g^2$ to their leading order respectively. As a consequence, eq. \eqref{eq:log0_sim}:
\bea
g\f{\partial\gamma_{F^2}}{\partial g}=-4\beta_0 g^2 = -\pi^2  \frac{4\beta_0}{\pi^2}g^2
\eea
is satisfied to order $g^2$.\par
Hence, substituting the lhs of eq. \eqref{eq:log0_sim} in the rhs of the LET in eq. \eqref{eq:LETgamzeroa} implies the following matching of the $1/\epsilon$ divergences to order $g^2$:
\be\label{eq:div0}
2\gamma_O(g)\bigg|_{\textrm{Order } g^2}=g\f{\partial\gamma_O}{\partial g}\bigg|_{\textrm{Order } g^2}
\ee
whose only solution is that the anomalous dimension, $\gamma_O(g)=-\gamma_0^{(O)}g^2$, is one-loop exact that is obviously consistent with our perturbative assumption to order $g^2$. Vice versa, since $\gamma_O(g)=-\gamma_0^{(O)}g^2$ in the conformal theory, no divergent contact term may arise in the $u=0$ scheme by the LET.\par
Hence, in the exactly conformal case to order $g^2$ in perturbation theory, the LET makes sense for $u=0$ and any $v_{F^2}$, with a finite contact term in the OPE of $F^2$ with itself.

\end{document}